\documentclass[useAMS,usenatbib,epsfig,epsf]{mnras}
\usepackage{times,epsfig,float}
\usepackage{natbib}
\usepackage{longtable}
\usepackage{pdflscape,hyperref}  
\usepackage[T1]{fontenc}
\usepackage{ae,aecompl}
\newcommand{\pgo}{PG\,0101$+$039}
\newcommand{\pgn}{PG\,0902$+$124}
\newcommand{\hzcnc}{HZ\,Cnc}
\newcommand{\ltcnc}{LT\,Cnc}
\newcommand{\pb}{PB\,6373}
\newcommand{\teff}{$T_{\rm eff}$}

\title[K2 subdwarf B pulsators + white dwarf companions]{K2 observations of five pulsating subdwarf-B stars with white dwarf companions}
\author[M.\,D.\,Reed et al.]{
	M.\,D.\,Reed$^{1,2}$\thanks{E-mail:MikeReed@missouristate.edu}, A.\,,S.\,Baran$^{1,2}$, J.\,H.\,Telting$^{2,3,4}$,  R.\,H.\,\O stensen$^{1,2,5}$, C.\,S.\,Jeffery$^{6}$, \newauthor Y.\,Gaibor$^{1,7}$\\
$^1$Department of Physics, Astronomy and Materials Science, Missouri State University, 901 S. National, Springfield, MO 65897, USA \\
$^2$ Ardastella Research Collaboration\\
$^3$Nordic Optical Telescope, Rambla Jos\'e Ana Fern\'andez P\'erez 7, 38711 Bre\~na Baja, Spain\\
$^4$  Department of Physics and Astronomy, Aarhus University, Ny Munkegade 120, DK-8000 Aarhus C, Denmark\\
$^5$Recogito AS, Storgaten 72, N-8200 Fauske, Norway\\
$^6$Armagh Observatory and Planetarium, College Hill, Armagh BT61 9DG, N. Ireland\\
$^7$MIT Kavli Institute for Astrophysics and Space Research, Massachusetts Institute of Technology, Massachusetts Avenue, Cambridge, \\ MA 02139\\
}

\date{Accepted
      Received }

\begin{document}

\maketitle

\begin{abstract}
We report seismic analyses of five pulsating subdwarf B (sdBV) stars observed during
\emph{Kepler's} K2 mission, each with a white dwarf companion. We find three of the five to be
	$g$-mode-dominated hybrid pulsators. For the other two, we only detect $g$ modes. We determine rotation
	periods from frequency multiplets for four stars and each rotates subsynchronously to its binary period, including 
	\pgo\, and \pgn\, both with binary periods near 0.57\,days and spin periods near 9\,days. We detect frequency multiplets in both
	$p$ and $g$ modes for \pgo\, and \ltcnc\, and determine that 
	\pgo\, rotates like a solid body while \ltcnc\, rotates
	differentially radially with the envelope spinning faster than deeper layers.  Mostly we find these five stars
	to be quite similar to one another, spectroscopically and seismically. We find the $p$ modes of the three hybrid
	pulsators to have gaps between regions of power, which we interpret as overtones and apply a technique to assign modes. 
	We examine their g mode period spacings and deviations thereof and again, find the
	stars to be similar with period spacings  near the average of 250\,s and deviations mostly under 25\,s. We compare
	\emph{Kepler}-observed sdBV stars of different binary types and likely-single pulsators.

\end{abstract}


\section{Introduction}
The \emph{Kepler} space telescope follow-on mission, K2 observed 20 fields 
in the ecliptic plane, ingeniously using solar pressure to assist with pointing stability \citep{howell14a}.
Each K2 campaign lasted 80-90\,days, producing four new pointings per year, with some overlap.
As part of our guest investigator programs, 185 subdwarf B (sdB) stars were observed during the five years of the K2
program and we estimate 44 of those to be pulsators. 

\begin{table*}
\label{tablist}
\caption{Properties of the sdBV stars analysed in this paper.}
\begin{tabular}{lcccc}
\hline\hline
Name & EPIC & Ra & Dec & K$_{\rm p}$ \\
\hline
\pgo   & 220376019 & 01:04:21.676 &     +04:13:37.06 & 12.11\\
\ltcnc & 211433013 & 09:10:25.434 & +12:08:27.08 & 14.02\\
\hzcnc & 211765471 & 08:53:23.658 & +16:49:35.24 & 14.04\\
\pgn   & 211437457 & 09:05:40.919 & +12:12:28.13 & 14.73\\
\pb    & 220188903 & 01:18:57.207 & -00:25:46.50 & 14.91\\
\hline
\end{tabular}
\end{table*}

Subdwarf B (sdB) stars are extreme horizontal branch stars that are at the blue edge of the horizontal branch.
Typical sdB stars are $\sim 0.5M_{\odot}$  with an envelope mass of $<0.01 M_{\odot}$,
likely stripped via binary evolution \citep{li2024,ge2024}.
They may pulsate in $p$ (pressure) or $g$ (gravity) modes, or both. Short-period variations,
with periods of a few minutes and amplitudes typically of a few part-per-thousand (ppt) or less are associated with
$p$ modes and longer-period variations with typical periods of a few hours and similar amplitudes are associated
with $g$ modes \citep[e.g.][]{fontaine03,charpinet00}. \emph{Kepler} and K2 data have observationally ``solved'' the
seismic properties of pulsating sdB (sdBV) stars in that $\approx 80$\% of periodicities could be associated with
pulsation modes, quantized as $n$ the radial overtone, $\ell$ the non-radial degree, and $m$ the azimuthal component(s).
The $g$-mode periodicities satisfy the asymptotic relation $n\gg\ell$ and therefore show nearly-even period spacings
as $P_{\ell ,n}\,=\,\frac{P_o}{\sqrt{\ell\left(\ell +1\right)}}n+\epsilon$ where $P_o$ and $\epsilon$ are constants 
with units of time \citep{bible,reed14}. The $\ell=1$ overtone spacing for sdBV stars is typically near 250\,s and
the $\ell=2$ overtone spacing is near 150\,s. Frequency splittings caused by rotation are also a useful tool for
associating periodicities with pulsation modes. The azimuthal $m$ index has $2\ell\,+\,1$ values ranging from $-\ell$
to $\ell$ in integer steps that separate as $\Delta\nu\,=\,\Delta m\Omega\left(1\,-\,C_{n,\ell}\right)$ with
$C_{n,\ell}\,\approx\,\frac{1}{\ell\left(\ell\,+\,1\right)}$ for $g$ modes and near zero for $p$ modes \citep{led51,charpinet01}.
For a review of sdB and sdO stars, see \citet{heber16} and for a review of sdBV pulsations
see \citet{ReedF7} and \citet{uzu24}. 

In this paper we complete analyses of sdBV stars observed during K2 which have binary companions we interpret
as being white dwarfs. Properties of the five stars are provided in Table\,\ref{tablist}.

\begin{figure*}
\centerline{\psfig{figure=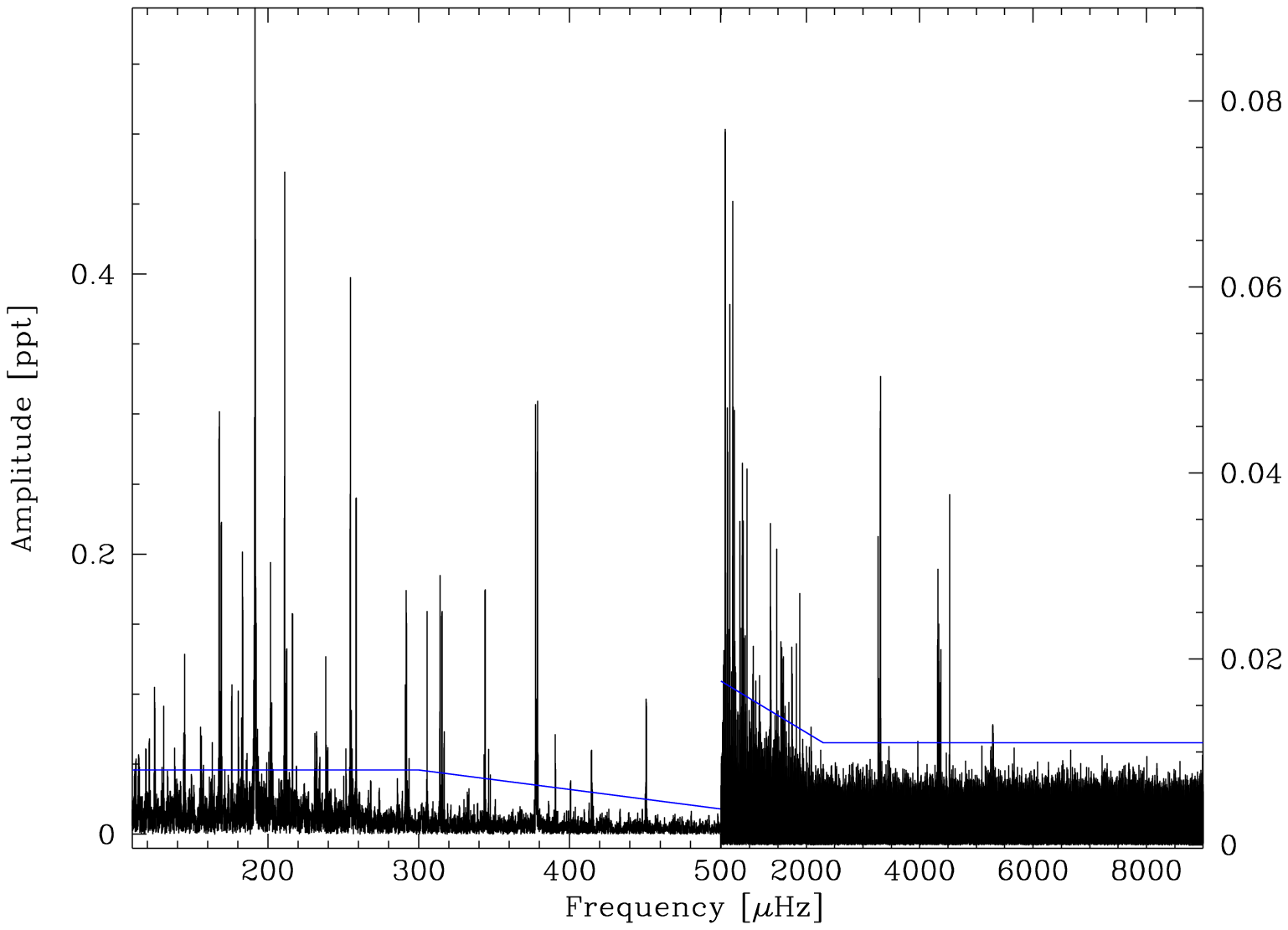,angle=-90,width=\textwidth}}
        \caption{Fourier transform (FT) of \pgo. The blue horizontal line indicates the detection limit.
        Note that the amplitude changes at 500\,$\mu$Hz to better show the higher-frequency pulations. Amplitudes
        are indicated on either side.}
    \label{pgoFT}
\end{figure*}

The following targets are included, in order of decreasing brightness:

EPIC\,220376019 (PG\,0101+039, Feige\,11)
is a well studied sdB star, and the second star recognized to be a $g$-mode variable \citep{green03}. It was detected to be in a binary system with a WD companion with an orbit of 0.57\,days by \citet{moran99}. It became the first sdB star to be observed from space when \citet{randall05} obtained 400\,h of photometry with the MOST satellite. They detected three $g$-mode pulsation frequencies as well as the first two harmonics of the orbital period.

EPIC\,211433013 (PG\,0907+123, LT\,Cnc) was recognized as a $g$-mode pulsator by \citet{koen10}. It was also known to be in a binary with a period of 6.1\,days \citep{morales-rueda03}.

EPIC\,211765471 (PG\,0850+170, HZ\,Cnc) is another star from the original set of $g$-mode pulsators presented in the discovery paper by \citet{green03}. It was also known to be a binary with a period of 27.8\,days from the study of \citet{morales-rueda03}.

EPIC\,211437457 (PG\,0902+124) is less well studied. Although included in the sample of \citet{vennes11} it was 
not recognized as a binary and was not previously known to be a pulsator.

EPIC\,220188903 (PB\,6373) was included in the RV survey of \citet{kupfer15} who found an orbital period of 
1.30\,days and an RV amplitude of 54.8\,$\pm$\,2.9 km s$^{-1}$. It was not previously known to be a pulsator.


\section{K2 observations and analysis}

All data were downloaded from MAST\footnote{Mikulski Archive for Space Telescopes \url{https://archive.stsci.edu/}}
as either pixel files, as EVEREST \citep[EPIC Variability Extraction and Removal for Exoplanet Science Targets;][]{ev1,ev2}
processed light curves, or both. For \pgo, \ltcnc, \hzcnc\, and \pgn, data are short cadence (SC, 59\,s) while \pb\,
only has long-cadence (LC, 30\,m) data. Only two reaction wheels were operating during the K2 mission so there was
drift with correctional thruster firings every six hours \citep{howell14a}. EVEREST light curves were processed to
correct those artefacts or pixel data had to be processed accordingly. For pixel data,
fluxes were extracted using
aperture photometry and spacecraft artefacts were removed using our
custom process described in \citet{baran16d,ketzerF2}. We then produce temporal spectra (Fourier transforms, FTs)
and sliding FTs (SFTs) to examine the pulsations. We calculate the frequency resolution as $1.5/T$ where $T$ is
the duration of the observations which is near to $0.15\,\mu$Hz. We determine the detection limit for a K2 
campaign as $4.2\sigma$ where $\sigma$ is
the average of the FT in the region of interest \citep[e.g.][]{ReedPG1142}. We also examined 
 TESS \citep[the Transiting Exoplanet Survey Telescope,]{tess1} data which has a single-sector resolution of about $0.42\,\mu$Hz.

\begin{figure*}
        \centerline{\psfig{figure=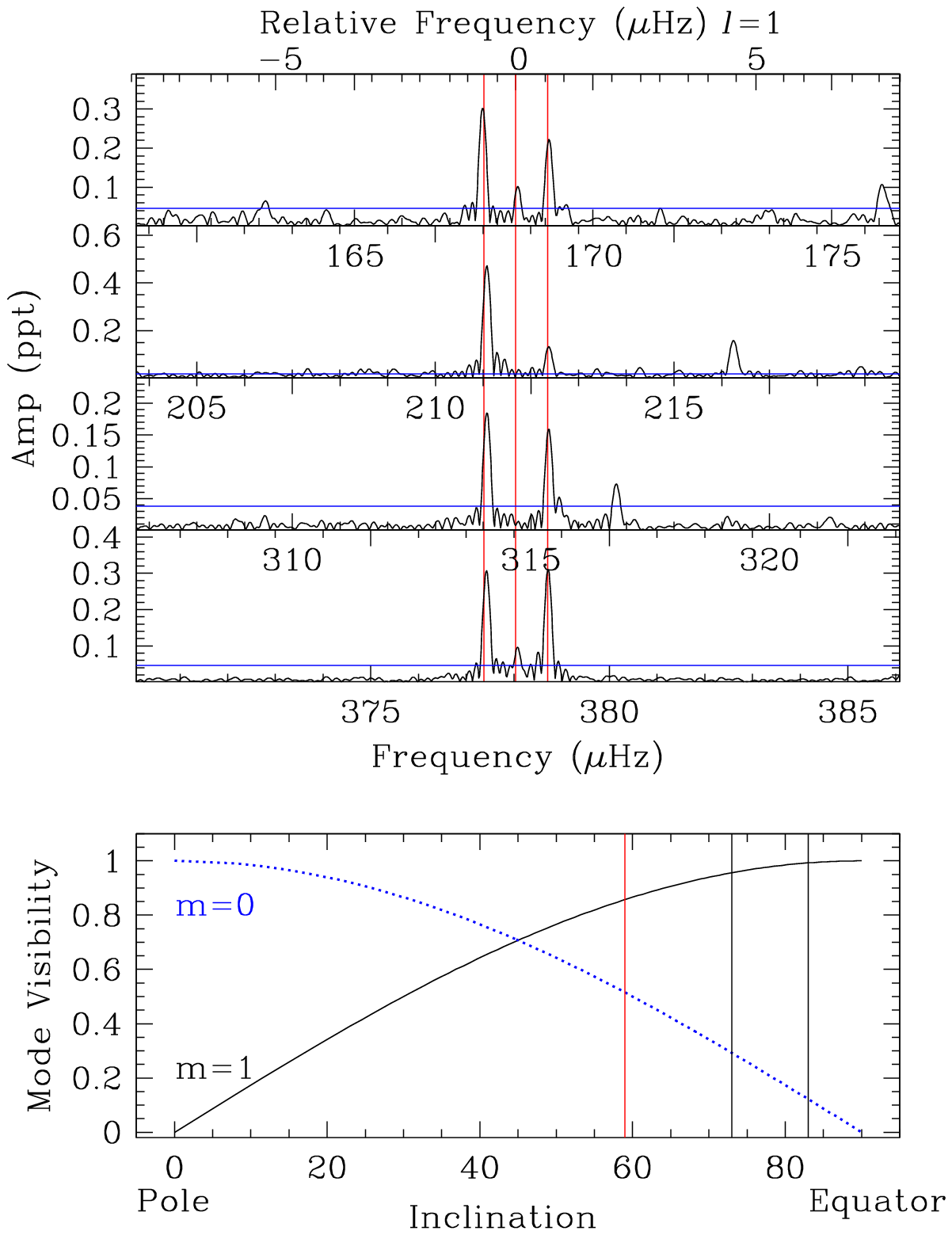,width=3.0in},\psfig{figure=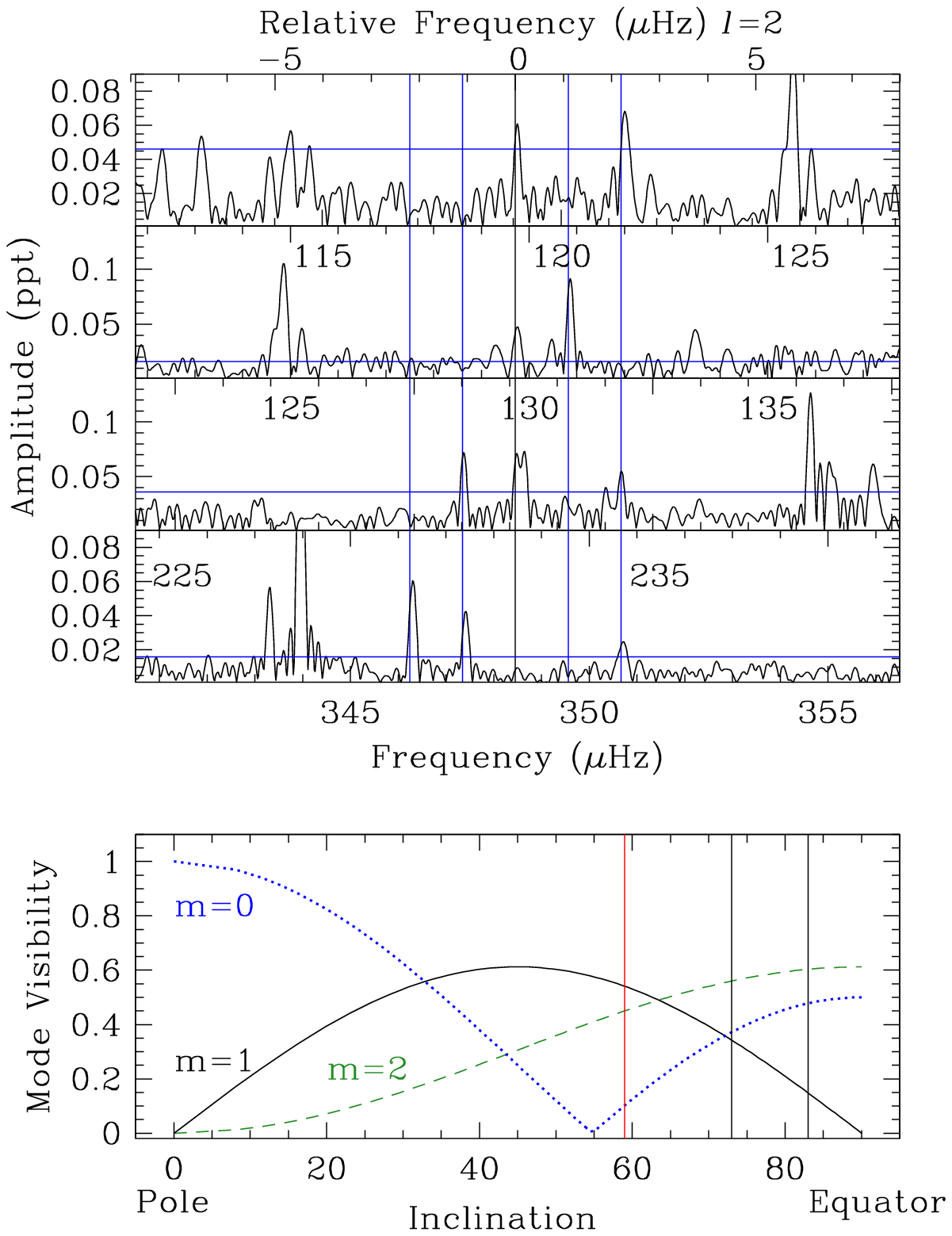,width=3.0in}}
        \centerline{\psfig{figure=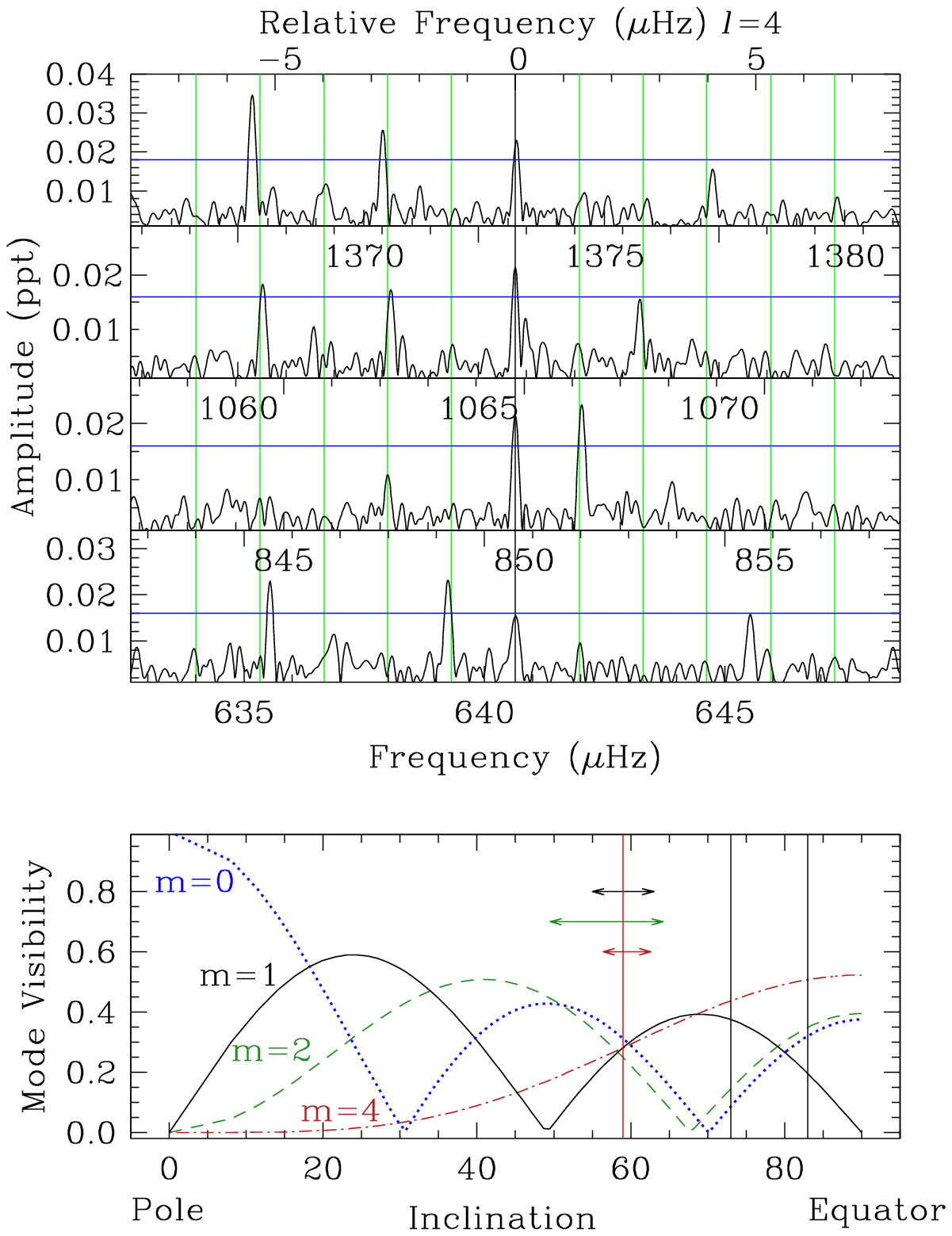,width=3.0in},\psfig{figure=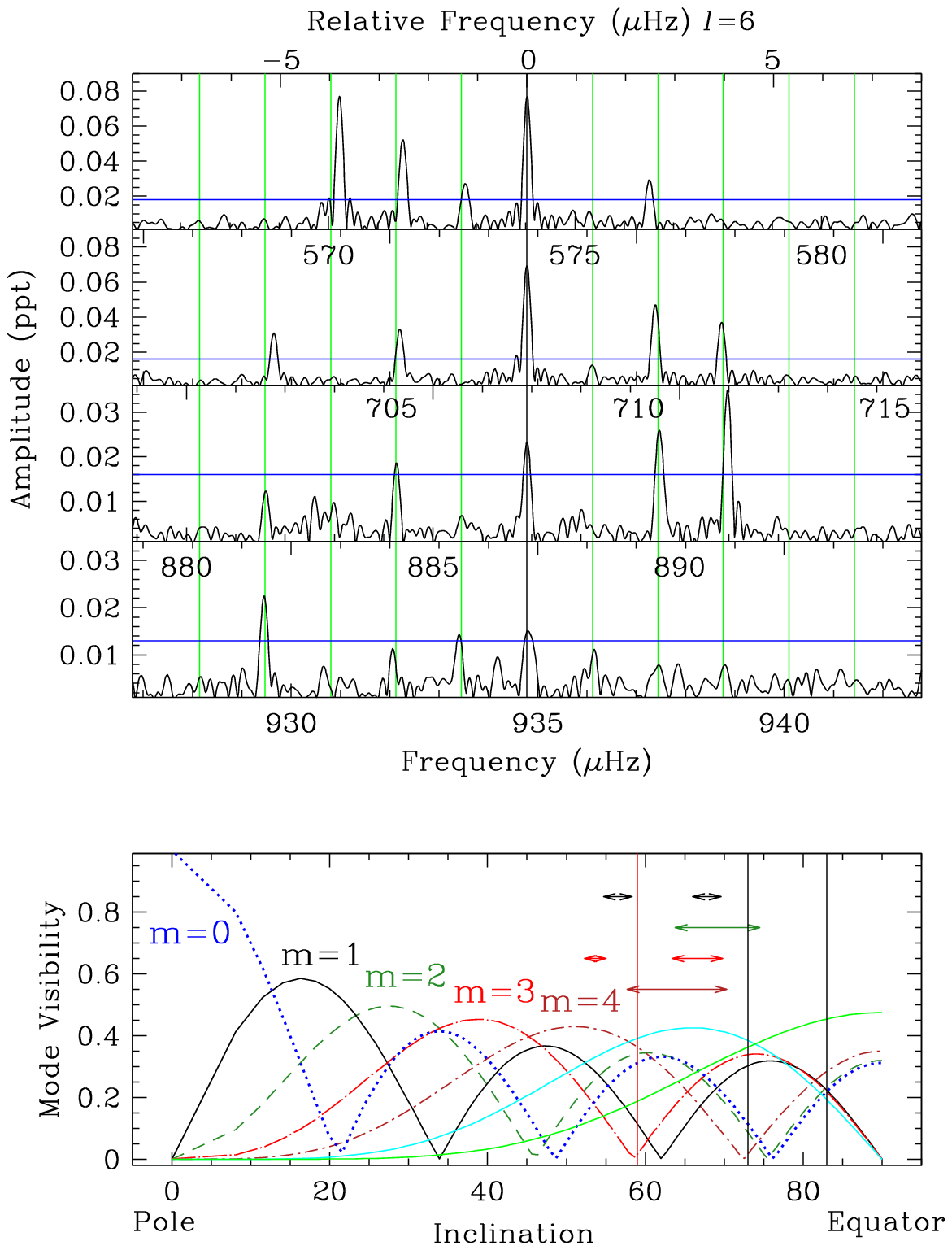,width=3.0in}}
	\caption{Inclination from multiplets of \pgo. Four examples of identified multiplets for $\ell=1$ (top left), 2 (top right),
	4 (bottom left), and 6 (bottom right) with their amplitude-inclination ratios in their bottom panels. 
In the  upper panels vertical lines indicate multiplet spacings for the panel degree.
	In the bottom panels the vertical lines indicate the $\ell=1$ inclination limits based on multiplet amplitudes (full triplets in black,
	doublets lower limit in red). Inclination lines are labeled and arrows indicate allowed inclinatinos for $\ell>1$ (color-coded 
	for $m$ value). } 
        \label{pgoInc}
\end{figure*}

\subsection{\pgo}
\pgo\ was observed during campaign 8 (3 Jan. -- 23 Mar, 2016).
\citet{ma23} had already published an analysis of these data. We include our own results for comparison.
Though we used the same Target Pixel Files, they
were processed using different software. \citet{ma23} used Lightkurve \citep{LK} and KEPSFF \citep{kepsff} whereas
we used our own custom software for flux extraction and pixel-thruster decorrelation \citep{baran16d,ketzerF2}.
The differences in processing led to detecting different pulsations at the low amplitude level; they found some
we did not, and we found some they did not. We provide our seismic findings in the appendix in Tables\,\ref{tabpgog}
and \ref{tabpgop} and show the FT in Fig.\,\ref{pgoFT}. For pulsations
detected in \citet{ma23} we use their IDs and for those not detected in \citet{ma23} we identify them with a
letter rather than a number.
We also examined TESS (TIC\,344719037) observations
obtained during Sector 42 (12 Oct -- 6 Nov., 2021) in 20\,s (USC) cadence.
In those data we detect only three frequencies: the 2016 highest-amplitude frequency, f1, f13 which is now the second
highest amplitude pulsation and f6 which is now the third highest amplitude pulsation. So while some of
the amplitudes have changed over the nearly five year difference in time, the highest-amplitude pulsation
has remained so.

In a comparison of our findings for \pgo\ to those of \citet{ma23}, we reach essentially the same conclusions
for the bulk seismic properties. They found a period spacing of $249.2\pm1.5$\,s and we find a spacing of 
$254.22\pm0.65$\,s, they find rotation periods of $8.81\pm0.6$ and $8.60\pm0.16$ days using $g$ and $p$-mode
multiplets, respectively, while we find $8.97^{+0.50}_{-0.45}$ and $8.82^{+0.40}_{-0.36}$ days. We infer that 
\pgo\, rotates like a solid body while \citet{ma23} gives a 60\% chance that the envelope spins just 
slightly faster.
While \citet{ma23} also include very interesting AM/FM (amplitude \& frequency modulation) analyses which
we do not, we have additions to our analyses that they do not. We use amplitude clues as part of our mode
assignments and to infer inclination angle 
and we do not infer combination frequencies. For the latter, it has been our experience that
combination frequencies only occur at exact frequencies of combinations from high-amplitude (relative for
the star)
pulsation frequencies. We do not think those conditions have been met. This creates no substantial changes
in results between \citet{ma23} and ourselves.

\subsubsection{$g$-mode regions}
The best $\ell=1$ multiplets are the triplets f7-f27-f9 and f6-f25-f5 and as the IDs were organized by \citet{ma23}
by decreasing amplitude, the $m=0$ components have lower amplitudes than
the $m=\pm1$ components (Fig.\,\ref{pgoInc}). 
Then there are several doublets, spaced appropriately for $\ell=1,\, \Delta m=2$
(e.g. f12-f14, f2-f19, f51-f59, and f32-fJ), which is further evidence that the $m=0$ component is
suppressed in amplitude compared to the $\pm$ components. As such, whenever we see a doublet
that looks like an $\ell=1,\Delta m=1$, we presume the lower amplitude component is $m=0$. Higher-degree 
odd $\ell$ modes should similarly have the $m=0$ component's amplitude suppressed whereas the 
even $\ell$ modes will have the $\pm1$ components suppressed \citep[see fig.\,A.5 of][]{charpinet11b}, 
as shown in Fig.\,\ref{pgoInc}.

\subsubsection{$p$-mode region}
We divide our $g$ and $p$ mode regions between 1068.35 (f180) and 1367.67$\mu$Hz (f82).
\citet{ma23} lists pulsations as \emph{mixed} from 1367.67 through 1888.43$\mu$Hz (f98)
and that is probably reasonable. They list f57-f167 as an $\ell=4$ multiplet,
however as f57 is the highest-amplitude pulsation in the $p-$mode region and somewhat
separated from the remaining multiplet members, we presume it is a
radial mode and the remaining members (f128, f177, and f167) are an $\ell=1$ triplet. Again,
this difference between us and \citet{ma23} is observationally arbitrary and only a detailed seismic
model would be able to discern which interpretation is correct. But we feel it is important to provide
our interpretation so that when models are sufficiently mature to distinguish between them,
modelers have both interpretations to work with.
We note that beginning at 3269.90$\mu$Hz (f85), the pulsations
occur in groups separated by roughly $1\,000\mu$Hz and we interpret this as overtone
spacings. That will be discussed in \S 4.1.

\begin{figure*}
\centerline{\psfig{figure=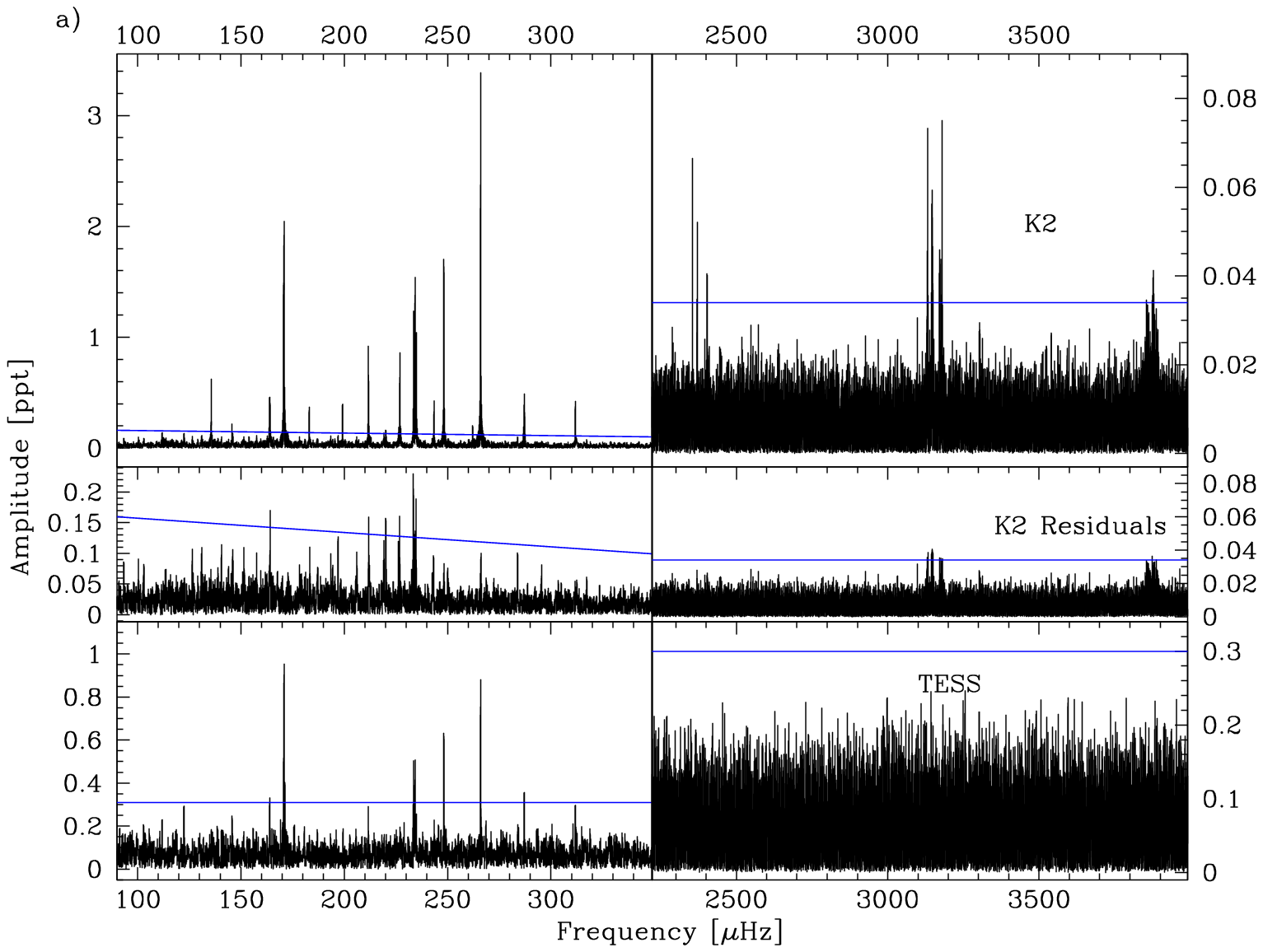,angle=-90,width=5.5 in}}
\centerline{\psfig{figure=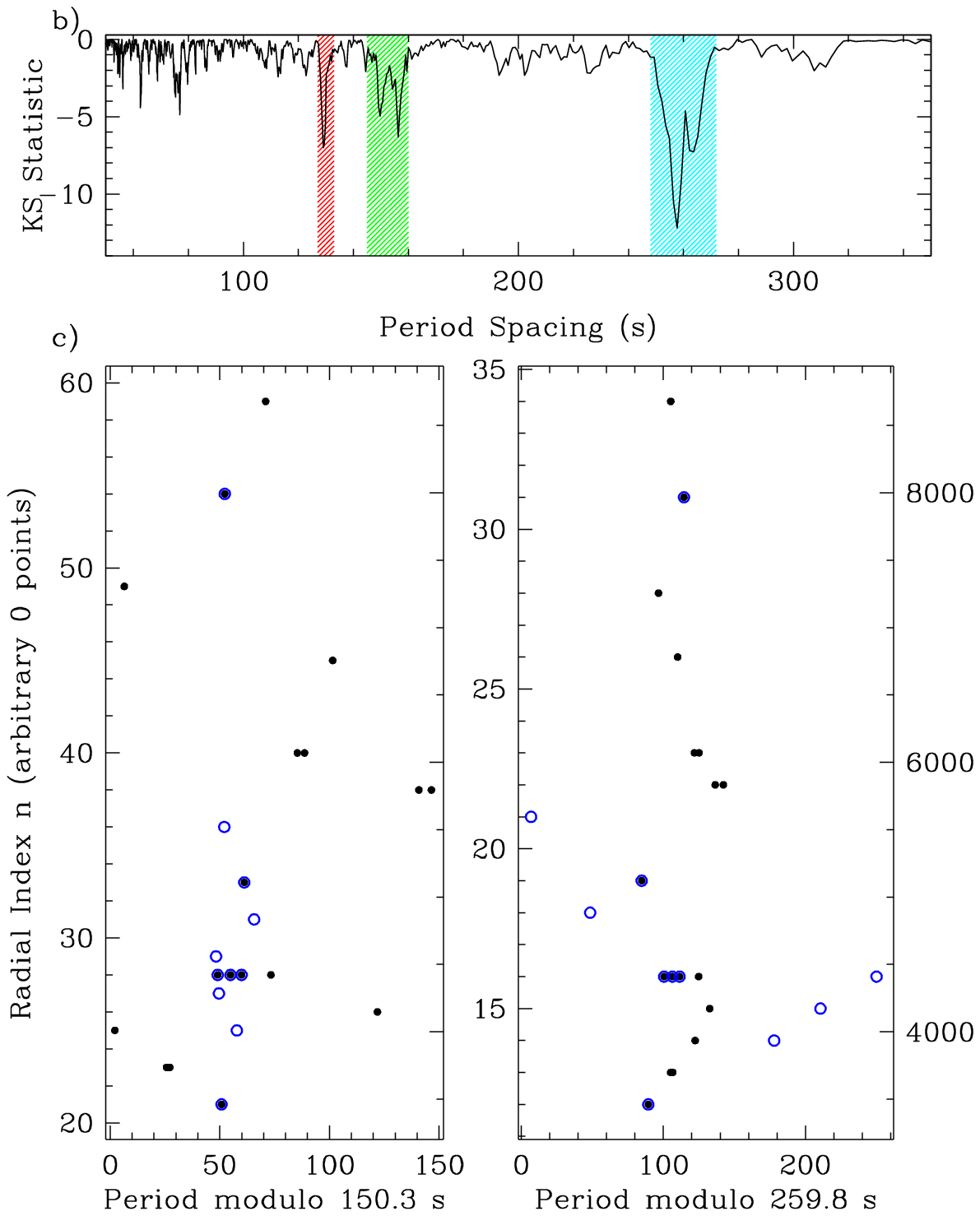,angle=0,width=\columnwidth}\psfig{figure=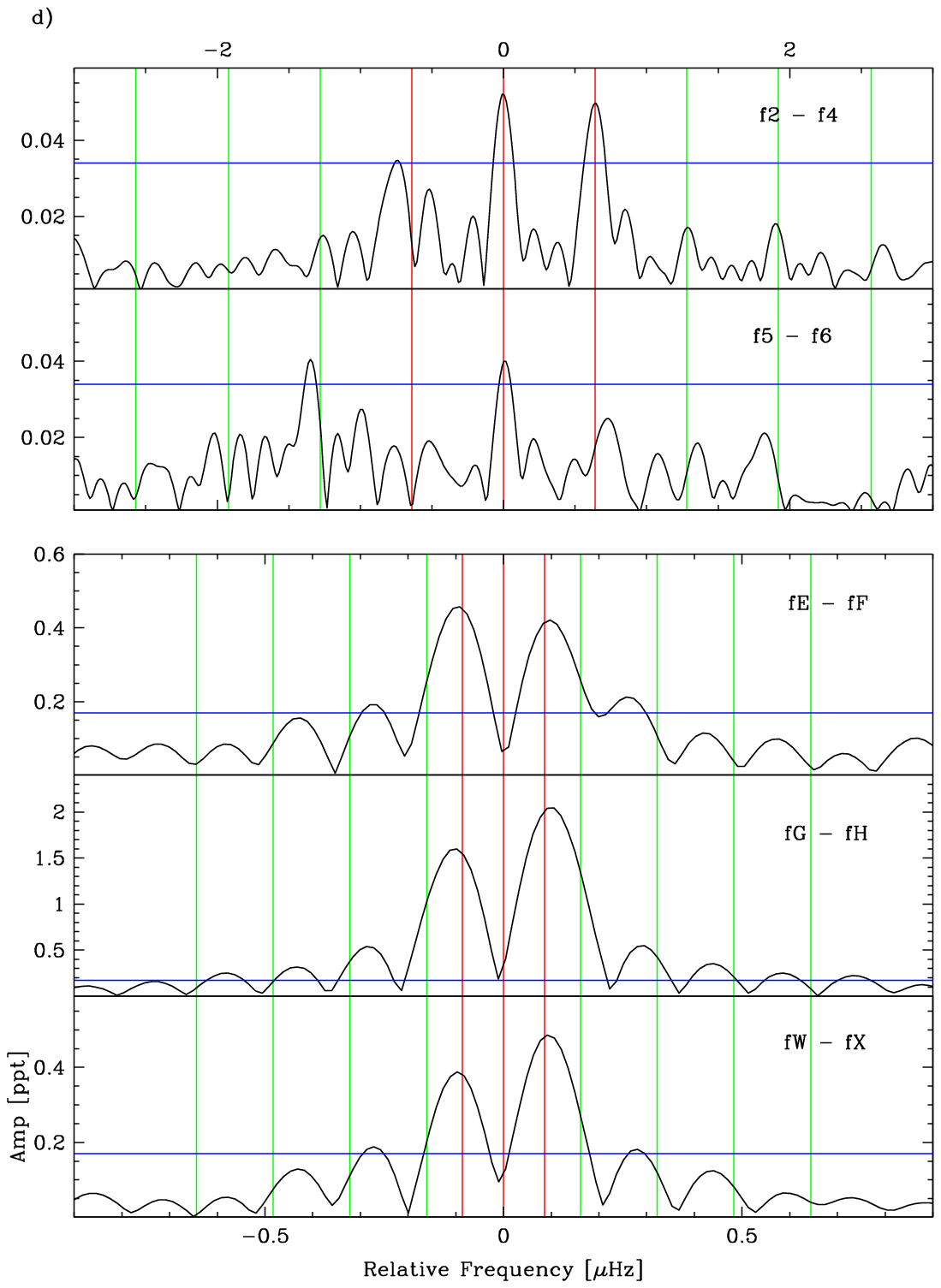,angle=0,width=\columnwidth}}
\caption{a) FT of \ltcnc\, with lower and higher frequencies plotted
at different amplitude scales. Horizontal blue lines indicate detection limits.
b) KS test with shaded regions indicating $\ell =1$ (cyan), 2 (green) spacings and the  $\ell =1$ overtone
(red). c) \'echelle diagrams for \ltcnc\, with black dots indicating $\ell =1$ modes and open blue
circles indicating  $\ell =2$ modes. d) Frequency multiplets for p modes
        (top two panels) and g modes (bottom two panels). $\ell =1$ (2) splittings are indicated
        with red (green) lines.}
    \label{ltFT}
\end{figure*}

\subsubsection{Pulsation inclination}
As we see an amplitude pattern in the frequency multiplets, we can use that to determine
an inclination. \emph{If} all multiplet members' amplitudes are intrinsically excited to the same
amplitude, then observed amplitudes are only affected by the inclination angle of the
pulsation axis \citep{bible}. While this has not been common in sdBV stars, it has previously been
observed \citep{charpinet11b,reed14,kern18}. We use this to estimate the inclination as shown in Fig.\,\ref{pgoInc}.
Sample multiplets are shown in the top panels from which amplitude ratios are used to determine inclination limits as shown
in the bottom panels. Normalized amplitudes with inclination-dependence are shown as curved lines with vertical lines 
indicating limits based on $\ell =1$ modes and arrows indicating limits for $\ell =4$ and 6 modes. $\ell =2$
multiplets do not provide constraints. For the most reliable $\ell=1$ (2) triplets the allowed inclination range
is $73\leq i\leq83^o$. If we add in all the $\ell =1$ doublets (4) the lower limit reduces to $59^o$.
The $\ell =4$ multiplets indicate an inclination of $56\leq i\leq 63^o$ (where all the arrows overlap) 
and that range is $66\leq i \leq 70^o$ for the $\ell =6$ multiplets, so those two sets are close to each other, but
not in complete agreement. From our results, we would predict an
inclination angle of roughly $60\leq i\leq70^o$. \citet{schaf23} measured
the binary inclination angle to be $89.4\pm0.6^o$ which does not agree with our multiplet results. It is
possible the pulsation and rotation axes do not align, yet we note that \citet{schaf23} also find a
rotation period of $0.85\pm0.09$\,days which is well short of the seismic rotation period near 9\,days.

\begin{figure*}
	\centerline{\psfig{figure=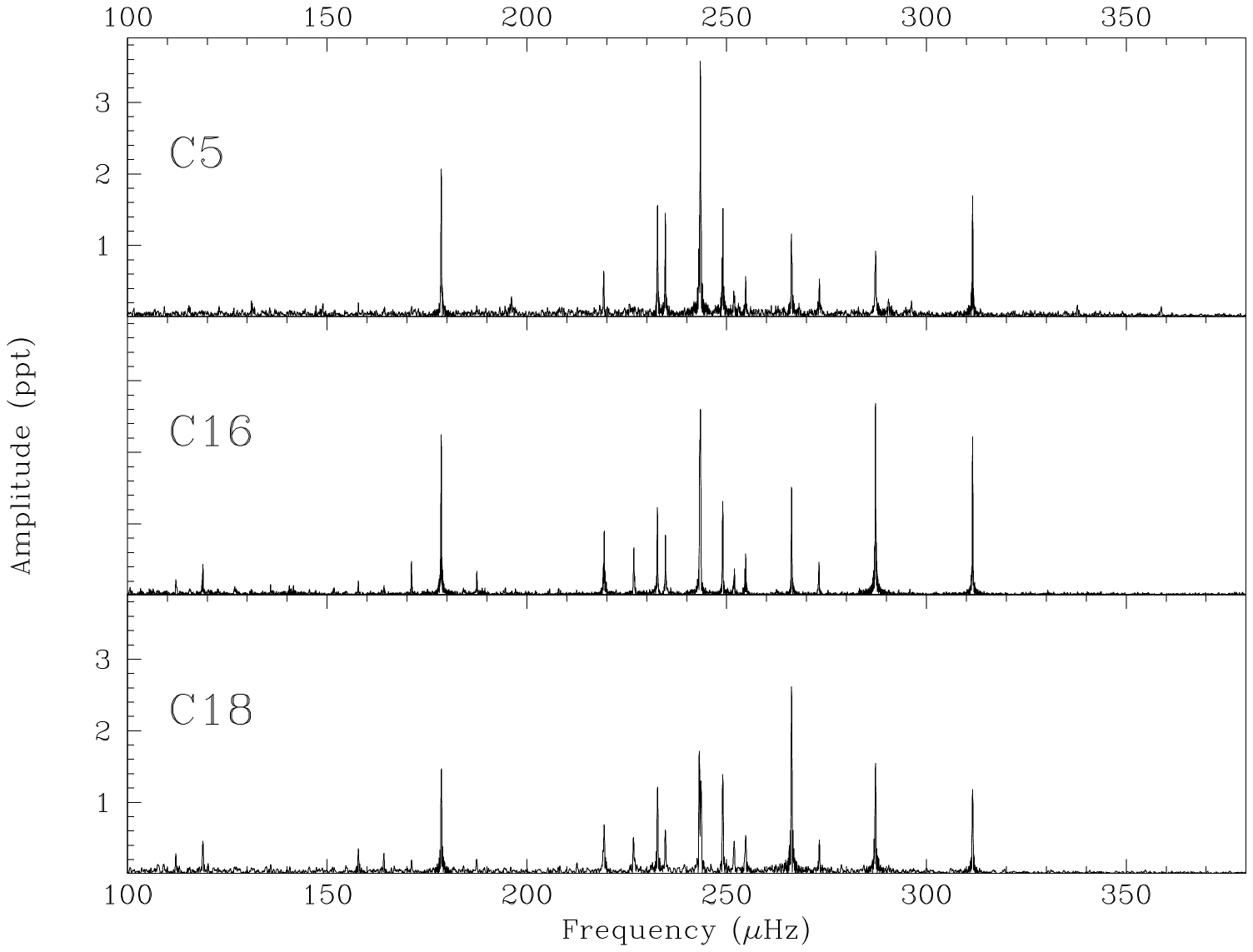,angle=-90,width=\columnwidth} \psfig{figure=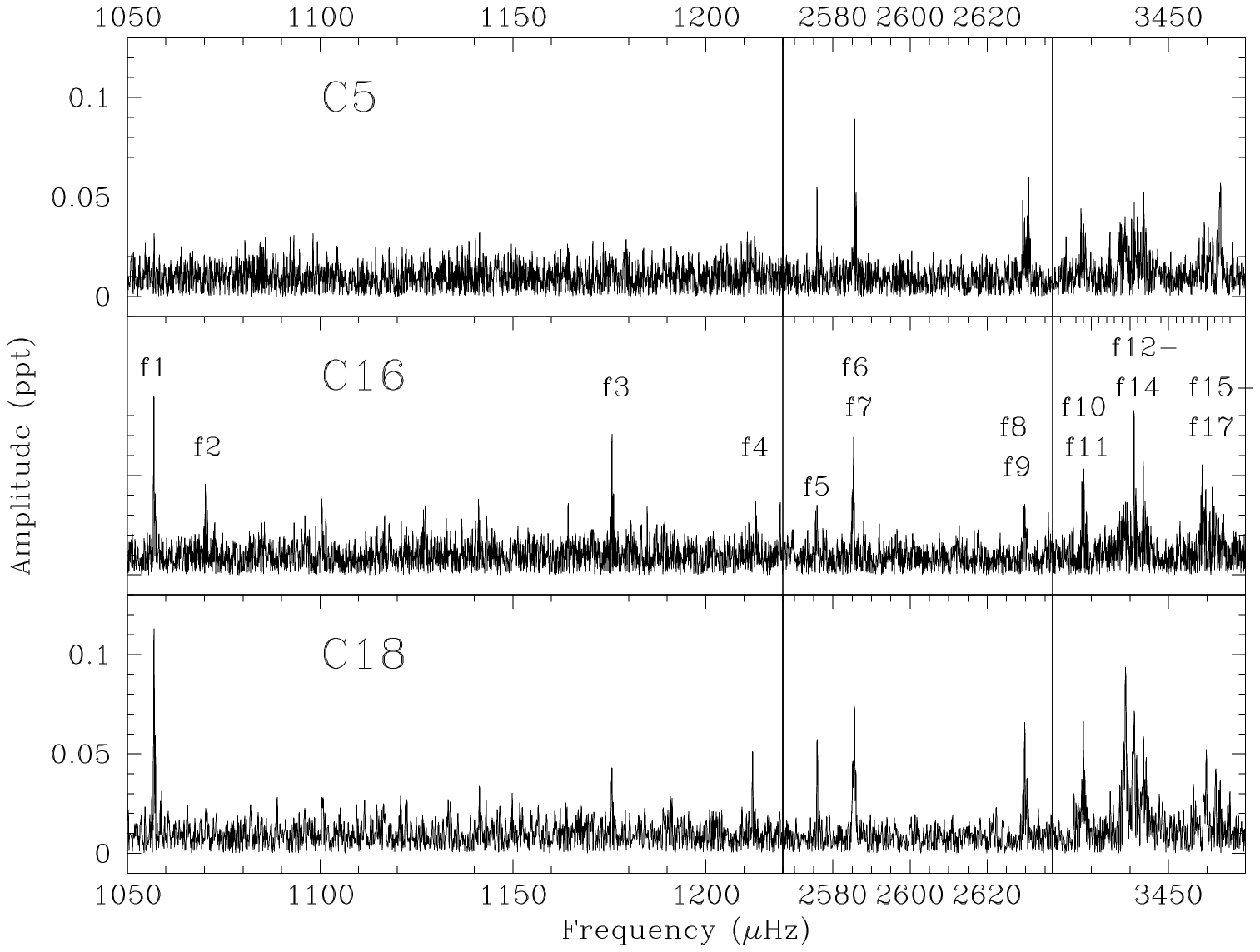,angle=-90,width=\columnwidth} }
        \caption{FT of \hzcnc\, for each K2 Campaign.}
    \label{hzFTg}
\end{figure*}

\subsection{\ltcnc}
\ltcnc\ was observed during K2's campaign 16 (7 Dec, 2017 -- 25 Feb. 2018). It is
a rich pulsator in both $p$ and $g$ mode regions. In total we detected 44 pulsations with
23 in a low-frequency region which we interpret as $g$ modes and 21 in a
high-frequency region which we interpret as $p$ modes (Tables\,\ref{tabltcncg}
and \ref{tabltcncp}). The low-frequency pulsations have
much higher amplitudes (see Fig.\,\ref{ltFT}a), so we consider it as a $g$-mode-dominated 
sdBV star. We determined detection limits in regions avoiding pulsations making a slight
decreasing slope across the low-frequency region. Prewhitening easily fitted, but did not
completely remove peaks in the FT, indicating some amplitude and/or frequency variations
during the observations. Another possibility would be unresolved, low-amplitude pulsations. 

TESS observed \ltcnc\, (TIC\,321287961) during Sectors 44 - 46 (12 Oct-–30 Dec, 2021)
in 20\,s cadence and those data show pulsations, though only
in the $g$-mode region. No new frequencies are detected. Eight K2 periodicities are 
recovered above the 0.31\,ppt detection threshold with
another five K2 periodicities detected below the detection threshold. Most of the amplitudes, relative
to each other are the same as in the K2 data (bottom panel Fig.\,\ref{ltFT}a), 
though the data are separated by nearly three years.
It appears the amplitude of fV has gone down slightly while that of fH has increased to have
the highest amplitude during the TESS observations. Since both of those have nearby frequencies,
it is possible the amplitude changes are related to unresolved pulsations.

A  Kolmogorov-Smirnov (KS) test (Fig.\,\ref{ltFT}b) finds a 
'normal' asymptotic $\ell =1$ period spacing sequence in 78\% of the periods
with a
linear regression fit of $\Delta P_{\ell =1}=259.77\pm 0.51$\,s. The $\ell =1$
sequence includes 11 periods, with 
six that fit both the $\ell =1$ and 2 sequences. The resulting $\ell =2$ linear-regression fit
of  $\Delta P_{\ell =2}=150.28\pm 0.22$\,s. \'Echelle diagrams are shown in  Fig.\,\ref{ltFT}c)
and seismic properties of the low-frequency
region are provided in Table\,\ref{tabltcncg}. Periods which could fit either sequence
are listed as such.

The high-frequency $p$ mode region contains frequency splittings we interpret as multiplets.
f2-f4 make a triplet and f5-f6 a doublet at twice the splitting.
f9 through f12's splittings are more complex as they appear to have $\Delta m=2$, but
even so they are a little large. f13 through f18 show similar complexities. f14-f15
are split by $0.62\mu$Hz, but f15-f17 would require  $\Delta m>2$. The average
of the splittings, with appropriate $\Delta m$s, is $0.64\pm0.08\mu$Hz. Assuming
a Ledoux constant $C_{n,\ell}$ of zero would give a rotation period of $18.0^{+2.1}_{-2.7}$\,days.

In the $g$ mode region, the $\ell=1$ splittings of fE-fF, fG-fH, and fW-fX average
to $0.160\pm0.037\,\mu$Hz and the $\ell=2$ triplet fP-fR average to $0.301\pm0.032\,\mu$Hz.
Using appropriate Ledoux constants, the $\ell=1$ and 2 multiplets give the same rotation
period, within the uncertainties, and this averages to $35.5\pm7.5\,$days.

Since we have both $g$- and $p$-mode multiplets (samples shown in Fig.\,\ref{ltFT}d), we can examine radially-differential
rotation as detected by \citet{mdr15}. It has been established that $g$ modes probe deeper
into the star than $p$ modes, with the latter being envelope modes \citep{charp14}.
As such, the $p$-mode multiplets indicate a spin period of $18.0^{+2.1}_{-2.7}$\,days
for the envelope and the $g$-mode multiplets indicate a spin period of
$35.5\pm7.5$\,days for the deep interior of \ltcnc . Additionally, \ltcnc\,
can be added to the list of subsynchronously-rotating sdB+WD binaries ($P_{\rm orbit}=6.1$\,days).

The $p$ modes in \ltcnc\, appear similar in spacing to the sdBV star
V585\,Peg \citep{mdr23}. The shortest-frequency group has a singlet
at the highest amplitude, a triplet, and another doublet, which we interpret as $\ell=$ 0, 1, and 2, respectively. 
Then there is a clear overtone gap, a
group with a lot of frequencies but no clear mode identifications, and then another overtone gap and a small group of
frequencies. Just like \pgo\, we can estimate radial overtone indices $n$ and include those in Table\,\ref{tabltcncp}.
They will be discussed further in \S 4.1.

\subsection{\hzcnc}
\hzcnc\ has the distinction of being the only sdBV star observed during three K2 campaigns. It was observed
during campaigns 5 (27 April -- 10 July, 2015), 16 (7 Dec, 2017 -- 25 Feb. 2018) and 18 (12 May -- 2 July, 2018).
\hzcnc\, was also observed by TESS during Sectors 44, 45, and 46 (12 Oct--30 Dec, 2021)
in 20\,s cadence.

\begin{figure*}
    \centerline{\psfig{figure=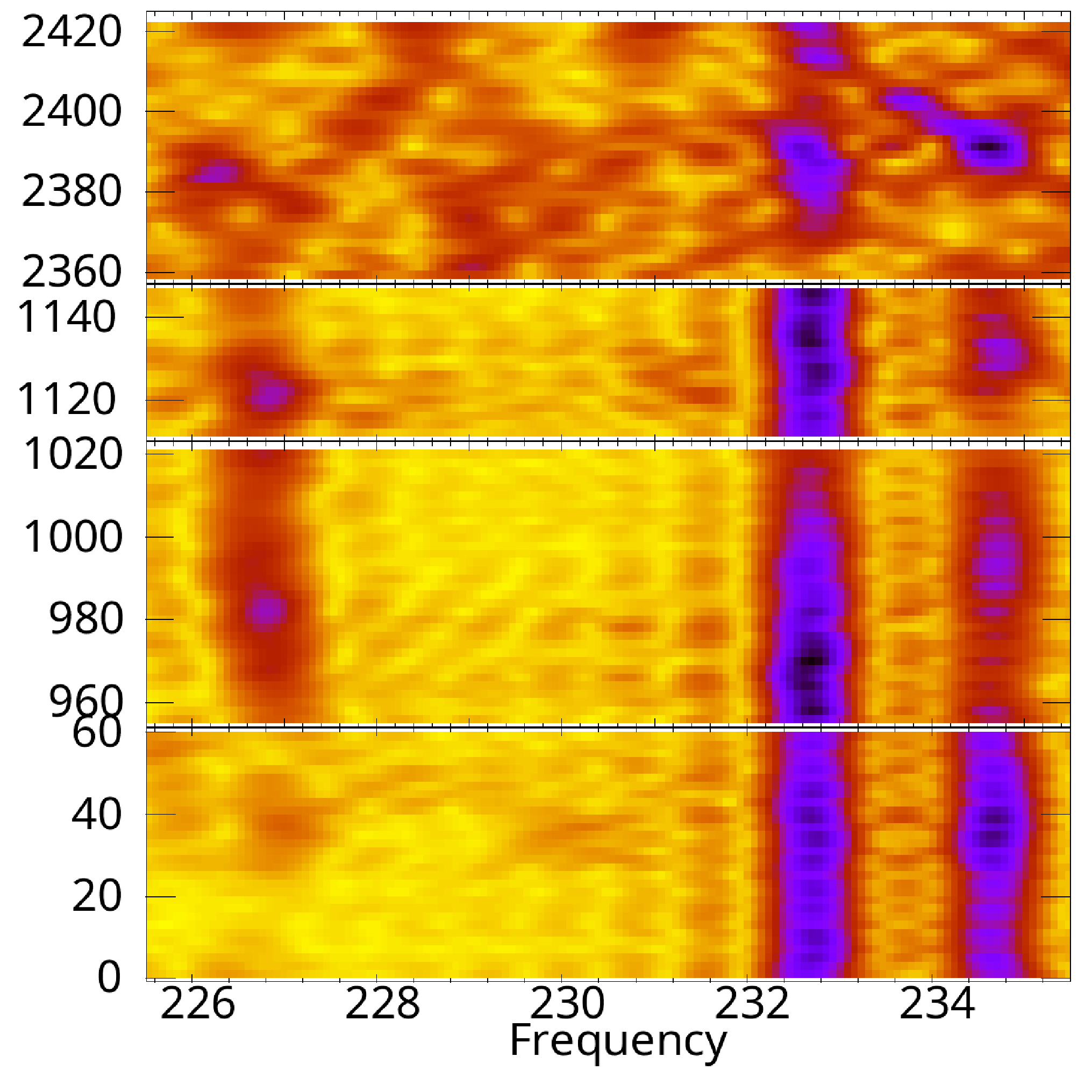,width=2.0in}\psfig{figure=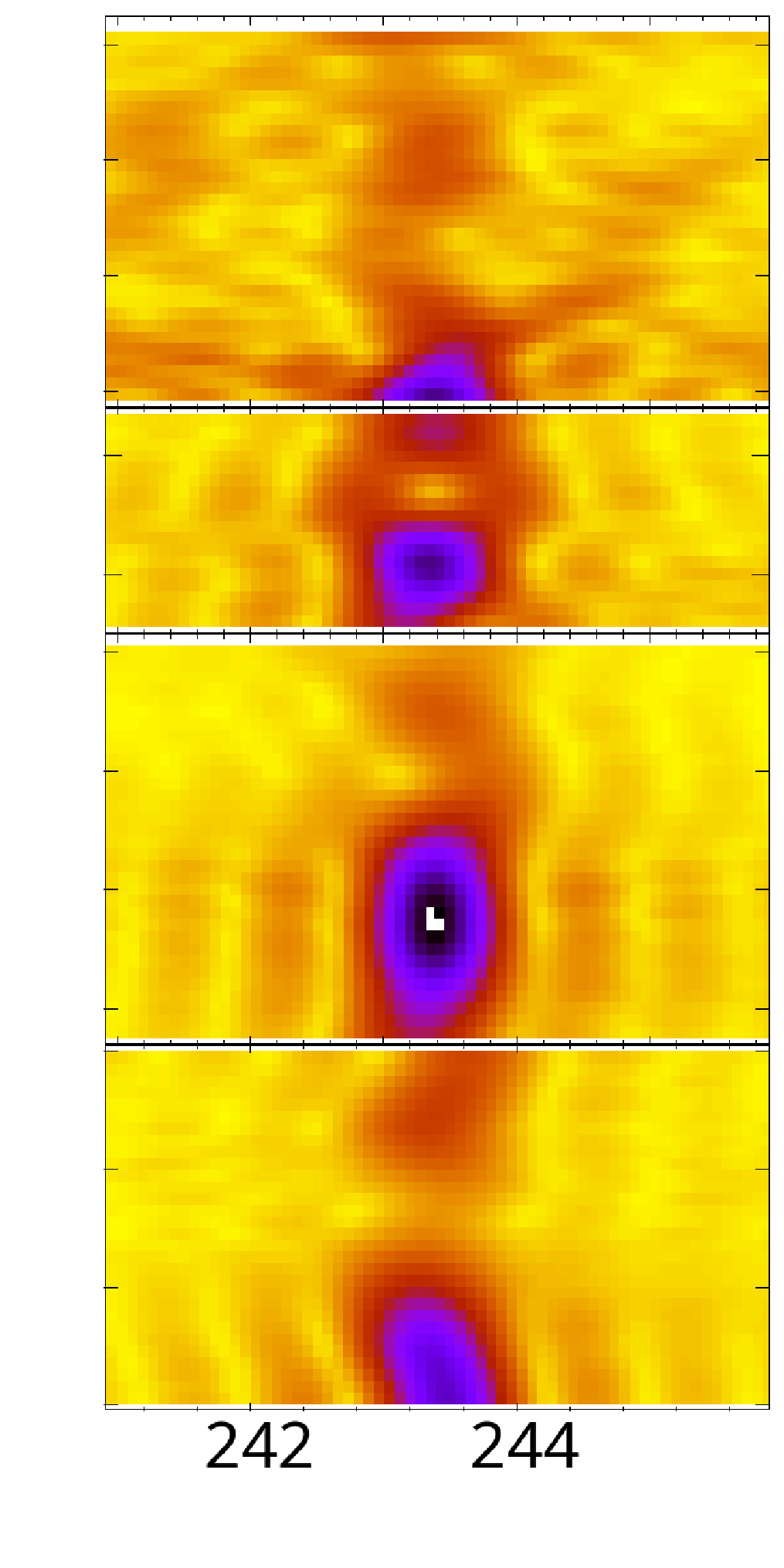,width=1.0in}\psfig{figure=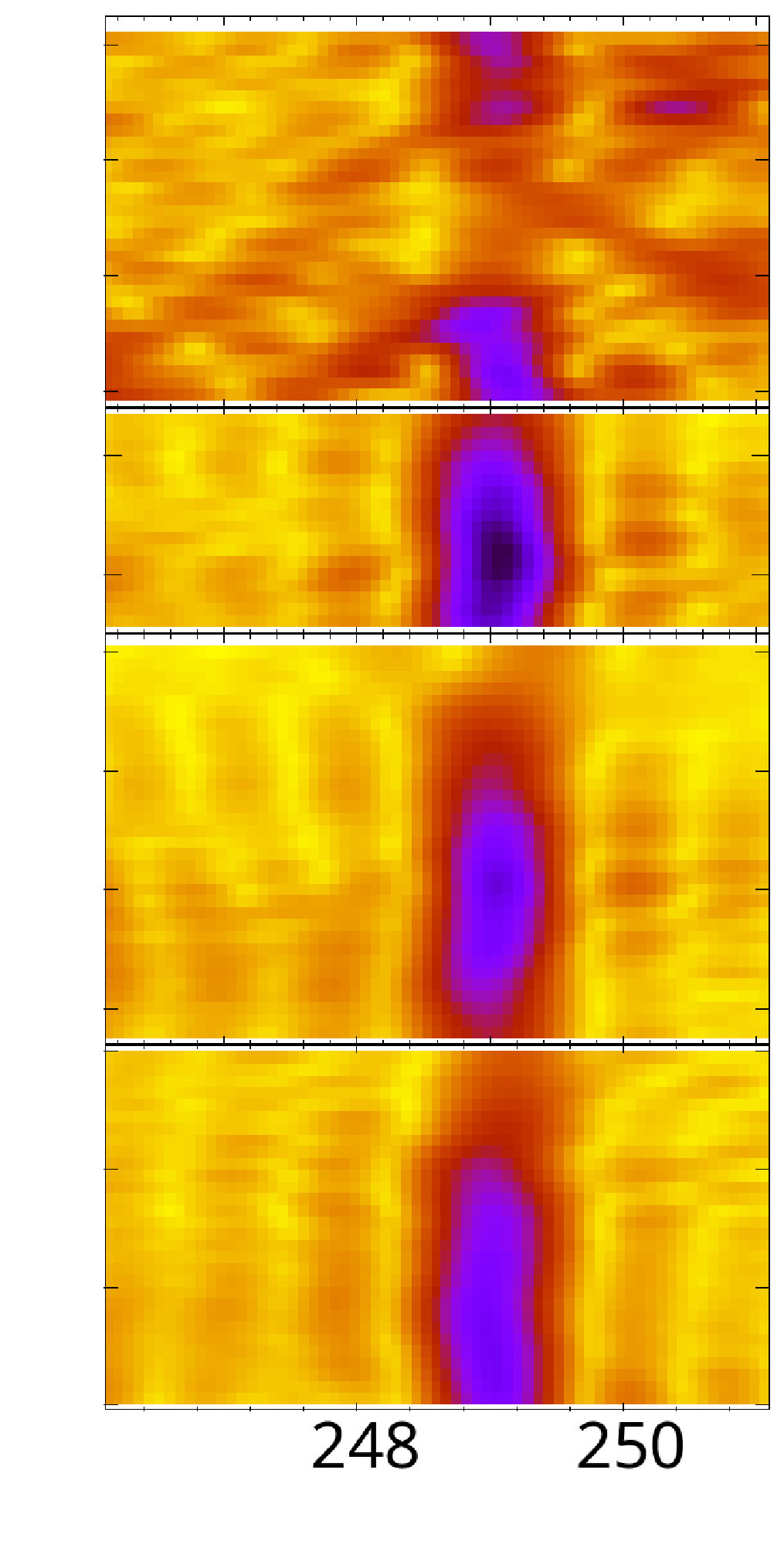,width=1.0in}\psfig{figure=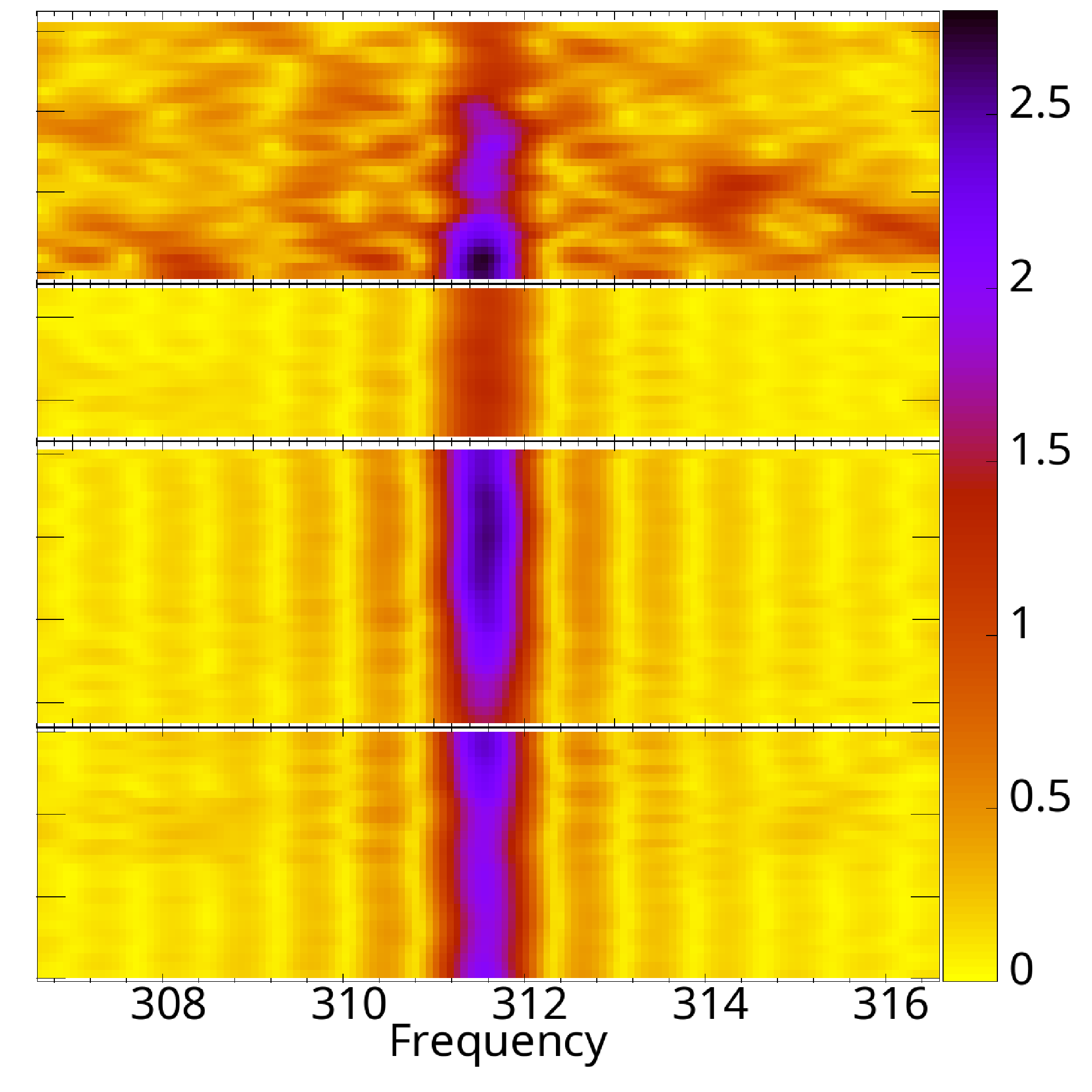,width=2.0in}}
    \caption{Sample sliding Fourier transforms of \hzcnc\, showing amplitude variations. Color indicates amplitude in ppt with the scale on the right.}
    \label{hzSFT}
\end{figure*}

Figures\,\ref{hzFTg} and \ref{hzSFT} show that the pulsations of \hzcnc\, undergo substantial amplitude
variations. One possibility is that the amplitude variations are caused by beating between unresolved multiplets, but as the 
times between C5 to C16 and C18 to S44 are so long, it was not possible to combine the data to increase resolution. 
However, \emph{if} the amplitudes were caused by unresolved frequency multiplets, we might be able to discern patterns in the
pulsation amplitudes that would indicate the beat period. 
Figure\,\ref{hzFTMO} shows FTs in TESS-length data segments and 
Fig.\,\ref{hzAmp} shows the amplitudes of nine higher-amplitude and resolvable pulsations.
There is no obvious pattern of amplitude variations so we conclude that the amplitude variability is intrinsic
to the modes themselves and is not an effect of unresolved multiplets caused by rotation. 
Unfortunately that means we cannot determine a rotation period.
As the binary period is 27.8\,days, rotationally split multiplets would likely be closely spaced, $0.4\,\mu$Hz for $\ell=2$ modes,
which, while resolvable during C5 and C16, would be difficult to separate from the amplitude variation. As such, we will only
presume that the rotation period is no shorter than the binary period.

\begin{figure}
\centerline{\psfig{figure=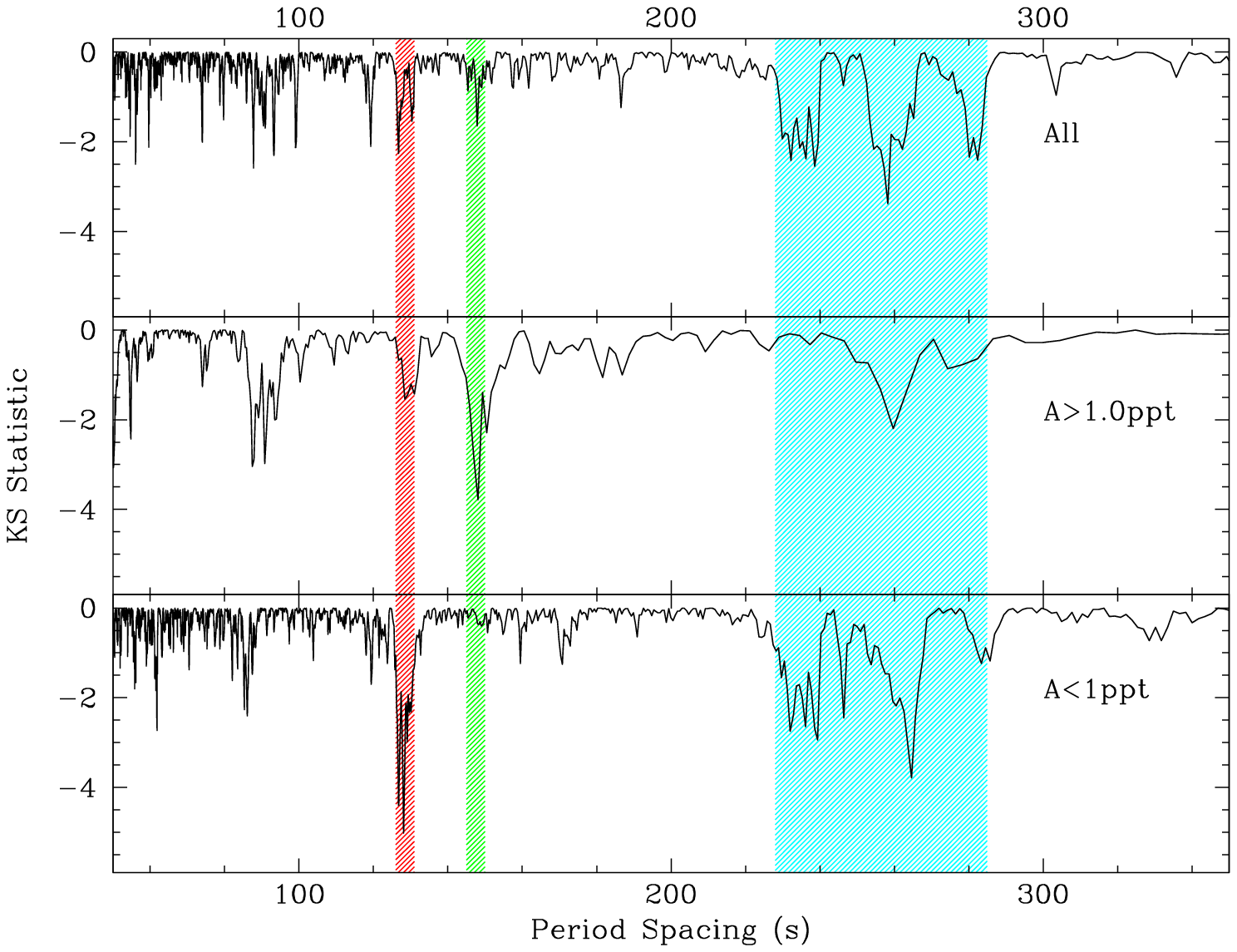,angle=-90,width=\columnwidth}}
\centerline{\psfig{figure=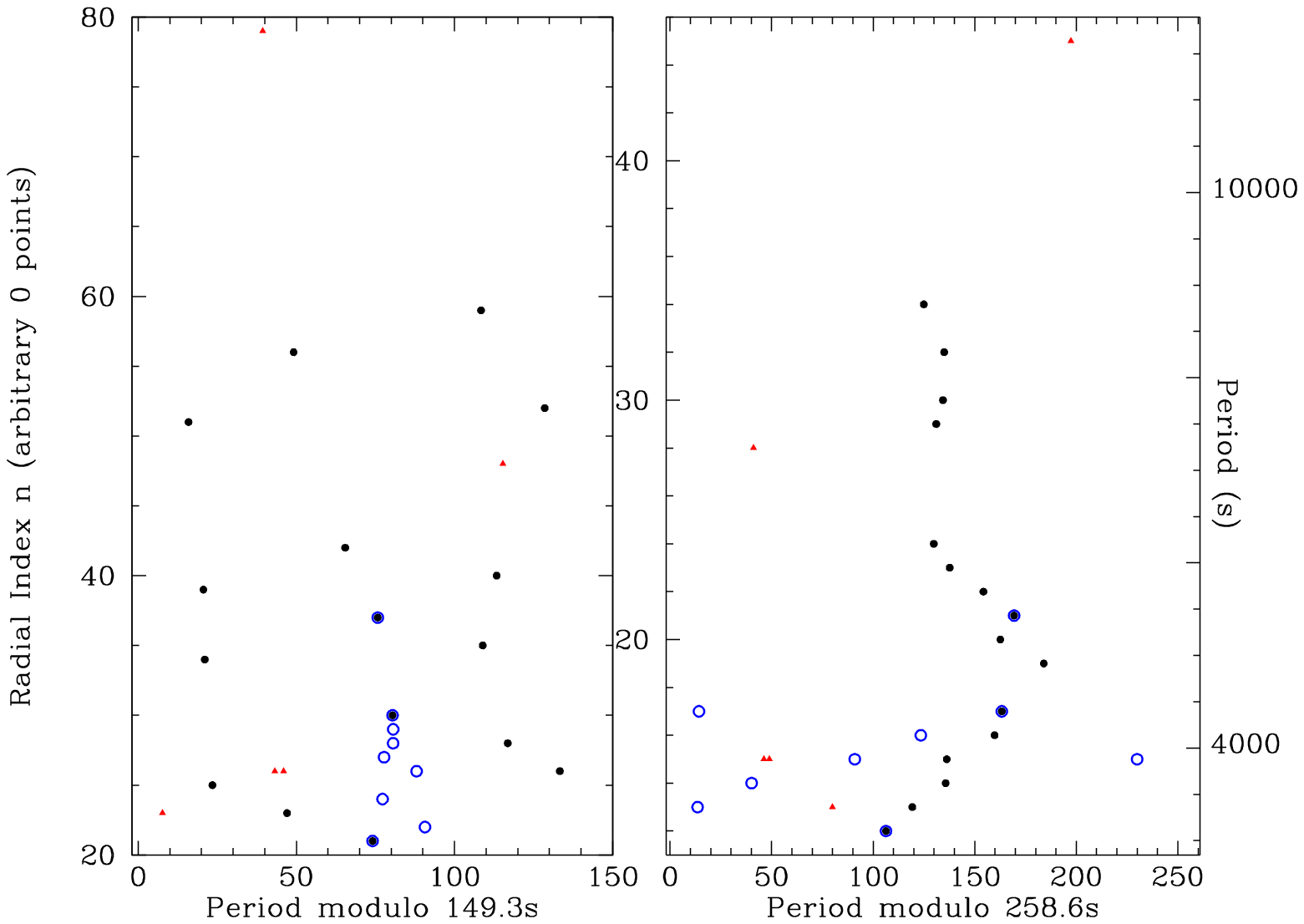,angle=-90,width=\columnwidth}}
	\caption{KS test and \'echelle diagram of \hzcnc's $g$-mode periods with same point/color
	coding as Fig.\,\ref{ltFT}.}
    \label{hzKS}
\end{figure}

For $g$-mode pulsations, as like the other stars in this paper, we can still look to identify modes using asymptotic period
spacings. In this case, we have success. Fig.\,\ref{hzKS} shows the results of KS tests using differing amounts
of data. The top panel shows results using all the periods and three separate troughs appear in the region
appropriate for $\ell=1$ and only a small, not significant trough in the $\ell=2$ region. As $\ell=1$ modes undergo the
least geometric cancellation \citep{bible}, sometimes an amplitude cut reveals the sequence. The middle panel shows the
KS test for only pulsations with amplitudes $>1.0$\,ppt and this shows a single  $\ell=1$ trough and a stronger 
$\ell=2$ trough as well. That provides a first-guess for the period spacings, allowing us to make an \'echelle 
diagram (Fig.\,\ref{hzKS}). On the \'echelle diagram we can pick out the $\ell=1$ sequence and do a linear regression
with a resultant period spacing of $257.32\pm0.46$\,s. Note that Fig.\,\ref{hzKS} has a modulo of 258.6\,s
and that is because there is quite a 'hook' feature with a long tail to shorter period. As such, the final fit begins
at the bend of the hook feature. Below the hook a linear regression gives a period spacing of $269.68\pm0.80$\,s for $\ell=1$.
The so-called \emph{hook} feature is not unique to \hzcnc\, as it has been observed for several other sdBV stars
\citep[e.g.][]{baran12c}.

\hzcnc's $p$-mode region also shows large variations between Campaigns 5, 16, and 18 and no $p$-modes were detected
in TESS data, as those data were not sufficiently sensitive. 
During C5 f1-f4 were not detected at all and then, during C16 and C18 f1 is the highest-amplitude pulsation
(Fig.\,\ref{hzFTg}). The region containing f5-f9 remains mostly consistent, as does the region covering f10-f17
although the pulsation 'mounds' of power are difficult to resolve into individual frequencies.
Similar to \pgo\, and
\ltcnc, \hzcnc\, shows clear radial overtones separated by $\sim1\,000\mu$HZ which will be discussed in \S 4.1.

\subsection{\pgn}
\begin{figure*}
\centerline{\psfig{figure=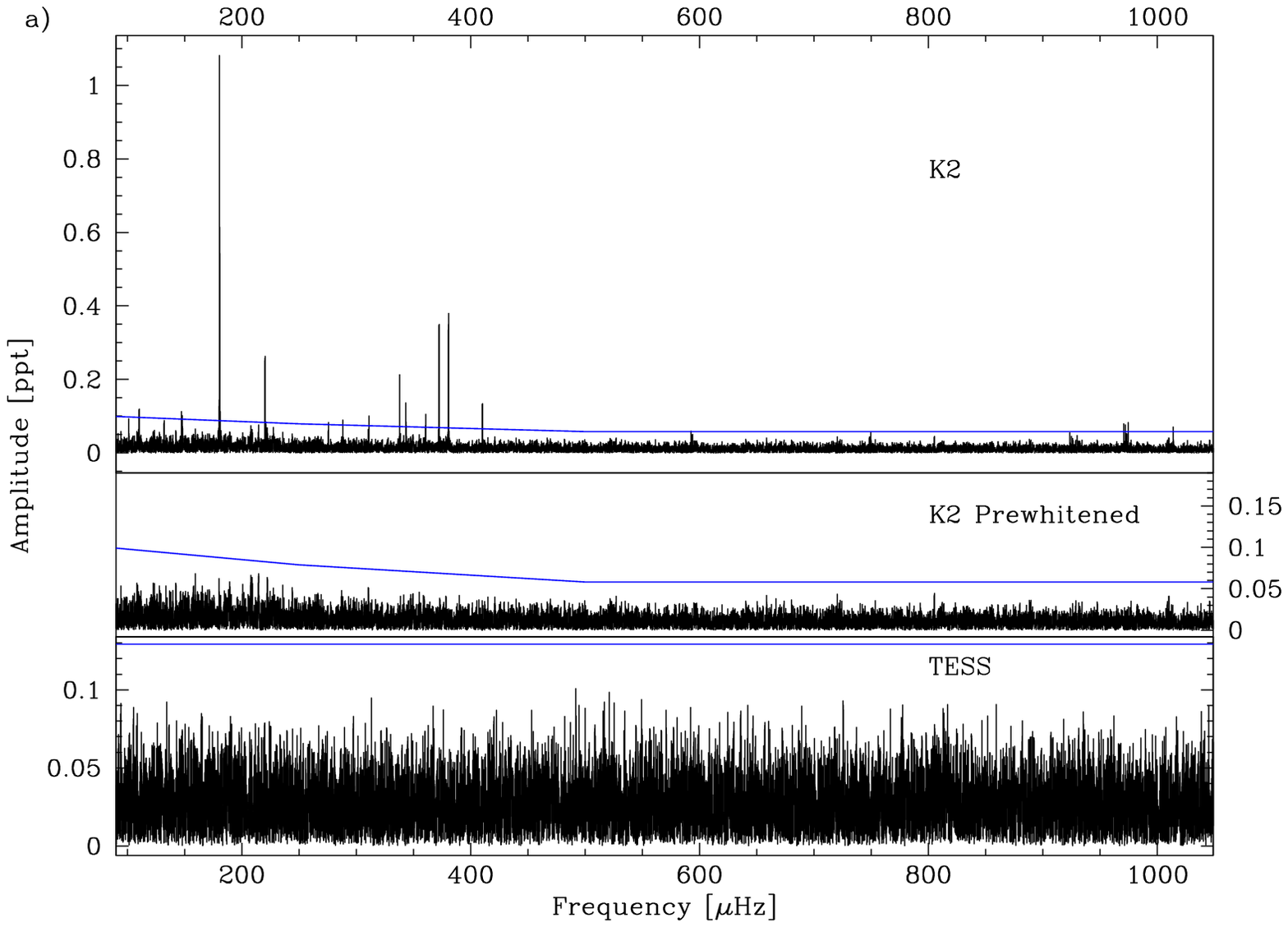,angle=-90,width=5.5 in}}
\centerline{\psfig{figure=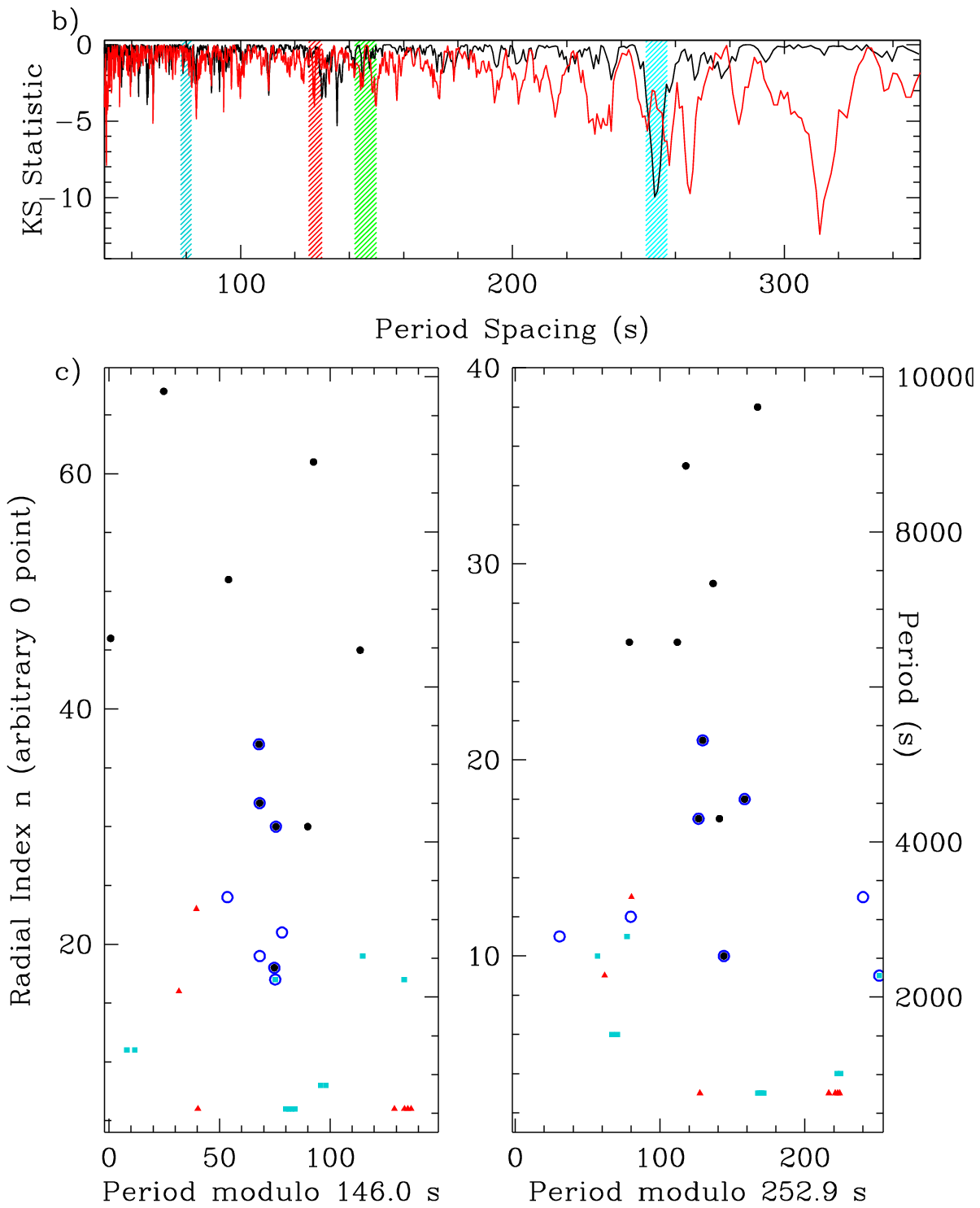,width=\columnwidth}\psfig{figure=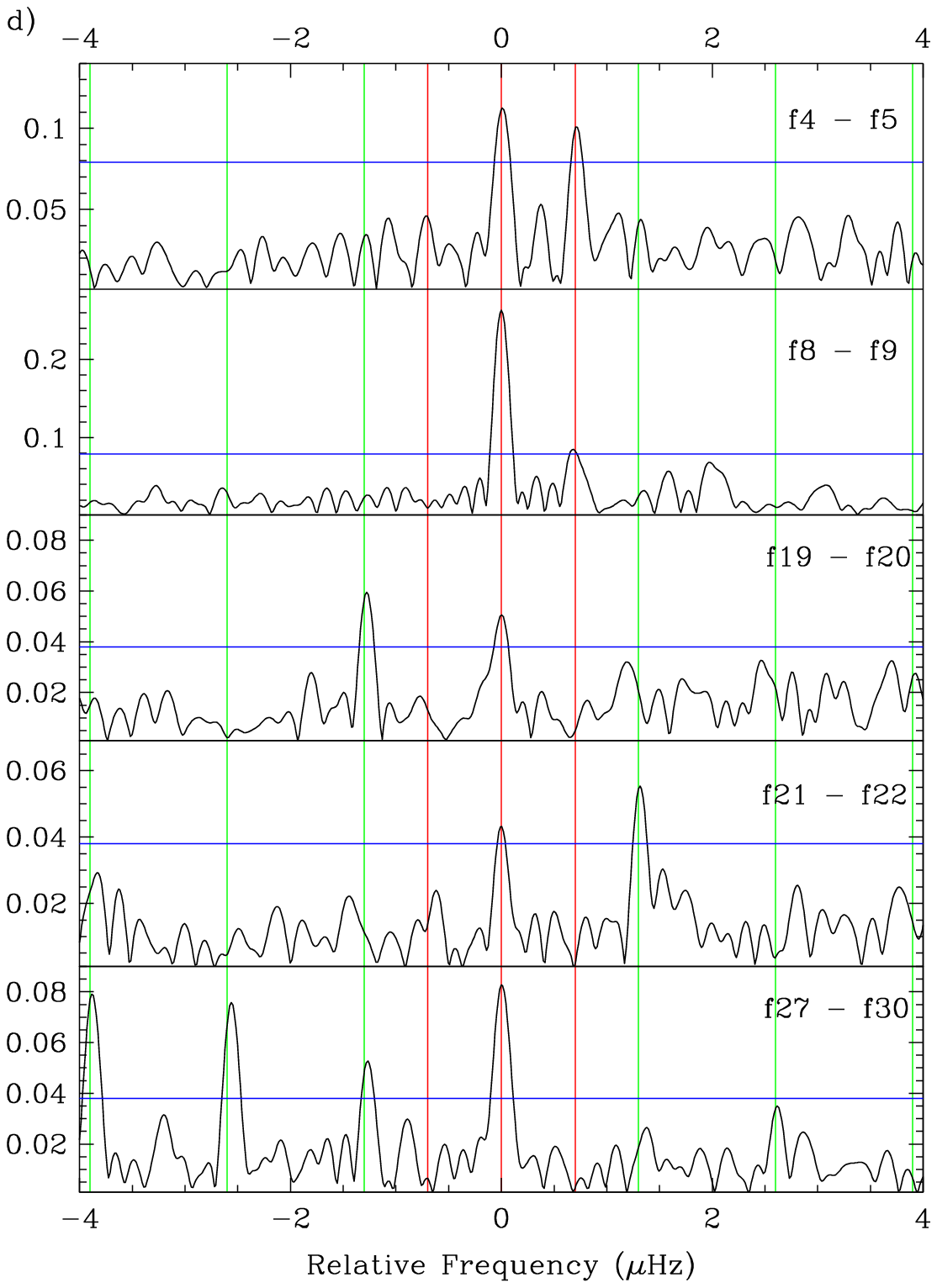,angle=0,width=\columnwidth}}
	\caption{Same as Fig.\,\ref{ltFT} for \pgn\, except in b) the black line excludes periods under 2\,000\,s
	while the red line includes all periods. }
    \label{pgnFT}
\end{figure*}

\pgn\, was observed during K2 campaign 16 (7 Dec, 2017 -- 25 Feb. 2018) and also TESS Sectors 44, 45, and 
46 (12 Oct -- 30 Dec, 2021), both of similar duration. 
While no pulsations were detected in the TESS data, 31 were detected in K2 data, all in the $g$-mode region
(Fig.\,\ref{pgnFT}). The lack of detections in TESS data is surprising as the amplitudes of five periodicities
during the K2 observations are above the detection limit of the TESS data and should have been detected. The \emph{Kepler}
and TESS passbands differ slightly but for the extremely blue sdB stars, no filter-induced amplitude changes
are expected. 

\begin{figure*}
\centerline{\psfig{figure=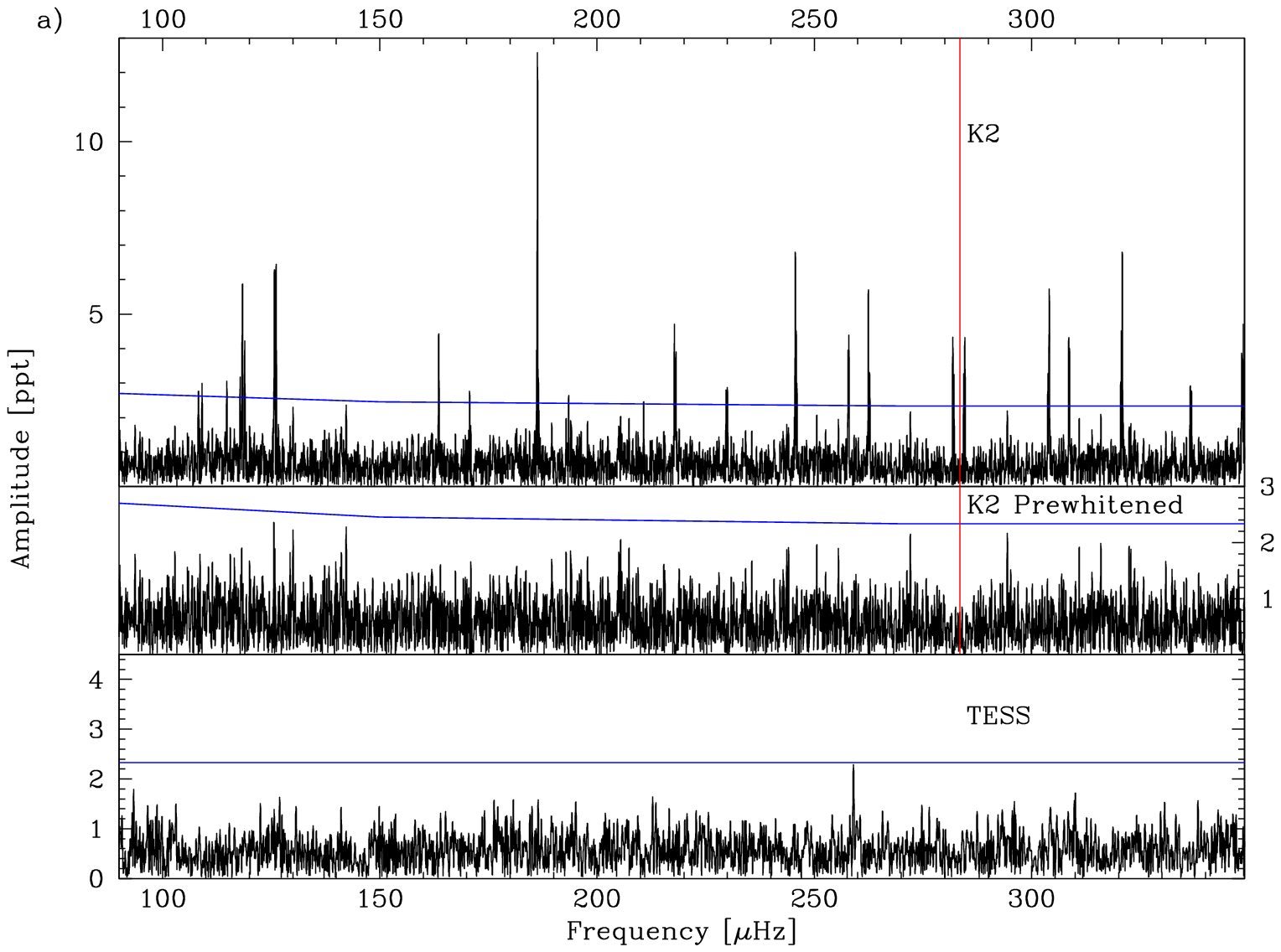,angle=-90,width=5.5 in}}
\centerline{\psfig{figure=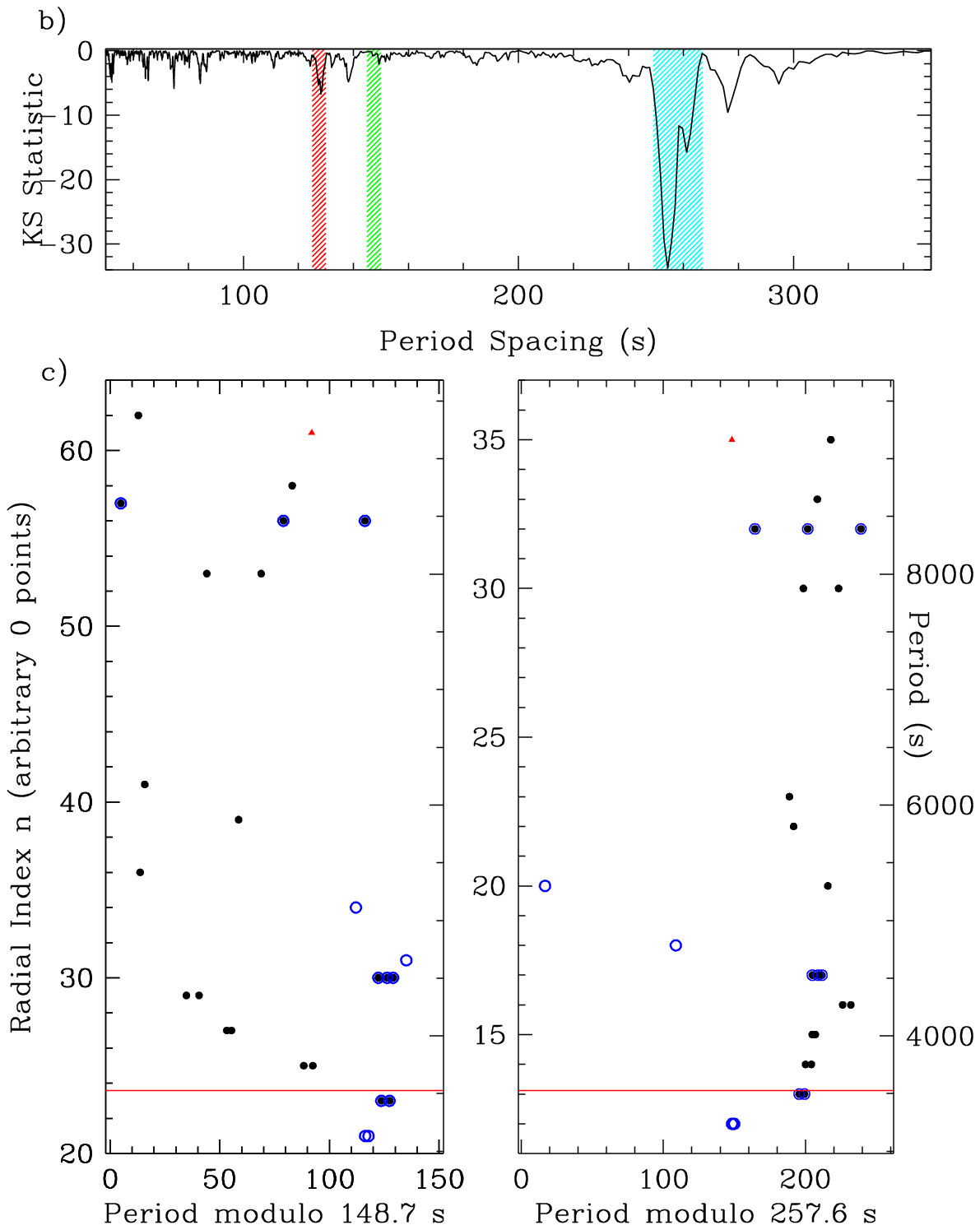,width=\columnwidth}\psfig{figure=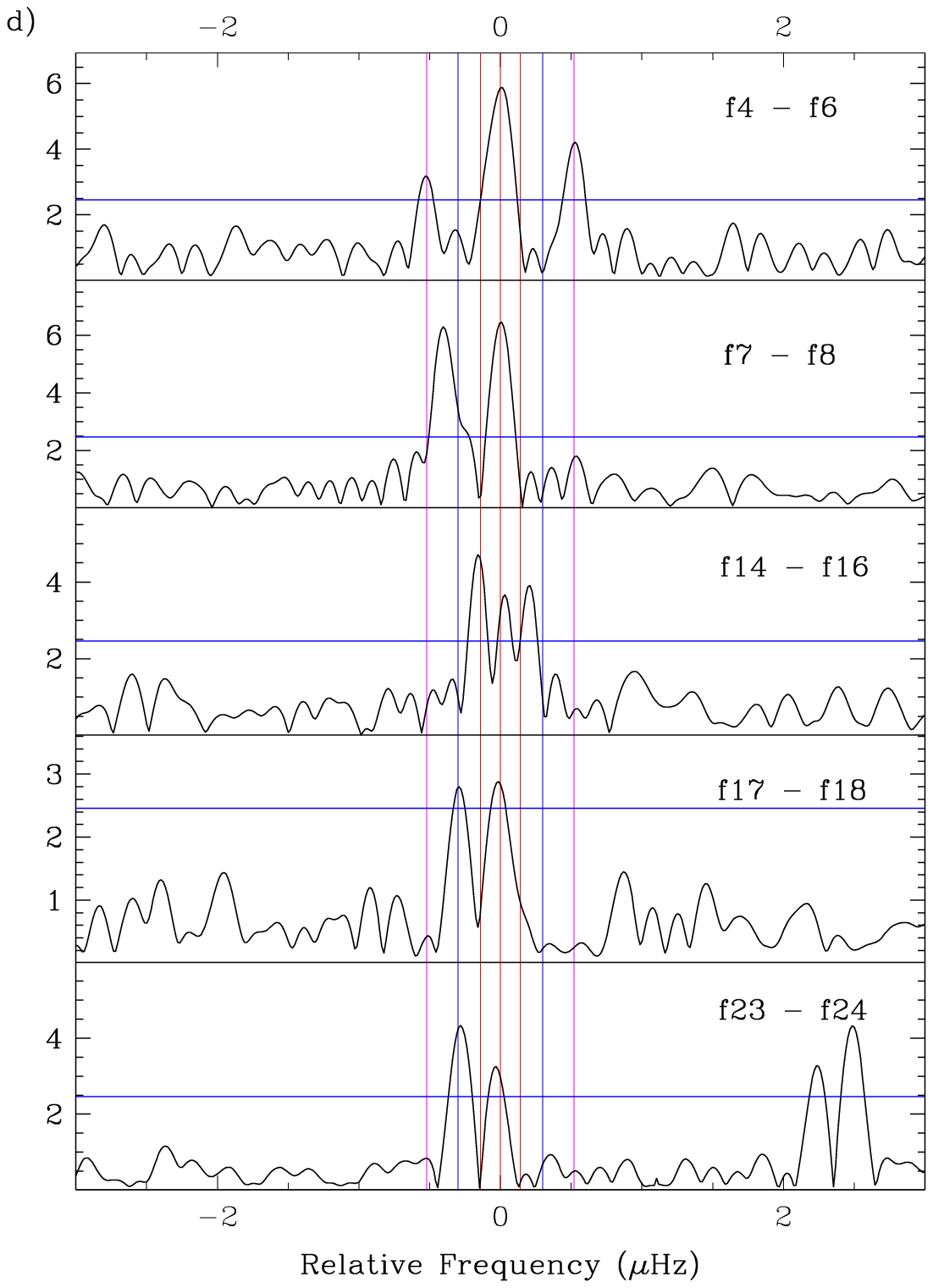,angle=0,width=\columnwidth}}
        \caption{Same as Fig.\,\ref{ltFT} for \pb\, except: Red lines (vertical in a, horizontal in c)
        indicate the Nyquist frequency for LC data. d) Vertical lines are all for $\ell =1$ multiplets but
         indicate the different splittings (also shown in Fig.\,\ref{pb_fsr}).}
    \label{pbFT}
\end{figure*}

\begin{figure}
\centerline{\psfig{figure=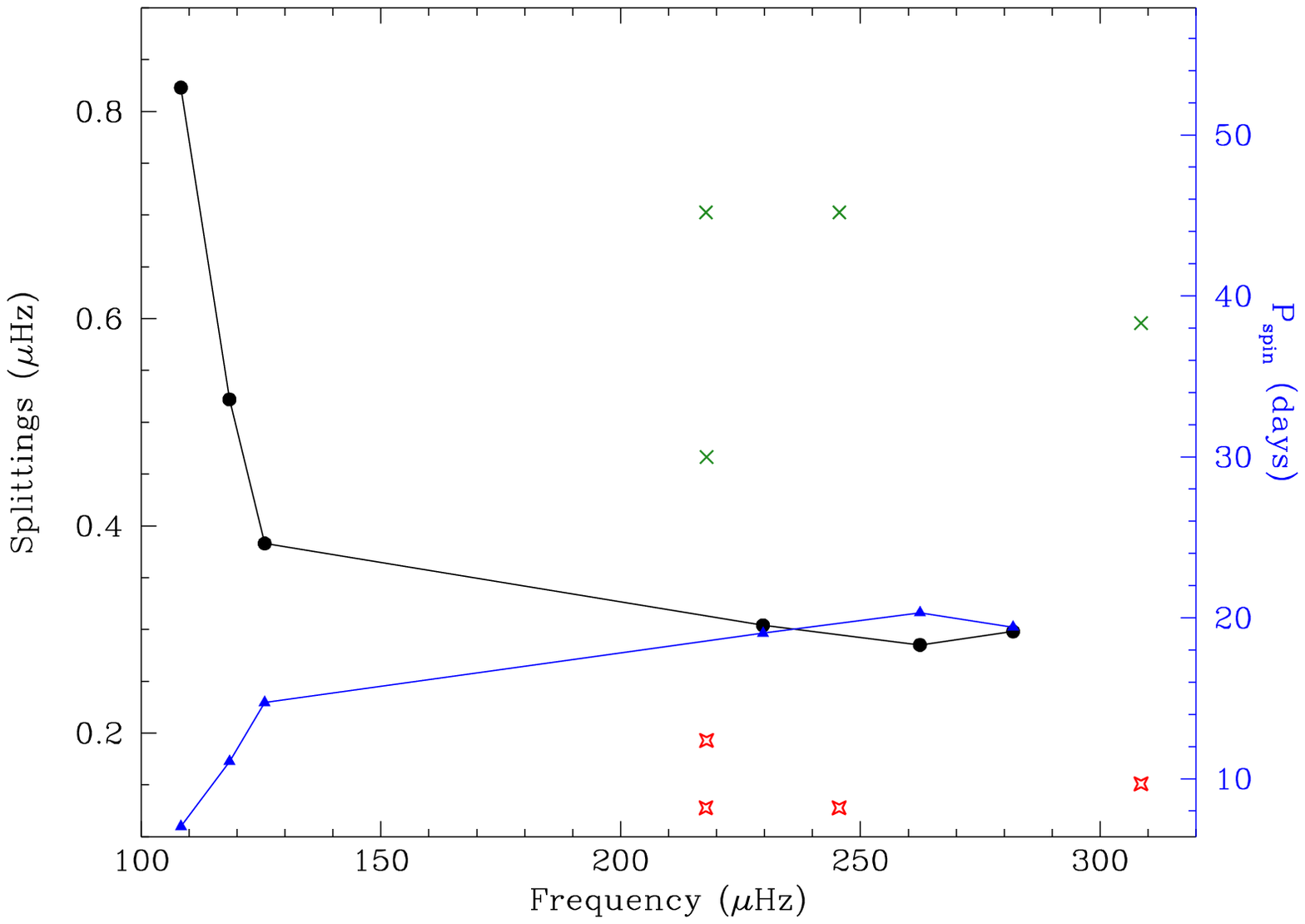,angle=-90,width=\columnwidth}}
        \caption{Frequency splittings detected in \pb.
        Left axis and black line/circles indicates frequency splittings and the right axis and blue line/triangles
	spin periods assuming $C_{n,\ell}=1/\ell\left(\ell+1\right)$.
The lines are for the fully-resolved splittings and the red squares (splittings)/green crosses (P$_{\rm spin}$) are for
        splittings $<0.2\mu$Hz.}
\label{pb_fsr}
\end{figure}

\pgn's period spacing sequence is not especially clear but frequency multiplets are, particularly in the higher-frequency
range. Those splittings indicate $\ell>2$ modes and so to find the $\ell =1$ period spacing sequence,
we calculated a KS test using only periods longer than 2\,000\,s (black line of Fig.\,\ref{pgnFT}b).
We then made an \'echelle diagram (Fig.\,\ref{pgnFT}c), 
folding over the period of the deepest trough, picked periods of the $\ell =1$
sequence, and fitted a linear regression. The resultant $\ell =1$ period spacing is $252.88\pm 0.74$\,s. We then
generated $\ell =2$, 3, and 4 sequences and identified periods that fit into those sequences. Our mode identifications are
provided in Table\,\ref{pgnlist}. 

There are two very nice frequency doublets in periods identified as $\ell =1$ with splittings of $0.7\mu$Hz
(Fig.\,\ref{pgnFT}d) and then
several splittings above $590\mu$Hz near $1.3\mu$Hz which we associate mostly with $\ell =4$ modes. Assuming Ledoux
constants of $C_{n,\ell}=\frac{1}{\ell\left(\ell+1\right)}$ we find rotation periods of $8.13\pm0.17$ and $9.45\pm0.18$\,days
for the $\ell =1$ and 4 modes, respectively. We do not consider the $\ell =1$ and 4 modes to be rotating at different
periods and so we deduce a rotation period for \pgn\, of $9.10\pm0.85$\,days. \citet{kern16,kern18} found that some
frequency multiplets were observed to vary their splittings over the three year duration 
of the original \emph{Kepler} mission, 
particularly the higher-degree modes, and so it is possible the $\ell =4$ splittings are
being affected by that same phenomenon. However, we have no way to deduce that from the data at hand. As none
of the multiplets are complete, in Table\,\ref{pgnlist} we arbitrarily assign the $m=0$ component as the one which best
matches the period spacing sequence, the smallest $\Delta P/P$.

\subsection{\pb}
\pb\, was observed during campaign 8 (3 Jan. -- 23 Mar, 2016) but only in long-cadence (30m) mode.
We determine the detection limit to be 2.3ppt and the 1.5/T resolution to be $0.22\mu$Hz.
It was also observed by TESS during Sectors 3, 30, 42, and 43 in short-cadence (2m) mode. Except
for Sectors 42 and 43, the observations are too separated to combine and no pulsations were
detected in any of the TESS observations except for one marginal detection in the combined
Sectors 42 and 43 data. That frequency is noted in Table\,\ref{pbFlist} as t01.
The Nyquist frequency for K2 LC observations is at $283.45\mu$Hz, which is in the range of
$g$-mode pulsations and so reflections across this is an issue (Fig.\,\ref{pbFT}a).  Phase smearing of the pulsations
caused by spacecraft motion can reduce the amplitude of the reflection \citep{baran12a,shiba13,F5_paper}. Unfortunately
the errors on the amplitudes are larger than the amplitude differences across the Nyquist. Fortunately, sdBV stars
are known to have asymptotic $g$-mode pulsations with dipole period spacings near 250s. For \pb\, the
sub-Nyquist region clearly has such a sequence while the super-Nyquist region does not. We determine
the $\ell=1$ sequence to be $257.60\pm 0.47$\,s (Fig.\,\ref{pbFT}b)
and using the relationship $\Delta$P$_2$/$\Delta$P$_1=\sqrt{3}$
the $\ell=2$ sequence would be 149\,s. We tested periods on both sides of the Nyquist and the only
super-Nyquist fit was for f25 and f26 as $\ell=2$ modes. For the two periods which did not fit either
sequence, we presume them to be sub-Nyquist since that is where most of the periods are.

The frequency splittings in \pb\, are quite tricky to interpret. The set f04-f05-f06 form a natural 
$\ell=1$ triplet with a splitting
of $0.52\mu$Hz, but then there are plenty of seemingly $\ell=1$ doublets; f01-f02, f07-f08, f17-f18, and f21-f22, but their 
splittings are inconsistent at 0.82, 0.39, 0.30, and 0.29$\mu$Hz, respectively. 
Then there are the multiplets which are within our formal
1.5/T resolution, but which were easily prewhitened as separate frequencies; f14-f15-f16, f19-f20, and f25-f26 with splittings of
0.13, 0.19, and 0.15$\mu$Hz, respectively. Figure\,\ref{pb_fsr} shows  the most direct interpretation with the frequency
splittings as measured and their rotation period counterparts. 
This shows a very steep trend to smaller splittings for the lowest frequencies
and then a shallower trend at the higher frequencies. The corresponding spin periods are 7\,days at the f01-f02 doublet and over
 19\,days at the f23-f24 doublet. Alternatively, we could presume the f01-f02 doublet is $\Delta m=2$ and average all the 
 resolved splittings 
together. The result would be a splitting of $0.39\pm0.10\mu$Hz or a spin period of $14.8^{+5.2}_{-3.1}$\,days.

\begin{table}
\caption{\label{specfit}\small
Spectroscopic model fit results.}
\centering
\renewcommand{\arraystretch}{1.0}
\begin{tabular}{llcc}
\hline\hline
Star & \teff & $\log g$ & $\log y$ \\
       & [K] & [cgs] & \\ \hline
\pgo   & 27145(111)& 5.502(16) & -2.750(32) \\
\ltcnc & 26032(83) & 5.275(11) & -2.801(41) \\
\hzcnc & 27557(85) & 5.430(14) & -2.811(25) \\
\pgn   & 27287(77) & 5.465(10) & -2.726(34) \\
\pb    & 26968(125)& 5.404(18) & -3.012(40) \\
\hline
\end{tabular}
\end{table}

\section{Spectroscopic observations and analysis}
As part of our follow-up spectroscopic survey \citep{telting12b} of \emph{Kepler}-observed sdBV stars, we used ALFOSC at the 2.56-m Nordic Optical Telescope \citep[][]{NOT}, with grism \#18 and a 0.5 arcsec slit providing R\,=\,2000 resolution or 2.2\,\AA\ to obtain multiple spectra of all targets. The observing log is provided in Table\,\ref{specobs}.

\begin{figure}
    \centerline{\psfig{figure=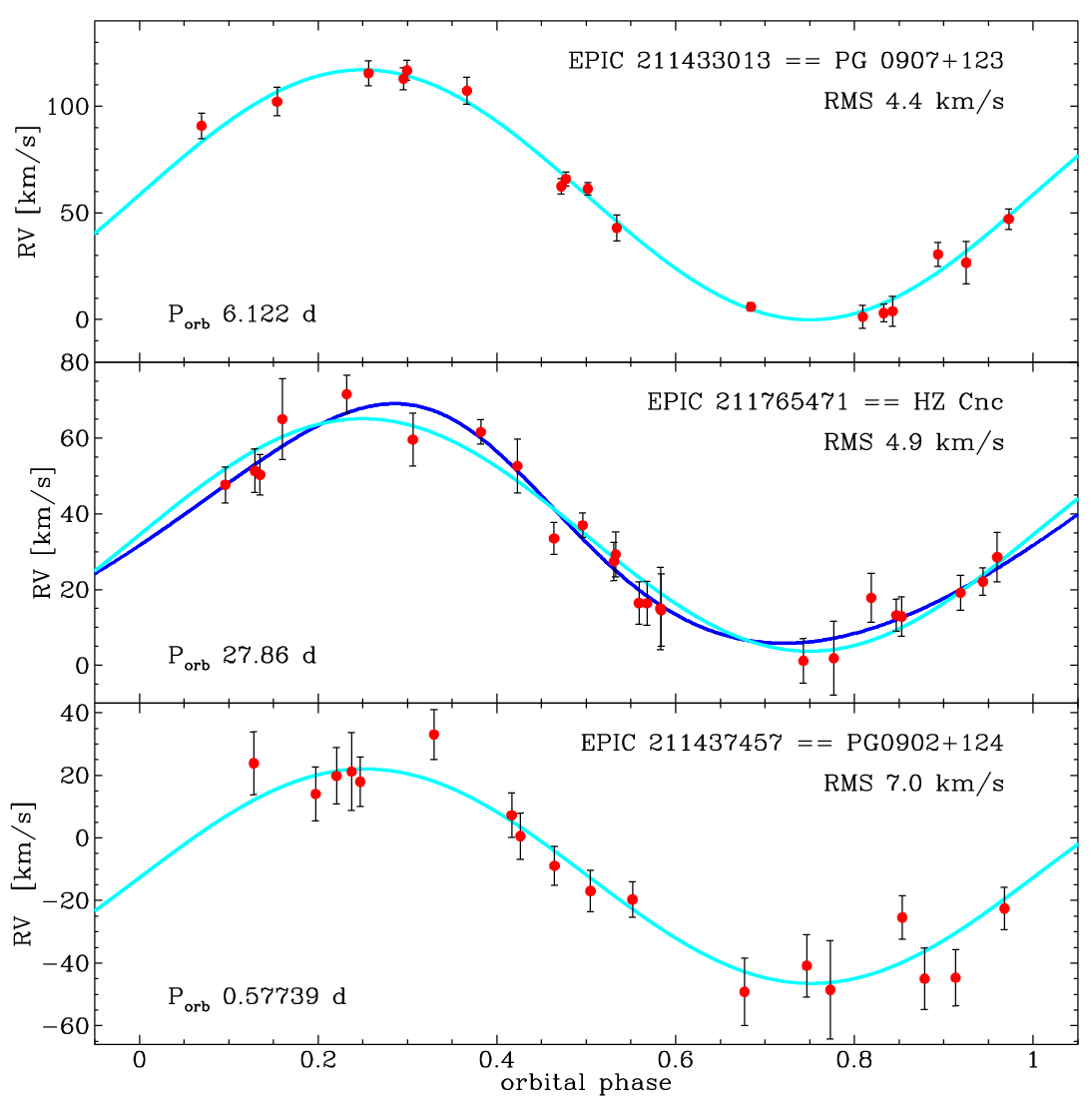,width=\columnwidth}}
    \caption{Binary orbital solutions, resulting from ALFOSC@NOT
spectrocopy.  Light-blue curves: circular-orbit fits.  Dark-blue curve:
        eccentric orbit fit.}
    \label{rv_k2sdB}
\end{figure}

We obtained new spectroscopy to allow us to derive the atmospheric
 parameters in a homogeneous way for the whole sample, and to
 verify or discover binarity.   As our access to ALFOSC@NOT
 allows near-optimal monitoring of long orbital periods, we
 gathered sufficient spectra to ensure proper sampling of
 the orbital periods. We did not obtain sufficient radial-velocity data to
 redetermine orbital parameters for \pb\, and \pgo.

The data were homogeneously reduced and analysed as described in \citet{telting12a}, \citet{ReedPG1142}, using IRAF for bias subtraction, removal of pixel-to-pixel sensitivity variations, optimal spectral extraction, and wavelength calibration based on arc-lamp spectra. The target spectra and the mid-exposure times were shifted to the barycentric frame of the Solar system.
The spectra were normalized to place the continuum at unity
by comparing with a model spectrum (see below) for a star with similar physical
parameters as we find for each target.

\begin{table*}
\caption{\label{specorbits}\small
Orbital RV fit results, assuming circular orbits. The last two columns are mass-function
results assuming $M_1$=0.47\,M$_{\odot}$ and orbital inclination $i$=90$^o$.}
\centering
\renewcommand{\arraystretch}{1.0}
\begin{tabular}{lcllll|rr}
\hline\hline
Star & N$_{\rm spec}$ & RV offset & $K_1$ & P & RMS & $M_{2, \rm min}$ & $(a_1+a_2)_{\rm min}$ \\
       &   & [km/s] & [km/s] & [d] & [km/s] & [M$_{\odot}$] & [R$_{\odot}$] \\ \hline
\ltcnc & 17 &   58.2 (1.3) & 59.2 (1.8) & 6.122 (0.004) & 4.4 & 0.50 & 14 \\
\hzcnc & 24 &  34.4 (1.1) & 30.6 (2.0) & 27.86 (0.03) & 4.9 & 0.40 & 37 \\
\pgn   & 18 & -12.2 (2.0) & 35.2 (3.1) & 0.57739 (0.00015) & 7.0 & 0.094 & 2.4\\
\hline
\end{tabular}
        \label{spectable}
\end{table*}

\subsection{Atmospheric parameters}
To determine atmospheric parameters we made averaged spectra
 after removing the orbital shifts for each individual spectrum.
We determined \teff\, and $\log g$ from the mean spectra using the H/He LTE grid of \citet{heber99b} for consistency with the 
\emph{Kepler} main field survey of \citet{roy10b,roy11b}. For the fit we used all the Balmer lines from H$\beta$ to H14 and the three strongest He\,{\sc i} lines as well as He\,{\sc ii} at 4686\,\AA\ (even when absent). The uncertainties listed on the measurements are the formal uncertainties of the fit, which reflect the S/N of the mean spectrum, or at the very high S/N levels here, the presence of `metal' lines, not explicitly included in the model spectra. These values and uncertainties are relative to the LTE model grid and do not reflect any systematic effects caused by the assumptions underlying those models. Such systematic effects depend strongly on the details of the model grid such as the assumed metal content and can be as large as 2.5\,kK in effective temperature and 0.2 in log g/cm s$^{-2}$.
See Table\,\ref{specfit}.

\begin{figure*}
        \centerline{\psfig{figure=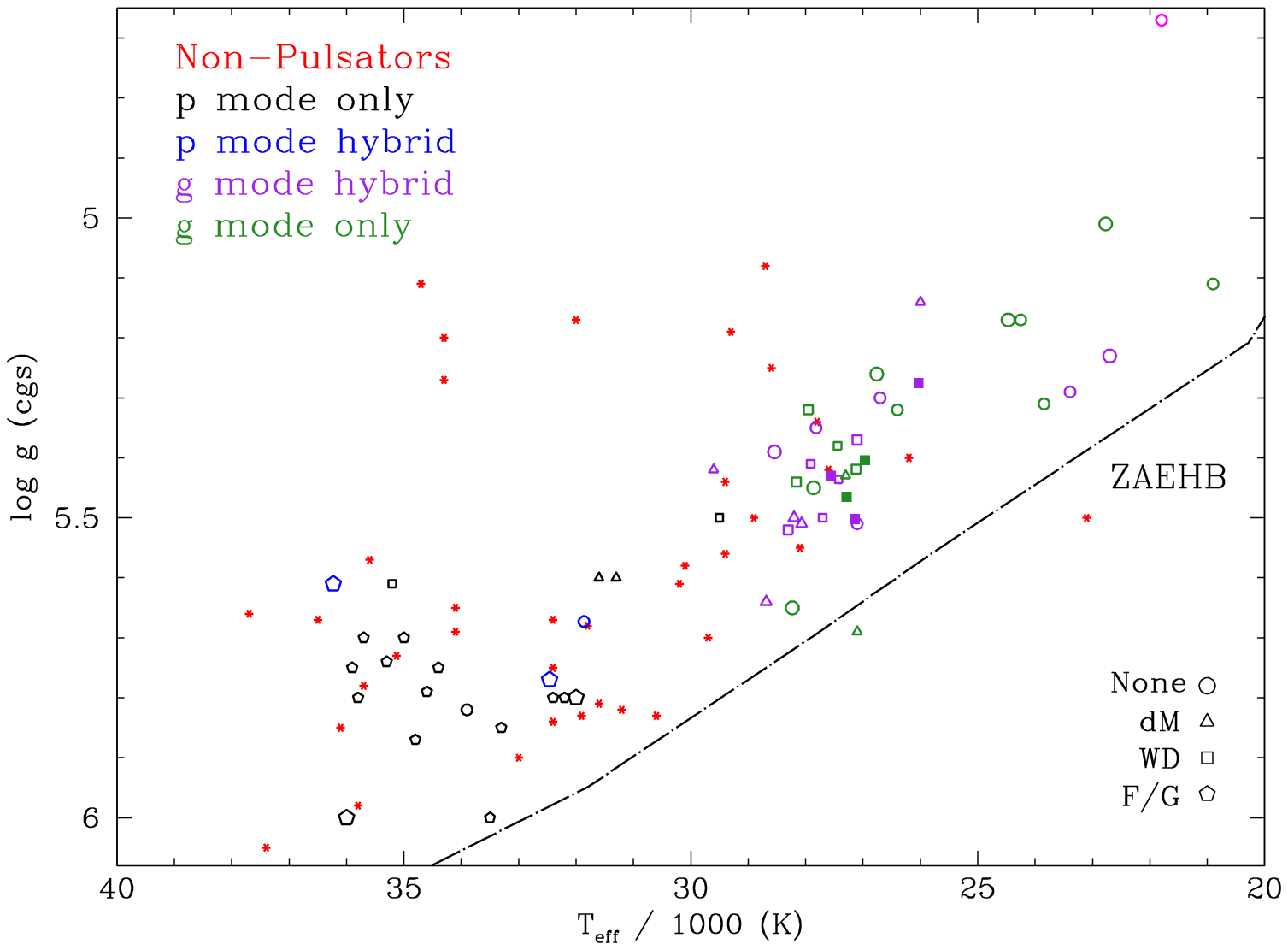, angle=-90,width=\textwidth}}
        \caption{Kiel diagram indicating the spectroscopic properties of non-pulsators (red asterisks), the stars in this paper
        (filled squares) and other \emph{Kepler}-observed sdB stars \citep[open points. Adapted from][]{,roy11b,ReedF7}.
Pulsations types indicated by color and companion type by point shape.}
\label{kiel}
\end{figure*}

\begin{figure*}
        \centerline{\psfig{figure=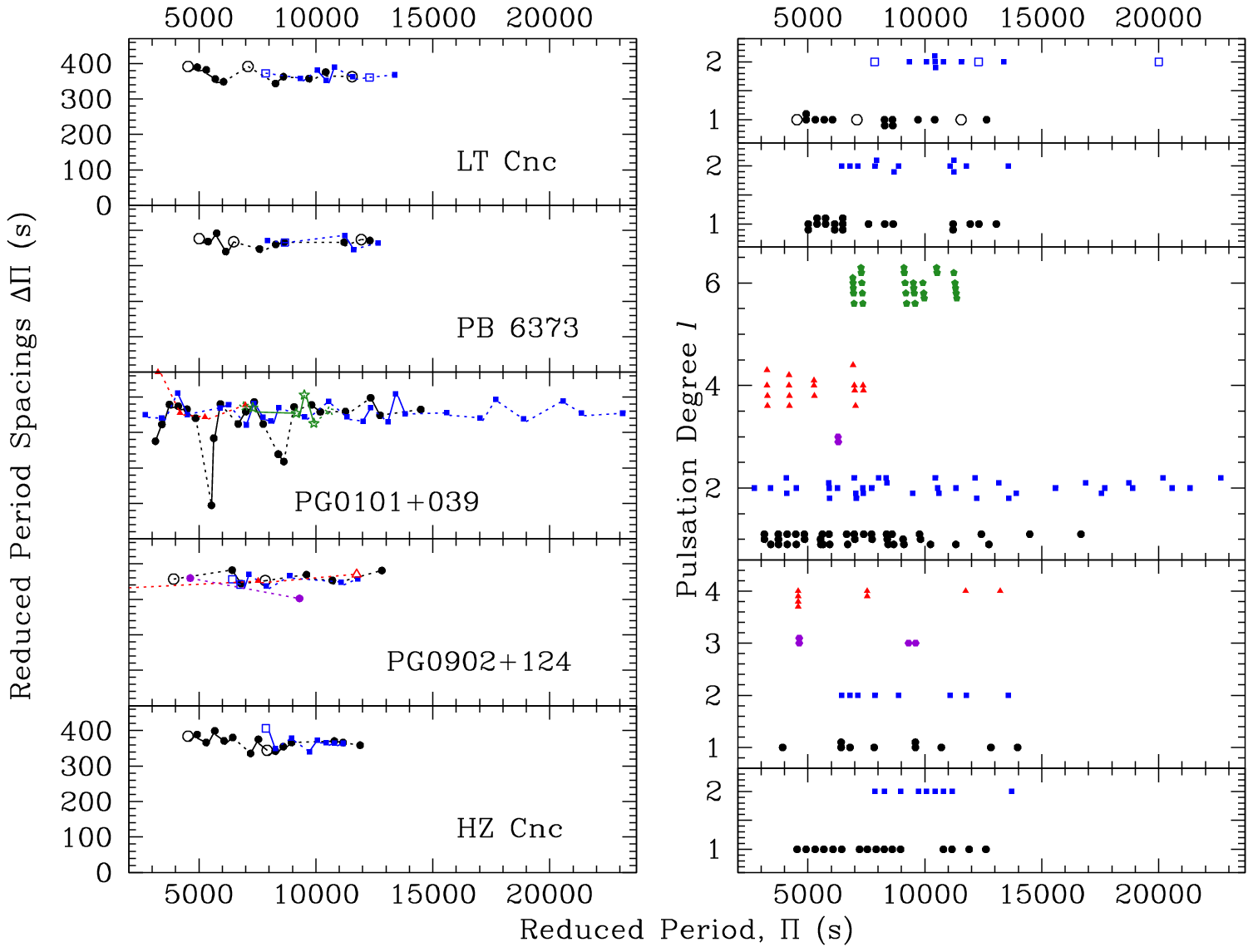,angle=-90,width=\textwidth}}
        \caption{Reduced period $\Pi = P\times\sqrt{\ell (\ell + 1)}$ diagrams compared with
        reduced period spacings $\Delta\Pi$ (left) and pulsation degrees $\ell$ (right). The left panel
        only has $m=0$ components while in the right panel azimuthal orders
are separated from $\ell$ as $0.1\times m$. $\ell = 1$ is in black,
        2 in blue, 3 in red, 4 in cyan, and 6 in green. Open points indicate periods which could have more than one
        degree and are therefore double-counted. Stars are organized by $T_{\rm eff}$ from coolest (LT\,Cnc)
        to hottest (HZ\,Cnc). In the left panel, solid lines indicate consecutive overtones and dotted lines
        indicate non-consecutive overtones.}
\label{redPAll}
\end{figure*}

\subsection{Radial velocities and orbits}
Radial velocities for the systems were derived with the FXCOR package in IRAF. We used
the H$\beta$, H$\gamma$, H$\delta$, H$\zeta$ and H$\eta$ lines to determine the
radial velocities (RVs).
Firstly, we used the average spectrum as a velocity template, and the resulting RVs and RV uncertainties 
were used for fitting the orbits.  Secondly, we determined the RV offset of the average spectrum 
from a cross-correlation against the spectral model fit (see above), and added that offset to each individual RV measurement.  
See Table~\ref{specobs} for the results.

The uncertainties reported by FXCOR are correct relative to each other, but may need
scaling depending on, amongst other things, the parameter settings and
the validity of the template as a model of the star.  Our initial fits
resulted in $\chi^2$-values close to unity.  For estimating the final fit parameters (see Table~\ref{rv_k2sdB}) we scaled
the RV uncertainties to obtain $\chi^2$-values of unity.

We used a sinusoidal orbit model, with a standard non-linear least-squares fitting method.
For our newly discovered binary, \pgn\, we used 
the highest peak in the discrete Fourier transform of our RV time series as a first guess for its orbital period.
For the known binaries, \ltcnc\, and \hzcnc\, we used literature values \citep{morales-rueda03} as first-guess period values.
We list the orbital fit results in Table\,\ref{spectable}. We note that although our
 sampling of the orbital phases of the long 28\,day orbit of \hzcnc\, is much
 improved with respect to the data presented by \citet{morales-rueda03}, our
 results for the orbital parameters are remarkably similar to theirs.

 \begin{table*}
\caption{Overtone spacings. Reduced period spacings $\Delta\Pi$ and their standard deviations $\sigma\Delta\Pi$ by degree $\ell$.
        \pgo\, has two rows including ($^{\dagger}$) and excluding ($^{\ddagger}$) trapped modes. P-mode overtone ratios $R$ and
        standard deviations, $\sigma R$.}
        \label{table_pspace}
\centering
\begin{tabular}{lcccccccccccc}
\hline\hline
        Star & \multicolumn{2}{c}{$\ell=1$} & \multicolumn{2}{c}{$\ell=2$} & \multicolumn{2}{c}{$\ell=3$} &
        \multicolumn{2}{c}{$\ell=4$} & \multicolumn{2}{c}{$\ell=6$} \\
        &  $\Delta\Pi$ & $\sigma\Delta\Pi$  &  $\Delta\Pi$ & $\sigma\Delta\Pi$  &  $\Delta\Pi$ & $\sigma\Delta\Pi$
         &  $\Delta\Pi$ & $\sigma\Delta\Pi$  &  $\Delta\Pi$ & $\sigma\Delta\Pi$ & $R$ & $\sigma R$ \\ \hline
        \ltcnc & 370 & 17 & 368 & 12 &-- &--&--&--&--&--&0.764 & 0.008 \\
        \pb & 366 & 14 & 367 & 14 & -- & -- & -- & -- & -- & -- & -- & --\\
        \pgo$^{\dagger}$ & 328 & 56 & 330 & 78 & -- & -- & 318 & 55 & -- & -- & -- & -- \\
        \pgo$^{\ddagger}$ & 366 & 22 & 363 & 22 & 359 & 1 & 345 & 12 & 369 & 26 & 0.792 & 0.005 \\
        \pgn   & 363 & 15 & 354 & 12 & 331 & 40 & 350 & 22 & -- & -- & -- & -- \\
        \hzcnc & 367 & 18 & 368 & 20 &-- &--&--&--&--&--&0.793 & 0.005 \\
\hline
\end{tabular}
\end{table*}

\begin{figure}
        \centerline{\psfig{figure=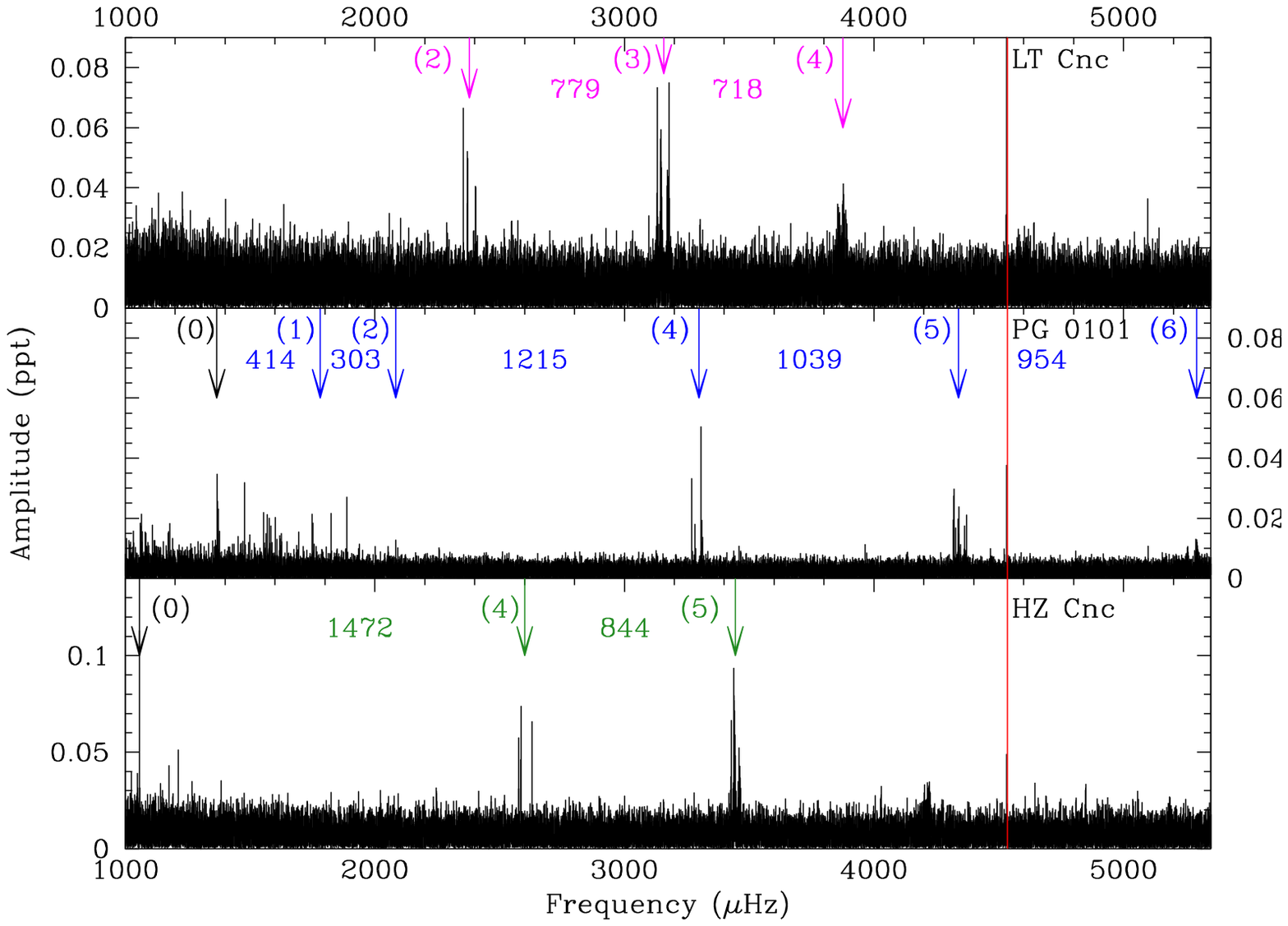, angle=-90,width=\columnwidth}}
        \centerline{\psfig{figure=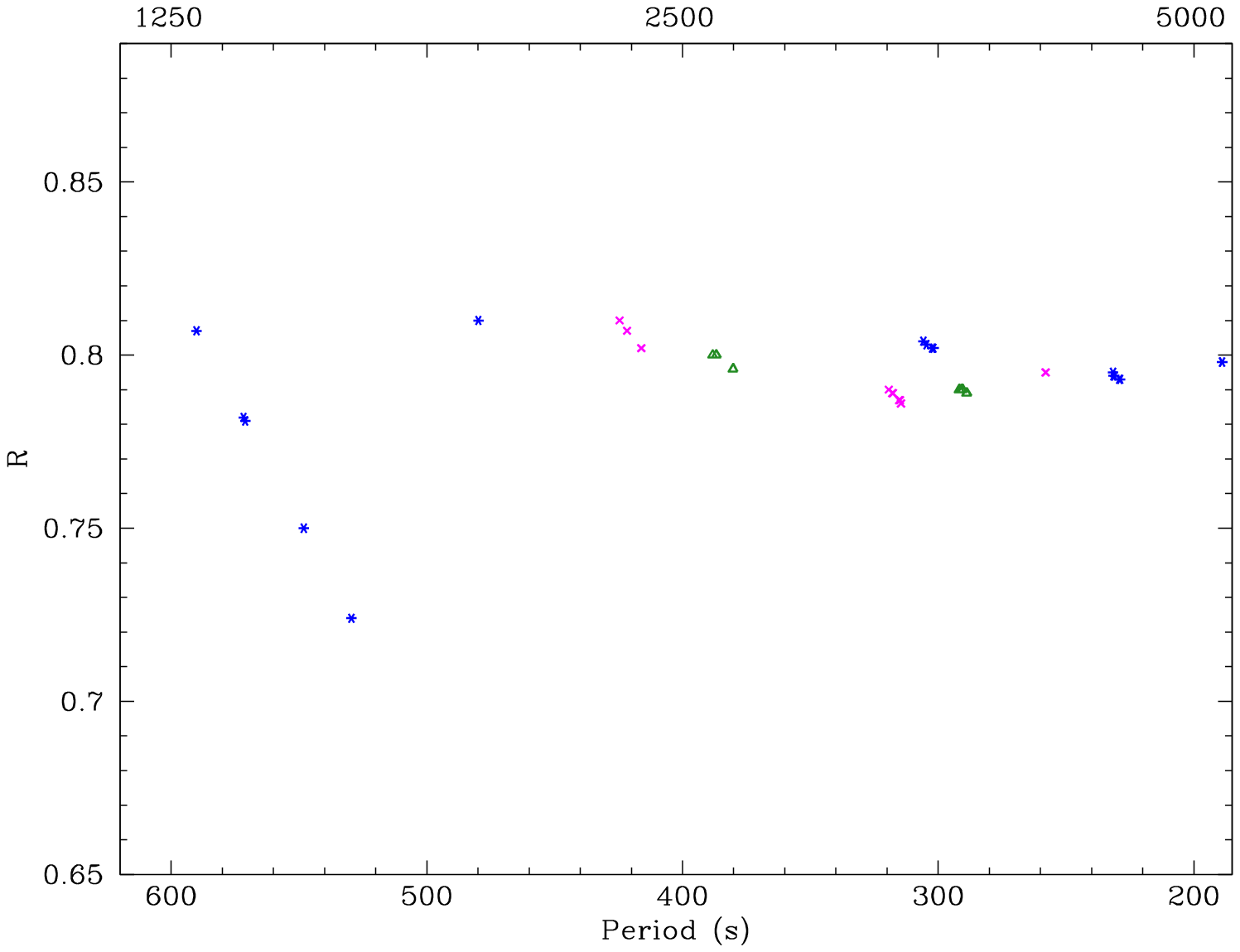, angle=-90,width=\columnwidth}}
	\caption{$p$-mode overtone spacings for stars in this paper organized by T$_{\rm eff}$. 
	Top: Fourier transforms with arrows indicating centers of average
	frequency groups with the spacings between them labeled and radial orders in parentheses. 
	Determined fundamental radial mode indicated with a black arrow
	(except for \ltcnc\, which is determined but not detected). Red vertical line indicates the Nyquist frequency.
	Bottom: Ratio of overtone periods relative to the radial fundamental: $R\approx (P_n/P_k)^{1/(n)}$. Blue asterisks for \pgo,
        magenta crosses for \ltcnc, and green triangles for \hzcnc. }
\label{pover}
\end{figure}

\begin{figure*}
\centerline{\psfig{figure=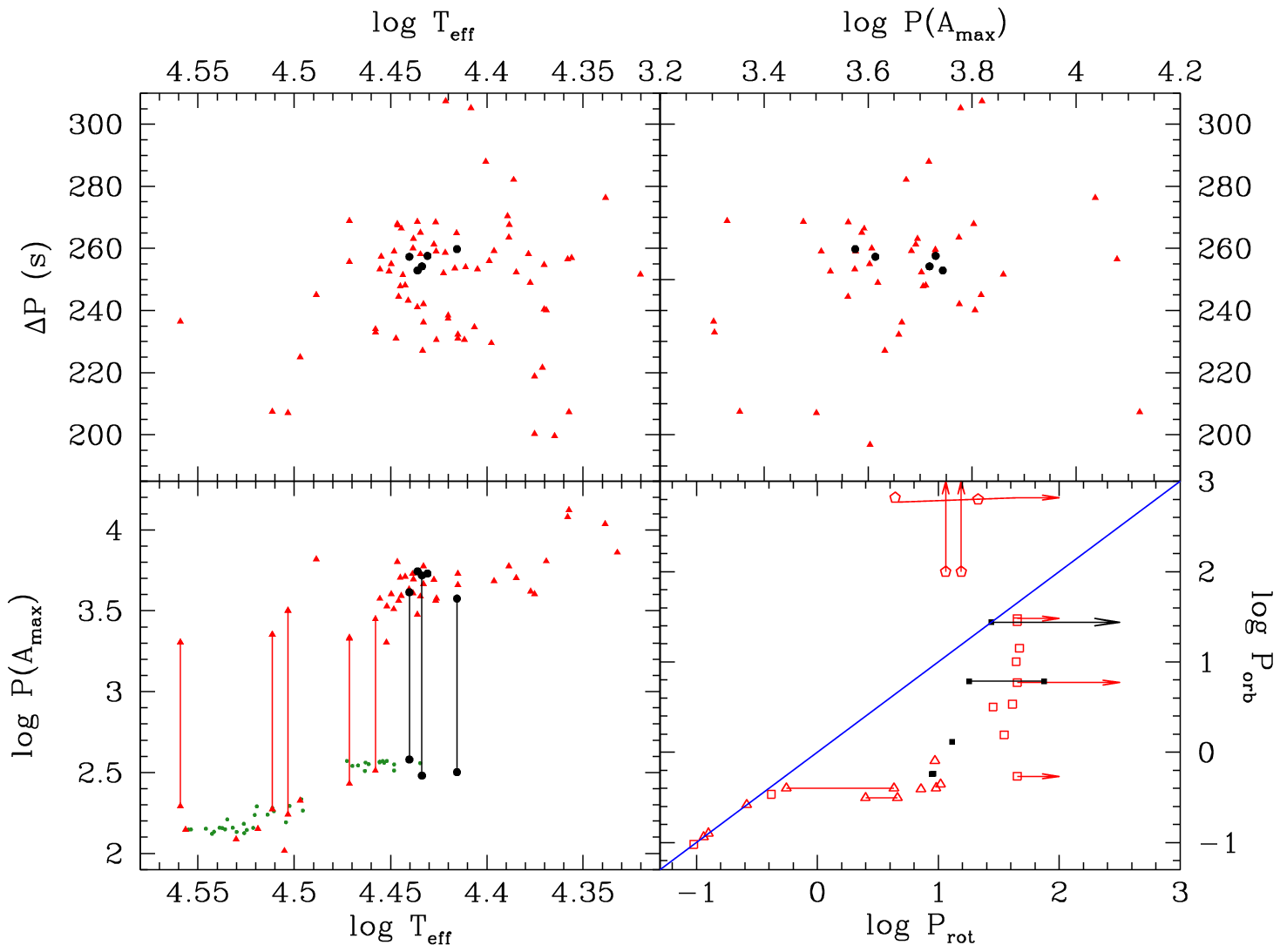, angle=-90,width=\textwidth}}
        \caption{Comparison of seismic properties of the five stars in this paper (black circles) with other
	sdBV stars.
Red triangles for other sdBV stars from \citet{ReedF7}. In the bottom-left panel, hybrid pulsators
have their highest-amplitude p- and g-mode periods connected with a line. Green points are TESS-observed stars from \citet{andy24}.
In the bottom-right panel, point types are as in Fig.\,\ref{kiel} and stars with differential radial
rotation have their rotation periods connected with a line and lower limits are represented with arrows.}
\label{4plot}
\end{figure*}

For the newly discovered binary \pgn, there are no
 signs of binarity in the spectra, other than the
 radial-velocity shifts. There is no brightness excess in the NIR bands.
 For lacking any signs of binarity other than the 14\,h radial-velocity
 orbit, we conclude that \pgn\, consists of a sdB+WD
 binary.

 Subdwarf B stars with main sequence companions heavier than
 M-type show eccentric orbits with periods of a few years or longer, and
 are thought to have undergone a Roche-lobe overflow evolutionary stage \citep{han02,vos13}.
 Given that the orbit of \hzcnc\,
 is the longest known for sdB+WD binaries, and that
 short-binary sdB+WD orbits are always found to be circular within the
 uncertainties, it is interesting to see if a long sdB+WD orbit shows signs
 of eccentricity.   In Fig.\,\ref{rv_k2sdB} we show a 6-parameter eccentric-orbit
 fit to the radial-velocity data of \hzcnc\, (dark-blue curve), together
 with the  circular-orbit fit (light-blue line).  Given the uncertainties of our
 radial-velocity measurements we find no statistical improvement for an eccentric orbit
 compared with the circular-orbit fit. 
 To differentiate between a circular orbit and a slightly eccentric one would require
 data with RV uncertainties to be an order of magnitude smaller than we have achieved.
 We can conclude that even the longest-period
 orbit, that results from binary evolution towards sdB+WD couples,
 is circular to within the uncertainties of our measurements.

 \begin{figure}
        \centerline{\psfig{figure=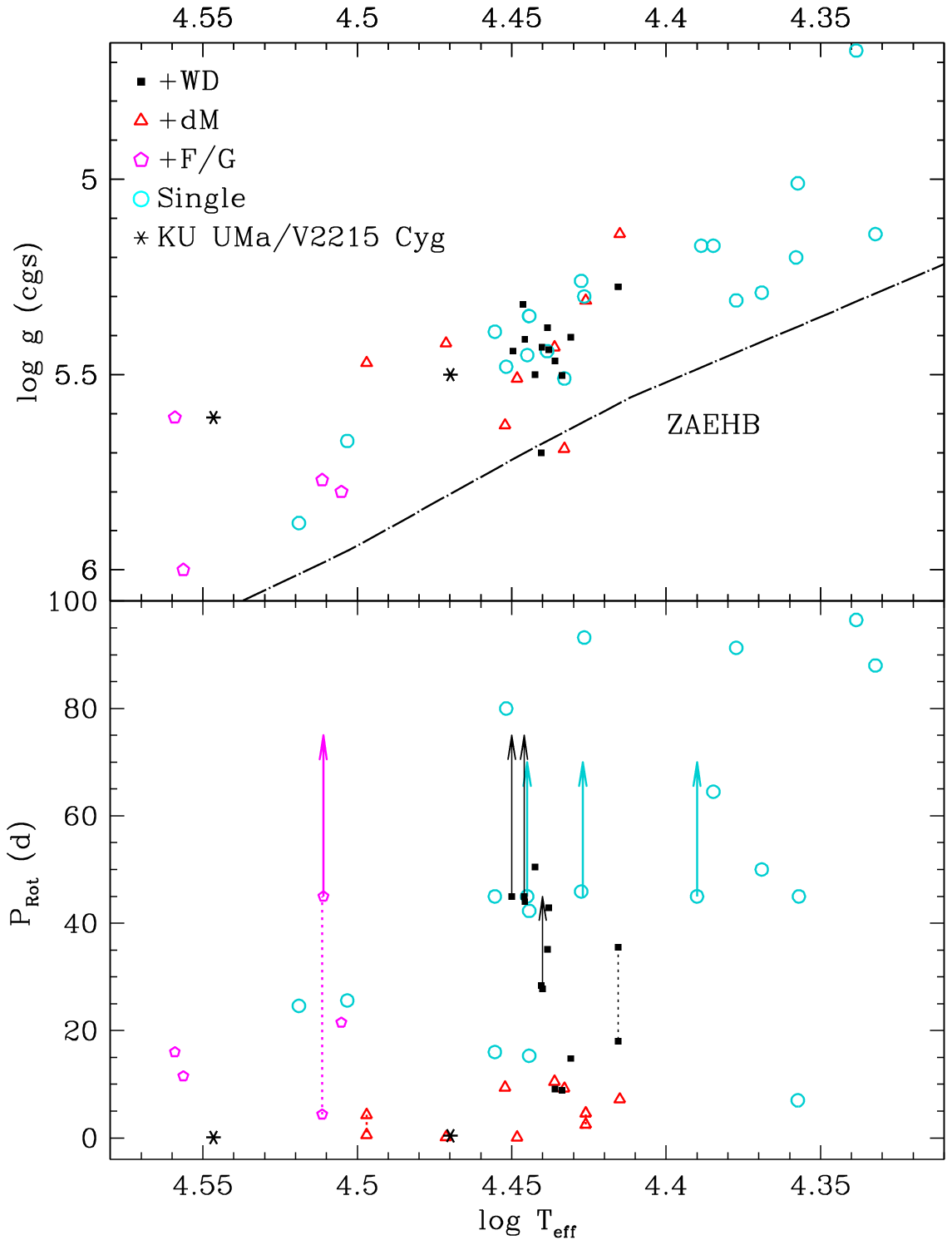,width=\columnwidth} }
        \caption{Comparisons of  \emph{Kepler}-observed stars by binary type. Binary type indicated in the top panel.
The stars KU\,UMa and V2214\,Cyg are ground-observed sdB+WD pulsators. }
        \label{6plotA}
\end{figure}

\section{Discussion}
We have analysed K2 observations of pulsations in five sdBV stars with white dwarf companions. To our knowledge, this completes
the K2-observed sdBV+WD binaries. All five stars are $g$-mode-dominated and three are hybrid
pulsators. \pb\, was only observed in long cadence (30\,m) while the others have short-cadence (58\,s) data.
The five stars are strikingly similar, spectroscopically (\teff\, and $\log g$, Fig.\,\ref{kiel}) and
seismically. The brightest star (\pgo) has the most pulsations detected at over 100, while the other four, with similar apparent
brightnesses, have similar numbers of pulsations detected. Their g-mode asymptotic period spacings fall very near the
sdBV average and only span eight seconds in differences and those with $p$ modes also show overtones near model predictions
(discussed below). 

We have been able to associate $>84$\% of the $g$-mode pulsations with modes and so we can convert them
to reduced periods and compare them between the stars. The left panel of Fig.\,\ref{redPAll} shows
the reduced period spacings and the right separates the pulsations by mode $\ell,m$. Reduced period spacings
are useful to indicate trapped modes and how close the period spacings are to asymptotic 
\citep[$\sigma\Delta\Pi$,][]{constantino15}. Our reduced period plot for \pgo\, differs substantially from that
of \citet{ma23} as they chose azimuthal values to make the $m=0$ component closest to the asymptotic sequence
whereas we used multiplet amplitude ratios as 
cues for our azimuthal choices. We do not specify which option is better, but provide our 
choices as a different option for modelers. It is also noteworthy that \pgo, being the brightest of the stars has
the most deviant spacings $\sigma\Delta\Pi$ which is likely correlated with the ability to detect more pulsations.
The right panel is an indicator of mode density, as a proxy for pulsation driving energy. Period regions where driving is stronger
will result in higher-amplitude pulsations and therefore more detections. Table\,\ref{table_pspace}
lists their reduced period spacings $\Delta\Pi$ and their standard deviations $\sigma\Delta\Pi$. As expected, excluding
the presumed trapped modes of \pgo, their standard deviations are quite similar with all, except the two $\ell=3$ values of
\pgn, under 22\,seconds. 

\subsection{$p$-mode overtones}
For the three stars with p-modes, we notice that there are groups of pulsations which are widely separated (top panel
of Fig.\,\ref{pover}).
This likely indicates the pulsations of each group share the same radial order $n$,  where $n$ is the number of radial nodes,
but have multiple values of $\ell$.
To establish the radial order for each p-mode frequency group we adopt an idea adapted from classical radial pulsation theory.
It is well known that the ratio between the periods $P$ of fundamental (F: $n=0$) and first overtone (1H: $n=1$) pulsations in double mode Cepheids and other pulsating variables is essentially constant for a given class.
Typically $P_1/P_0 \approx 0.70$ for classical Cepheids, $\approx 0.76$ for RR\,Lyr variables, and $\approx 0.77$ for $\delta$\,Sct variables \citep[e.g.][]{jcd93}.
\citep{jeffery2025} has recently shown this value to be $\approx0.81$ for blue large amplitude pulsators (BLAPs).
He also found that, independent of mass $M \leq 0.70M_{\odot}$ and luminosity, the periods of radial modes of order $n\leq 6$ can be related to the fundamental or other radial mode ($k$) by the relation
\begin{equation}
        (P_n/P_k)^{1/(n-k)} \approx R,
\end{equation}
with $R\approx 0.81$.
The relation is valid over a substantial volume of parameter space for stars with similar $T_{\rm eff}$ but lower $g$ than our sdBVs.
Even if the BLAP value for $R$ is not quite correct for sdBVs, we can use the same physics to deduce that  $l=0$ modes in our stars should lie at periods which satisfy  Eqn.\,1 for some value of $R$.
On the basis that the observed low-order p-mode frequencies cluster, we assume that modes belonging to the same cluster share the same radial order $n$ , with frequencies increasing with degree $\ell$.
For sufficiently high $n$,  we expect Eqn.\,1 to be replaced by the asymptotic solution with a uniform spacing in frequency: \(\nu_{n,\ell} \approx \Delta\nu \left( n + \frac{\ell}{2} + \epsilon \right). \)

To deduce mode radial orders in our observed stars we need to identify the period $P_{n,\ell} =P_{0,0}$ of the F mode.
One approximation is to assume that this is the longest-period $p$-mode.
Another is to identify a radial mode with $n>0$ and deduce $P_{0,0} \approx P_{n,0} / 0.81^n$.
Another is to use the mean period $\bar{P}$  for a tightly-spaced cluster of excited modes.
Radial orders can then be estimated for all other modes from $\log(P/P_{0,0})/\log(0.81)$ rounded to the nearest integer.
In either of the latter cases a suitable estimate for $n$ is required.
This can be partly verified by computing the ratio $R$ for all modes, and doing a consistency check.
An integer uncertainty may persist if insufficient low-order modes are present to identify $P_{0,0}$ securely,  but the relative order numbers should be correct.
Confirmation can be had if there is some other means to estimate the stellar mean density, since this can be used to provide P$_{0,0}$  independently.
 The values for $n$ so deduced for our targets are provided in the last columns of Tables\,\ref{tabpgop}, \ref{tabltcncp}, and
\ref{hzplist}.

\begin{figure*}
\centerline{\psfig{figure=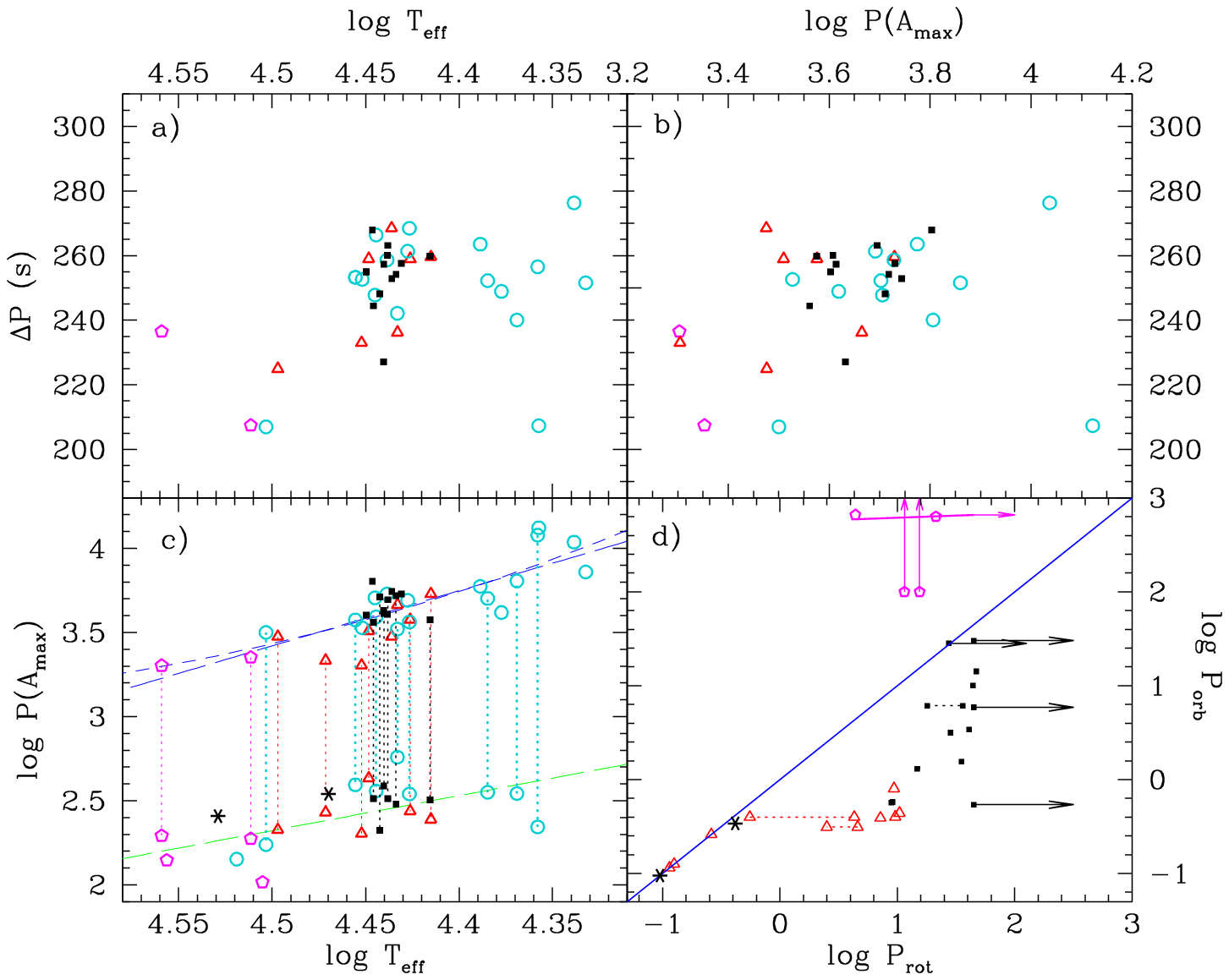,angle=-90,width=\textwidth}}
        \caption{More comparisons of  \emph{Kepler}-observed stars by binary type. Point types/colors same as Fig.\,\ref{6plotA}}
\label{6plotB}
\end{figure*}

The bottom panel of Fig.\,\ref{pover} shows the $R$ value for all the p-mode frequencies for each star.
For modes with $n>1$, periods cluster in groups as the FTs demonstrate.
In the case of PG0101+039, the observed $R$ values for $n = 1$ show a large scatter, presumably corresponding to a range of $\ell$ and the largest departures from asymptotic behaviour.
Otherwise,  $R$  values are similar, regardless of $\ell$ (or $m$ as rotation in these stars is slow).
In this case, we obtain the same identifications for $n$ whether we incorrectly use f82 to represent the F mode (f82 is part of an $\ell=4$ multiplet) or experimentally use f57 as the $n=4, \ell=0$ mode.
For LT\,Cnc the longest period p-mode observed is too short to be the F mode; using the same principle as above we estimate $P_{0,0}\approx 647$\,s from the remainder of the p-mode spectrum.
For modes identified as $n \geq 2$ we obtain tight solutions $0.786 \leq R \leq 0.810$, and mean values $R=0.794\pm0.008, 0.793\pm0.005, 0.792\pm0.005$ for LT\,Cnc, Hz\,Cnc and PG\,0101+039 respectively.
Given the higher gravity of sdBVs, these results are consistent with BLAP theory  (Jeffery 2025) and substantially validate the prediction of Eqn.\,1.

Using the surface gravities given in Table 2 and assuming a classical mass of 0.475M$_{\odot}$, we estimate
$P_{0,0}$ to be 9.09\,min (LT\,Cnc), 6.11\,min (PG0101+039)  and 6.93\,min (HZ\,Cnc) from the period-mean density relation ($\pm 0.17$ min in all cases).
These should be compared with periods estimated from the observed frequencies of 10.72, 12.19 and 15.77\, min respectively.
The result for LT\,Cnc is encouraging. The higher surface gravities for  PG0101+039 and HZ\,Cnc are not consistent
with the observed longest p-mode periods being longer than that of LT\,Cnc.

\subsection{Comparison of seismic properties}
We can compare these five similar stars to each other and also to published group analyses.
Figure\,\ref{4plot} compares seismic, spectroscopic, and rotation-binary properties. It could be
expected that the asymptotic period spacing, $\Delta P$, would be a measure of the resonant cavity
and that would change with changing core and/or envelope mass, which would produce a trend, most likely
with $T_{\rm eff}$ or $\log g$. However, as the  panel a shows, there is no obvious trend, and that
is also true between $\Delta P$ and the period of maximum amplitude, $P\left(A_{\rm max}\right)$ (panel b). 
However, a trend has
been found between $P\left(A_{\rm max}\right)$ and $T_{\rm eff}$ \citep{ReedF7} and that is shown
in  panel c. The five stars in this paper do not deviate from that trend and the hybrid ones supply
cool points for the $p$ modes. Adding the nearly 100 $p$-mode stars from \citet{andy24}  (green points) 
not only continues the trend to
shorter periods, but reveals a gap, dividing the $p$-mode group into two separate groups.

Another trend that these five stars adds to is subsynchronous binary rotation  (panel d). It is possible that \hzcnc\,
rotates synchronous with its orbit, but the other four stars rotate slower than their binary period. \pgo\, and
\pgn\, appear nearly identical, with short binary periods just over half a day and spin periods of 8.9 and
9.1\,days, respectively. \pgo\, also appears to spin like a solid-body object, with $p$ and $g$ modes indicating a nearly
identical spin period while \ltcnc\, appears to rotate differentially radially with the $p$-mode multiplets indicating
a spin period of $18\pm2$\,days and the $g$-mode multiplets indicating a spin period of $75\pm15$\,days. This differential
rotation is in agreement with other sdBV stars where the core rotates more slowly than the envelope \citep{ReedF7}.
A surprise are the multiplets
of \pb, which vary, decreasing in separation with increasing frequency. Taken at face value, they would indicate
a doubling of spin period, but as they are all $g$ modes, which should roughly sample the same region of the star,
this would be extremely surprising. Similar to the cases of converging and diverging multiplets observed in two
\emph{Kepler}-observed stars \citep{kern16,kern18}, there is no obvious physical explanation, but rotation is not likely to
be the cause.

\subsubsection{Subdwarf $+$ white dwarf binaries}
This work completes observational seismic analyses of all \emph{Kepler}-observed sdBV stars which are in binaries with
white dwarf companions. In total, the \emph{Kepler} and K2 missions observed 12 sdB$+$WD binaries, all predominantly
g-mode pulsators. 
Of the five sdB+WD binaries with hybrid pulsations, \ltcnc\, is the only one where p- and g-mode multiplets indicate
radially differential rotation. It is also has the longest orbital period of those five binary systems.

In Figs.\,\ref{6plotA} and \ref{6plotB} and Table\,\ref{table_props}
we compare some features of published \emph{Kepler}-observed stars. To date, besides the sdB+WD
pulsators, this includes nine sdB+dM, four sdB+F/G, and 18 stars for which no companion has been detected. Those
figures reveal some possible selection effects. In the $T_{\rm eff}$ panels, the sdB+F/G binaries only appear at the hot end and
that may be related to those being the brightest companions and so brighter sdB stars are needed to detect them. This
is also indicated in column 1 of Table\,\ref{table_props}, where they have the highest average $T_{\rm eff}$.
They also have shorter $\Delta P$  (Fig.\,\ref{6plotB} panels a \& c) 
while the other types are near the ``canonical'' value of 250\,s.
The sdB+WD pulsators cluster tightly in  their spectroscopic and seismic properties. However, there are two known
sdBs+WD companions detected with ground-based and TESS data \citep[KU\,UMa and V\,2214\,Cyg;][]{reed04b,mdr11} which are
indicated by black asterisks. Both of those are p-mode-only pulsators, are substantially hotter, and with
the shortest orbital periods. Perhaps that is causal of their differences, but with only two examples, it is not possible
to tell.
In (Fig.\,\ref{6plotB} panel d
the sdB+dM only appear at the shortest orbital periods and this is likely because they are detected via the reflection effect
and with more distant separations, the dM star is not visible. The sdB+WD binary periods are also short, as the WD spectroscopically
looks very similar to the sdB star, but is much dimmer. Those binaries are often detected via ellipsoidal variations or Doppler
boosting \citep[e.g.][]{telting12a}, which again prefers shorter orbital periods. So some currently-unknown fraction of the
apparently single sdB stars are likely binaries, either with orbits too separated for current detections or at inclinations
minimizing those effects.

\begin{table*}
	\caption{Comparison of properties between sdBV stars with companions.}
        \label{table_props}
\centering
\begin{tabular}{l|cc|cc|cc|cc}
\hline\hline
 & \multicolumn{2}{c}{sdB+WD} & \multicolumn{2}{c}{sdB+dM} & \multicolumn{2}{c}{sdB+F/G} &  \multicolumn{2}{c}{single}  \\
No. & \multicolumn{2}{c}{12} & \multicolumn{2}{c}{8} & \multicolumn{2}{c}{4} &  \multicolumn{2}{c}{18}  \\ 
&  Ave & $\sigma$ &  Ave & $\sigma$ &  Ave & $\sigma$ &  Ave & $\sigma$ \\ \hline
	$T_{\rm eff}$ [K] & 28\,430 & 560 & 28\,060 & 1740 & 34\,170 & 2\,250 & 26\,120 & 3\,270 \\
	$\log g$ [cgs] & 5.44 & 0.11 & 5.45 & 0.17 & 5.80 & 0.16 & 5.32 & 0.26 \\
	$\Delta P$ [s]& 254 & 11 & 249 & 17 & 222 & 21 & 250 & 19 \\
	$f_{A_{max}}$ [$\mu Hz$] & 218 & 38 & 326 & 109 & 471 & 38 & 203 & 77 \\ \hline
\hline
\end{tabular}
\end{table*} 

There are still unpublished K2-observed sdBV stars, which will be presented in a future
paper and an analysis of pulsation statistics, which will also be published at a future date.

\section{Summary of results}
This paper examined five sdBV+WD binaries, completing analyses of \emph{Kepler}-observed stars of
this type. \\
$\bullet$ In total, \emph{Kepler} observed 12 sdBV+WD binaries with orbital periods between 
0.5 and 28\,d. \\
$\bullet$ All (12) are g-mode dominated though there are two known ground-observed p-mode sdBV+WD stars (KU\,UMa
and V2214\,Cyg). Six (50\%) are hybrid pulsators. \\
$\bullet$ Frequency multiplets indicate all are rotating subsynchronous to their binary periods. We detect
p- and g-mode multiplets in two stars, one we determine to be rotating like a solid-body (\pgo) and the
other indicates radially differential rotation with the envelope
rotating faster (\ltcnc).\\
$\bullet$ For the five stars of this paper, period spacings ($\Delta P$) of $\ell =1$ modes range from 227 to 268\,s, averaging to
$254\pm 11$\,s. \\
$\bullet$ Spectroscopy indicates very similar effective temperatures, $26\,000<T_{\rm eff}<28\,200$, averaging to
$27\,430\pm560$\,K. The full range of \emph{Kepler}-observed sdBV stars is $21\,490<T_{\rm eff}<36\,230$.\\
$\bullet$ We apply a new method to determine p-mode radial orders.

\smallskip

{\bf Data Availability Statement:} The photometric data underlying this article are available in the
Mikulski Archive for Space Telescopes (MAST). \url{https://archive.stsci.edu/}
Data obtained with the Nordic Optical Telescope are available after
a one-year proprietary period. \url{http://www.not.iac.es/archive/}
\smallskip

ACKNOWLEDGMENTS: This paper includes data obtained by the \emph{Kepler} mission. Funding for the
 \emph{Kepler} mission is provided by the NASA Science Mission directorate.
Data presented in this paper were obtained from the Mikulski Archive for
Space Telescopes (MAST). STScI is operated by the Association of Universities
for Research in Astronomy, Inc., under NASA contract NAS5-26555. Support
for MAST for non-HST data is provided by the NASA Office of Space Science via
grant NNX13AC07G and by other grants and contracts.

The spectroscopic observations used in this work were obtained with the
Nordic Optical Telescope at the Observatorio del Roque de los Muchachos
and operated jointly by Denmark, Finland, Iceland, Norway, and
Sweden.

YG was supported by the Missouri Space Grant which is funded by NASA.

\hzcnc\, has also been analyzed by Xiaoyu Ma and will appear in her Ph.D thesis. 

\bibliography{sdbrefsMNRAS}

\appendix
\section{Seismic properties and list of spectroscopic observations.}
\begin{table*} \caption{Seismic properties of $g$-mode pulsations in \pgo .  Column four provides either the mode identifications $\ell ,m$ or C indicative of combination, or N indicative of not detected from \citet{ma23}. Column six indicates mode identifications based on frequency multiplets and seven based on period spacings with deviations from equal period spacings in column eight. Column nine gives our best estimate for the mode. $^*$Note that all $n$ values are relative to an arbitrary $n_0$.} \label{tabpgog}
\begin{tabular}{llccccccc} ID & Frequency & period & MA & Amp & Mode (multiplet) & Mode ($\Delta$P) & $\Delta$P/P & mode \\
	- & [$\mu$Hz] & [s] & - & [ppt] & l,m & l,n$^*$ & [\%] & n$^*$,l,m \\ \hline
	f74 & 84.1489 & 11883.7 & 1,-1 & 0.046 & - & 1,42 & 2.6 & 42,1,+1 \\
	fI & 96.9634 & 10313.2 & N & 0.046 & - & 1,36 & -8.7 & 36,1,+1 \\ 
	f52 & 98.7149 & 10130.2 & C & 0.057 & - & 2,64 & -1.9 & 64,2,-1 \\ 
	fF & 108.137 & 9247.5 & N & 0.050 & - & 2,60 & 2.4 & 60,2,+2 \\ 
	f62 & 111.640 & 8957.4 & Y & 0.057 & 2,-1 & 1,31 & 4.3 & 31,1,-1 \\ 
	f54 & 112.461 & 8892.0 & 2,-2 & 0.058 & 2,0 & - & - & - \\ 
	fM & 113.902 & 8779.5 & N & 0.042 & - & 1,30 & -6.9 & 30,1,+1 \\ 
	f63 & 114.345 & 8745.5 & 2,-2 & 0.057 & - & - & - & - \\ 
	fG & 114.744 & 8715.1 & N & 0.048 & - & 2,55 & 6.6 & 55,2,0 \\ 
	f53 & 119.089 & 8397.1 & 1,-1 & 0.052 & 2,0 & 2,53 & -10.9 & 53,2,0 \\ 
	f49 & 121.351 & 8240.6 & 1,+1 & 0.061 & 2,+2 & 2,53 & - & 53,2,+2 \\ 
	f26 & 124.849 & 8009.7 & Y & 0.094 & 2,+3 & 1,27 & -7.6 & 27,1,-1 \\ 
	fH & 129.756 & 7706.8 & N & 0.047 & 2,0 & - & - & 48,2,0 \\ 
	f29 & 130.848 & 7642.5 & C & 0.086 & 2,+1 & - & - & 48,2,+1 \\ 
	fK & 133.472 & 7492.2 & N & 0.043 & - & - & - & - \\ 
	fC & 138.086 & 7241.9 & N & 0.059 &   & 1,24 & -7.8 & 24,1,-1 \\ 
	f61 & 138.426 & 7224.1 & 2,-1 & 0.048 & 2,0 & - & - & -,2,0 \\ 
	fL & 139.519 & 7167.5 & N & 0.043 & 2,+1 & - & - & -,2,+1 \\ 
	f44 & 144.087 & 6940.3 & 1,0 & 0.063 & 1,0 & 1,23 & -13.3 & 23,1,0 \\ 
	f20 & 144.717 & 6910.0 & 1,+1 & 0.116 & 1,+1 & 1,23 & -12.8 & 23,1,+1 \\ 
	fA & 145.081 & 6892.7 & N & 0.072 & - & 2,43 & -3.8 & 43,2,+1 \\ 
	f30 & 155.283 & 6439.9 & 1,0 & 0.088 & 1,-1 &   & - & -,1,-1 \\ 
	f37 & 155.914 & 6413.8 & 1,+1 & 0.072 & 1,0 or 2,0 & - & - & -,1,0 \\ 
	f56 & 157.129 & 6364.2 & Y & 0.051 & - & 2,39 & -1.4 & 39,2,0 \\ 
	f32 & 163.128 & 6130.2 & 2,-1 & 0.078 & 1,-1 &   & - & -,1,-1 \\ 
	fJ & 164.415 & 6082.2 & N & 0.044 & 1,+1 &   & - & -,1,+1 \\ 
	f7 & 167.695 & 5963.2 & 1,-1 & 0.303 & 1,-1 & 1,19 & 0.7 & 19,1,-1 \\ 
	f27 & 168.447 & 5936.6 & 1,0 & 0.090 & 1,0 & 1,19 & -9.8 & 19,1,0 \\ 
	f9 & 169.080 & 5914.4 & 1,+1 & 0.212 & 1,+1 & 1,19 & -18.6 & 19,1,+1 \\
f23 & 176.059 & 5679.9 & 1,0/2,-2 & 0.103 & - & 2,34 & 6.4 & 34,2,-1 \\ 
	f34 & 180.361 & 5544.4 & 2,-2 & 0.079 & - & 2,33 & -7.8 & 33,2,-2 \\ 
fN & 181.718 & 5503.0 & N & 0.038 & 1,-1 & 1,17 & 15.9 & 17,1,-1 \\ 
	f35 & 182.532 & 5478.5 & 1,-1 & 0.071 & 1,0 & 1,17 & 9.3 & 17,1,0 \\ 
f11 & 183.163 & 5459.6 & 1,0 & 0.201 & 1,+1 & 1,17 & 1.8 & 17,1,+1 \\ 
	f42 & 185.664 & 5386.1 & 2,-2 & 0.060 & - & - & - & - \\ 
fE & 186.063 & 5374.5 & N & 0.057 & - & 2,32 & 0.1 & 32,2,+1 \\ 
	f1 & 191.436 & 5223.7 & 1,0/2,+2 & 1.536 & 1, & 1,16 & 1.6 & 16,1,+1 \\ 
f66 & 200.578 & 4985.6 & 2,-1 & 0.047 & 2,-1 & 2,30 & -6.5 & 30,2,-2 \\ 
	f13 & 201.647 & 4959.2 & 2,0 & 0.191 & 2,0 & 2,29 & 0.9  & 29,2,+2 \\ 
fB & 201.935 & 4952.1 & N & 0.063 & 1,0 & 1,15 & 1.4 & 15,1,0 \\ 
	f40 & 202.536 & 4937.4 & 2,1 & 0.076 & 1,+1 & 1,15 & -4.4 & 15,1,+1 \\ 
f2 & 211.078 & 4737.6 & 2,0 & 0.476 & 1,-1 & 1,14 & 16.6 & 14,1,-1 \\ 
	f19 & 212.393 & 4708.3 & 1,0 & 0.134 & 1,+1 & 1,14 & 5.1 & 14,1,+1 \\ 
f18 & 216.251 & 4624.3 & Y & 0.155 & - & 2,27 & 8.5/-7.9 & 27,2,0/+1 \\ 
	f43 & 231.048 & 4328.1 & 2,-1 & 0.064 & 2,-1 & 2,25 & 6.0 & 25,2,-1 \\ 
f46 & 232.156 & 4307.4 & 2,0 & 0.060 & 2,0 & 2,25 & -8.2 & 25,2,0 \\ 
f64 & 234.354 & 4267.0 & 2,+2 & 0.049 & 2,+2 & 2,25 & -35.8 & 25,2,+2 \\ 
	f21 & 238.295 & 4196.5 & 1,0 & 0.106 & 1,-1/>2,0 & 1,12/2,24 & -1.7,15.9 & 12,1,-1 \\ 
f45 & 239.599 & 4173.6 & 2,1 & 0.060 & 1,+1/>2,1 & 1,12/2,24 & -1.5/0.3 & 12,1,+1 \\ 
	f51 & 250.332 & 3994.7 & Y & 0.058 & 1,-1 & - & - & -,1,-1 \\ 
f59 & 251.735 & 3972.4 & C & 0.046 & 1,+1 & - & - & -,1,+1 \\ 
	f4 & 254.637 & 3927.2 & 1,0 & 0.405 & 1,-1 & 1,11  & -3.4 & 11,1,-1 \\ 
f28 & 255.285 & 3917.2 & 1,+1 & 0.089 & 1,0 & 1,11  & -7.4 & 11,1,0 \\ 
	f10 & 258.462 & 3869.0 & Y & 0.215 & - & 2,22 & 3.3 & 22,2,-1 \\ 
f22 & 291.065 & 3435.7 & 1,-1 & 0.103 & 1,0 & 1,9 & 2.4 & 9,1,0 \\ 
	f17 & 291.737 & 3427.7 & 1,0 & 0.153 & 1,+1 & 1,9 & -0.7 & 9,1,+1 \\ 
fO & 292.417 & 3419.8 & N & 0.038 & 2,+1 & 2,19 & -6.5  & 19,2,+1 \\ 
	f47 & 293.531 & 3406.8 & 2,+2 & 0.059 & 2,0 & 2,19 & -6.6  & 19,2,+2 \\ 
f16 & 305.513 & 3273.2 & 2,-2 & 0.151 & - & 2,18 & 0.7 & 18,2,+2 \\ 
f12 & 314.080 & 3183.9 & 1,0 & 0.185 & 1,-1 & 1,8 & 3.0 & 8,1,-1 \\ \hline \end{tabular} \end{table*}

\begin{table*} \contcaption{}     \begin{tabular}{llccccccc} ID & Frequency & period & MA & Amp & Mode (multiplet) & Mode ($\Delta$P) & $\Delta$P/P & mode \\ 
	- & [$\mu$Hz] & [s] & - & [ppt] & l,m & l,n$^*$ & [\%] & n$^*$,l,m \\ \hline 
	f14 & 315.379 & 3170.8 & Y & 0.173 & 1,+1 & 1,8 & -2.2 & 8,1,+1 \\ 
	f31 & 316.798 & 3156.6 & Y & 0.077 & - & 2,17 & 4.6 & 17,2,0 \\ 
	f84 & 332.196 & 3010.3 & 2,0 & 0.033 & 2,-1 & 2,16 & 4.6 & 16,2,-1 \\ 
	f94 & 333.26 & 3000.7 & 2,+1 & 0.025 & 2,0 & 2,16 & -2.0 & 16,2,0 \\ 
	f69 & 343.320 & 2912.7 & 1,-1 & 0.049 & 1,0 & 1,7 & -4.1 & 7,1,0 \\ 
	f15 & 343.956 & 2907.3 & 1,0 & 0.168 & 1,+1 & 1,7 & -6.2 & 7,1,+1 \\ 
	f48 & 346.303 & 2887.6 & C & 0.060 & 2,-2 & 2,15 & 20.7 & 15,2,-2 \\ 
	f76 & 347.399 & 2878.5 & C & 0.039 & 2,-1 & 2,15 & 14.5 & 15,2,-1 \\ 
	f78 & 350.706 & 2851.4 & 2,-2 & 0.025 & 2,+2 & 2,15 & -4.1 & 15,2,+2 \\ 
	f6 & 377.420 & 2649.6 & 1,-1 & 0.303 & 1,-1 & 1,6 & -8.0 & 6,1,-1 \\ 
	f25 & 378.085 & 2644.9 & 1,-1 & 0.090 & 1,0 & 1,6 & -9.9 & 6,1,0 \\ 
	f5 & 378.718 & 2640.5 & 1,+1 & 0.304 & 1,+1 & 1,6 & -11.6 & 6,1,+1 \\ 
	f39 & 390.534 & 2560.6 & 2,-2 & 0.066 &   & 2,13 & -3.0 & 13,2,0 \\ 
	f75 & 400.614 & 2496.2 & C & 0.041 & - & - & - & - \\ 
	f125 & 412.861 & 2422.1 & Y & 0.020 & >2,-1 & 1,5/2,12 & -3.0/2.3 & 12,2,-2 \\ 
	f41 & 414.494 & 2412.6 & 1,0 & 0.067 & 2,-1 & 1,5/2,12 & -0.1/-4.3 & 5,1,-1 \\ 
	f77 & 414.969 & 2409.8 & C & 0.039 & >2,0 & 1,5/2,12 & -2.7/-6.2 & 12,2,0 \\ 
	fP & 415.659 & 2405.8 & N & 0.020 & 2,0 & 1,5/2,12 & -1.8/-4.5 & 12,2,+1 \\ 
	f111 & 450.112 & 2221.7 & 1,-1 & 0.023 & 1,0 & -  & - & -,1,0 \\ 
	f24 & 450.807 & 2218.2 & 1,0 & 0.092 & 1,+1 & -  & - & -,1,+1 \\ 
	f105 & 544.112 & 1837.9 & Y & 0.022 & - & 2,8 & 2.6 & 8,2,0 \\ 
	f142 & 549.607 & 1819.5 & Y & 0.019 & >2,-1 & 3,13 & 5.8 & 13,3,-1 \\ 
	f109 & 550.775 & 1815.6 & C & 0.021 & >2,0 & 3,13 & 2.1 & 13,3,0 \\ 
	f33 & 570.219 & 1753.7 & 4,-3 & 0.077 & >2,-3 & 6,27 & 19.6 & 27,6,-3 \\ 
	f60 & 571.511 & 1749.7 & 4,-2 & 0.053 & >2,-2 & 6,27 & 12.4 & 27,6,-2 \\ 
	f102 & 572.773 & 1745.9 & 4,-1 & 0.026 & >2,-1 & 6,27 & 5.4 & 27,6,-1 \\ 
	f36 & 574.025 & 1742.1 & 4,0 & 0.078 & >2,0 & 6,27 & -1.5 & 27,6,0 \\ 
	f99 & 576.509 & 1734.6 & 4,2 & 0.027 & >2,+2 & 6,27 & -15.1 & 27,6,+2 \\ 
	f112 & 597.734 & 1673.0 & C & 0.022 & 2,-1 & 2,7 & -12.2 & 7,2,-1 \\ 
	f146 & 601.516 & 1662.5 & Y & 0.018 & 2,+2 & 2,7 & -13.1 & 7,2,+2 \\ 
	fZ & 606.478 & 1648.9 & N & 0.013 & >2,-1 & 4,16 & 4.8 & 16,4,-1 \\ 
	f65 & 607.784 & 1645.3 & Y & 0.047 & >2,0 & 4,16 & 0.4 & 16,4,0 \\ 
	fY & 615.812 & 1623.9 & N & 0.014 & >2,-1 & 6,25 & -15.4 & 25,6,+2 \\ 
	f73 & 617.205 & 1620.2 & Y & 0.042 & >2,0 & 6,25 & -22.1 & 25,6,+3 \\ 
	f104 & 635.539 & 1573.5 & Y & 0.024 & >2,-4 & 4,15 & 10.7 & 15,4,-4 \\ 
	f119 & 639.264 & 1564.3 & C & 0.021 & >2,-1 & 4,15 & -0.8 & 15,4,-1 \\ 
fW & 640.645 & 1560.9 & N & 0.015 & >2,0 & 4,15 & -5.0 & 15,4,0 \\ 
	fT & 645.532 & 1549.1 & N & 0.016 & >2,+4 & 4,15 & -19.7 & 15,4,+4 \\ 
f137 & 649.594 & 1539.4 & C & 0.021 & >2,-1 & 6,23 & 31.7 & 23,6,-3 \\ 
	f79 & 651.395 & 1535.2 & 2,-1 & 0.035 & >2,0 & 6,23 & 24.0 & 23,6,-2 \\ 
f50 & 653.535 & 1530.1 & 2,+1 & 0.060 & >2,+2 & 6,23 & 14.9 & 23,6,0 \\ 
	fX & 676.177 & 1478.9 & N & 0.015 & >2,-2 & 6,22 & 22.2 & 22,6,-4 \\ 
f148 & 678.782 & 1473.2 & C & 0.016 & >2,0 & 6,22 & 12.0 & 22,6,-2 \\ 
	f136 & 680.449 & 1469.6 & C & 0.018 & >2,+1 & 6,22 & 5.5 & 22,6,-1 \\ 
f144 & 681.42 & 1467.5 & C & 0.018 & >2,+2 & 6,22 & 1.7 & 22,6,0 \\ 
	f87 & 702.651 & 1423.2 & 6 & 0.031 & 4,-4 & 6,21 & 21.3 & 21,6,-4 \\ 
f88 & 705.207 & 1418.0 & 6 & 0.031 & 4,-2 & 6,21 & 12.0 & 21,6,-2 \\ 
	f38 & 707.780 & 1412.9 & 6/8 & 0.071 & 4,0 & 6,21 & 2.7 & 21,6,0 \\ 
f67 & 710.387 & 1407.7 & 8 & 0.046 & 4,+2 & 6,21 & -6.7 & 21,6,+2 \\ 
	f80 & 711.718 & 1405.1 & 8 & 0.036 & 4,+3 & 6,21 & -11.5 & 21,6,+3 \\ 
f132 & 718.571 & 1391.7 & C & 0.020 & - & 2,5 & -2.6 & 5,2,0 \\ 
	f169 & 830.064 & 1204.7 & C & 0.017 & >2,0 or >2,2 & - & - & - \\ 
f83 & 831.463 & 1202.7 & C & 0.034 & >2,+1 or >2,3 & - & - & - \\ 
	fGG & 847.155 & 1180.4 & N & 0.011 & >2,0 & 4,10 & 19.8 & 10,4,-4 \\ 
f120 & 849.818 & 1176.7 & Y & 0.022 & >2,+1 & 4,10 & 15.1 & 10,4,-2 \\ 
	f126 & 851.194 & 1174.8 & C & 0.023 & >2,1 or >2,3 & 4,10 & 12.8 & 10,4,-1 \\ 
f93 & 871.462 & 1147.5 & Y & 0.031 & - & *  & -  & - \\ 
	f108 & 871.911 & 1146.9 & Y & 0.024 & - & -  & -  & - \\ 
f71 & 872.834 & 1145.7 & Y & 0.041 & - & -  & -  & - \\ 
fFF & 881.608 & 1134.3 & N & 0.012 & >2,-4 & 6,16 & -1.5 & 16,6,-4 \\ 
f131 & 884.258 & 1130.9 & $\geq 8$ & 0.019 & >2,-2 & 6,16 & -7.6 & 16,6,-2 \\ 
f115 & 886.905 & 1127.5 & $\geq 8$ & 0.023 & >2,0 & 6,16 & -13.7 & 16,6,0 \\ \hline \end{tabular} \end{table*} 

\begin{table*} \contcaption{}     \begin{tabular}{llccccccc} ID & Frequency & period & MA & Amp & Mode (multiplet) & Mode ($\Delta$P) & $\Delta$P/P & mode \\ 
	- & [$\mu$Hz] & [s] & - & [ppt] & l,m & l,n$^*$ & [\%] & n$^*$,l,m \\ \hline 
f96 & 889.585 & 1124.1 & $\geq 8$ & 0.027 & >2,+2 & 6,16 & -19.9 & 16,6+2 \\ 
	f86 & 890.971 & 1122.4 & $\geq 8$ & 0.034 & >2,+3 & 6,16 & -23.0 & 16,6,+3 \\ 
f118 & 904.254 & 1105.9 & C & 0.022 & - & 2,3 & 2.0 & 3,2,0 \\ 
	f127 & 929.456 & 1075.9 & $\geq 8$ & 0.022 & >2,-3 & 6,15 & -7.2 & 15,6,-4 \\ 
f179 & 932.054 & 1072.9 & $\geq 8$ & 0.012 & >2,-1 & 6,15 & -12.7 & 15,6,-2 \\ 
	f170 & 933.407 & 1071.3 & $\geq 8$ & 0.015 & >2,0 & 6,15 & -15.5 & 15,6,-1 \\ 
f163 & 934.783 & 1069.8 & $\geq 8$ & 0.015 & >2,+1  & 6,15 & -18.3 & 15,6,0 \\ 
	f178 & 936.127 & 1068.2 & $\geq 8$ & 0.012 & >2,+2 & 6,15 & -21.1 & 15,6,+1 \\ 
f92 & 953.600 & 1048.7 & Y & 0.031 & >2,-1 & - & - & -,-,-1 \\ 
	f72 & 955.040 & 1047.1 & Y & 0.040 & >2,0 & - & - & -,-,0 \\ 
f135 & 1060.51 & 942.9 & 8 & 0.018 & 4,-4 & 4,7 & 23.1 & 7,4,-4 \\ 
	f133 & 1063.18 & 940.6 & 8 & 0.018 & 4,-2 & 4,7 & 20.2 & 7,4,-2 \\ 
f106 & 1065.77 & 938.3 & 8 & 0.021 & 4,0 & 4,7 & 17.3 & 7,4,0 \\ 
f180 & 1068.35 & 936.0 & 8 & 0.015 & 4,+2 & 4,7 & 14.5 & 7,4,+2 \\ \hline \end{tabular} \end{table*}

\begin{table*} \caption{Seismic properties of $p$-mode pulsations in \pgo . Column four provides either the mode identifications $\ell ,m$ or C indicative of combination, or N indicative of not detected from \citet{ma23}. Column six indicates mode identifcations based on frequency multiplets for $\ell$ and $m$ and estimated $n$ values based on where the radial fundamental mode would likely occur.} \label{tabpgop}     \begin{tabular}{llccccccc} MA ID & Frequency & period & MA & Amp & Mode \\
	- & [$\mu$Hz] & [s] & - & [ppt] & n,l,m \\ \hline 
		f82 & 1367.67 & 731.2 & 4 & 0.035 & 0,4,-4 \\ 
		f107 & 1370.39 & 729.7 & 4 & 0.025 & 0,4,-2 \\ 
		f101 & 1373.15 & 728.3 & 4 & 0.023 & 0,4,0 \\ 
		f152 & 1377.25 & 726.1 & C & 0.015 & 0,4,+3 \\ 
		f91 & 1478.31 & 676.45 & Y & 0.031 & 0,-,- \\
		f130 & 1555.00 & 643.09 & C & 0.021 & 0,-,- \\
		f117 & 1568.38 & 637.60 & Y & 0.022 & 0,-,- \\
		f138 & 1577.09 & 634.08 & C & 0.020 & 0,-,- \\
		f165 & 1601.09 & 624.57 & C & 0.021 & 0,-,- \\
		f173 & 1694.55 & 590.13 & C & 0.015 & 1,-,- \\
		f116 & 1749.14 & 571.71 & C & 0.021 & 1,-,- \\
		f134 & 1751.41 & 570.97 & Y & 0.018 & 1,-,- \\
		f110 & 1824.39 & 548.13 & Y & 0.022 & 1,-,- \\
		f98 & 1888.43 & 529.54 & Y & 0.027 & 1,-,- \\
		fAA & 2084.28 & 479.78 & N & 0.013 & 2,-,- \\
		f85 & 3269.90 & 305.82 & Y & 0.033 & 4,-,- \\ 
		f121 & 3282.50 & 304.65 & C & 0.018 & 4,-,- \\ 
		f57 & 3306.93 & 302.40 & 4 & 0.050 & 4,0,0 \\ 
		f128 & 3310.76 & 302.05 & 4 & 0.019 & 4,1,-1 \\ 
		f177 & 3312.22 & 301.91 & 4 & 0.012 & 4,1,0 \\ 
		f167 & 3313.59 & 301.79 & 4 & 0.014 & 4,1,+1 \\ 
		fS & 4318.14 & 231.58 & N & 0.017 & 5,-,- \\ 
		fQ & 4318.32 & 231.57 & N & 0.019 & 5,-,- \\ 
		f90 & 4318.71 & 231.55 & Y & 0.028 & 5,-,- \\ 
		fR & 4320.38 & 231.46 & N & 0.018 & 5,-,- \\ 
		f89 & 4320.73 & 231.44 & Y & 0.030 & 5,-,- \\ 
		f171 & 4326.70 & 231.12 & Y & 0.017 & 5,-,- \\ 
		f151 & 4362.38 & 229.23 & Y & 0.012 & 5,-,- \\ 
		fU & 4362.63 & 229.22 & N & 0.016 & 5,-,- \\ 
		fV & 4363.38 & 229.18 & N & 0.016 & 5,-,- \\ 
		f213 & 4371.20 & 228.77 & Y & 0.021 & 5,-,- \\ 
		fBB & 5290.01 & 189.04 & N & 0.013 & 6,2,-2 \\ 
	fDD & 5292.61 & 188.94 & N & 0.012 & 6,2,0 \\ 
	fCC & 5292.84 & 188.93 & N & 0.013 & 6,-,- \\ 
	fEE & 5295.17 & 188.85 & N & 0.012 & 6,2,+2 \\ \hline     
\end{tabular} 
\end{table*} 

\begin{table*}
	\caption{Seismic properties of $g$-mode pulsations in \ltcnc . Errors on list digit(s) are
in parentheses. Column 5 provides the mode identifications from period spacings with n relative
to an arbitrary n$_o$. Column 6 provides the deviation from a linear period
spacing sequence and Column 7 lists frequency differences. Column 8 provides our best estimate
for the mode identification.
$^a$ indicates the frequency was Lorentzian fitted.}
\label{tabltcncg}
    \begin{tabular}{llcccccc}
            ID &   Freq & Per & Amp & Mode ($\Delta P$)& $\Delta P/P$& $\delta$Freq & Mode\\
            & $\mu$Hz & s & ppt     & $\ell$,n       &  \%       & $\mu$Hz & $n,\ell,m$\\ \hline
            fA &    111.889 (5) & 8937.42 (44) & 0.136 (9) & 1,34& -1.7& & 34,1,-\\
            fB & 122.438 (7) & 8167.38 (44) & 0.112 (9) & 1,31/2,54& 1.8/3.5& & 31,1,-\\
            fC &  135.682 (1) & 7370.15 (7) & 0.617 (9) & 1,28& -5.1& & 28,1,-\\
            fD & 145.686 (3) & 6864.07 (16) & 0.225 (9) & 1,26& 0.1& & 26,1,-\\
            fE$^a$ &  163.888 (1) & 6101.73 (31) & 0.497 (76) & 1,23& 5.9& 0.194 & 23,1,-1\\
            fF$^a$ & 164.082 (14) & 6094.51 (33) & 0.470 (76) & 1,23& 4.7& & 23,1,0\\
            fG & 170.731 (1) & 5857.17 (5) & 1.298 (14) & 1,22& 12.5& 0.166 & 22,1,-1\\
            fH &170.897 (1) & 5851.46 (3) & 1.855 (14) & 1,22& 10.3& &22,1,0\\
            fJ &  183.08 (2) & 5462.10 (6) & 0.377 (9) & 2,36& -0.3& &36,2,-\\
            fK & 199.187 (2) & 5020.40 (5) & 0.388 (9) & 1,19/2,33& -9.6/5.2& & 33,2,-\\
            fL &  211.669 (1) & 4724.35 (2) & 0.938 (9) & 2,31& 7.8& &31,2,-\\
            fM & 226.943 (1) & 4406.39 (2) & 0.870 (9) & 2,29& -4.2& &29,2,-\\
            fN & 233.582 (1) & 4281.15 (2) & 1.233 (18) & 1,16& 5.8& 0.732 & 16,1,-\\
            fP & 234.314 (1) & 4267.77 (2) & 1.514 (18) & 1,16/2,28& 0.7/3.3& 0.278 & 28,2,-1\\
            fQ &234.592 (2) & 4262.72 (4) & 0.704 (18) & 1,16/2,28& 1.3/0.0& 0.323 &28,2,0\\
            fR & 234.915 (2) & 4256.86 (3) & 0.959 (18) & 1,16/2,28& -3.5/-3.9&  &28,2,+1\\
            fS &243.478 (3) & 4107.14 (6) & 0.421 (18) & 2,27& -3.8& &27,2,-\\
            fT &248.189 (1) & 4029.19 (1) & 1.710 (18) & 1,15& 8.8& &15,1,-\\
            fU & 262.137 (8) & 3814.79 (12) & 0.177 (18) & 2,25& 1.3& &25,2,-\\
            fV & 266.019 (1) & 3759.13 (1) & 3.395 (18) & 1,14& 4.9&  &14,1,-\\      
            fW & 287.058 (11) & 3483.61 (13) & 0.390 (46) & 1,13& -1.2& 0.12 & 13,1,0\\
            fX & 287.178 (8) & 3482.16 (10) & 0.515 (46) & 1,13& -1.8& &13,1,+1\\
            fY & 311.849 (3) & 3206.69 (3) & 0.428 (18) & 1,12/2,21& -7.8/-4.1& &12,1,-\\ \hline
\end{tabular}
\end{table*}

\begin{table*}
\caption{Seismic properties of $p$-mode pulsations in \ltcnc .}
\label{tabltcncp}
    \begin{tabular}{llcccc}
ID &   Freq & Per & Amp& $\delta$Freq & Mode\\
   &  $\mu$Hz  & s & ppt & $\mu$Hz & $n,\ell ,m$ \\ \hline
f1 &2355.183 (9) & 424.5954 (17) & 0.0667 (79) &  &2,0,0\\
f2 &2370.535 (19) & 421.8457 (35) & 0.0328 (79) & 0.68  & 2,1,-1\\
f3 &2371.215 (12) & 421.7247 (21) & 0.0549 (79) & 0.64 & 2,1,0\\
f4 &2371.855 (12) & 421.6109 (22) & 0.0530 (79) &  & 2,1,+1\\
f5 &2402.737 (18) & 416.1920 (31) & 0.0349 (79) & 1.323 & 2,2,-2\\
f6 &2404.060 (16) & 415.9629 (27) & 0.0408 (79) & & 2,2,0\\
f7 &3131.563 (90) & 319.3294 (9) & 0.0750 (80) & 0.186 & 3,-,-\\
f8 &3131.749 (10) & 319.3104 (10) & 0.0620 (80) & & 3,-,-\\
f9 &3145.101 (15) & 317.9548 (15) & 0.0433 (79) & 0.9 & 3,-,-\\
f10 &3146.001 (11) & 317.8638 (11) & 0.0580 (79) & 1.406 & 3,2,-2\\
f11 &3147.407 (9) & 317.7219 (9) & 0.0740 (79) & 1.599 & 3,2,0\\
f12 &3149.006 (14) & 317.5605 (14) & 0.0467 (79) &  & 3,2,+2\\
f13 &3170.789 (15) & 315.3789 (15) & 0.0409 (79) & 0.958 & 3,-,-\\
f14 &3171.747 (15) & 315.2837 (14) & 0.0436 (79) & 0.62 & 3,-,0\\
f15 &3172.367 (16) & 315.2221 (16) & 0.0406 (79) & 3.733 & 3,-,+1\\
f16 &3176.100 (15) & 314.8515 (15) & 0.0409 (79) & 2.965 & 3,-,-\\
f17 &3179.065 (16) & 314.5579 (16) & 0.0414 (80) & 0.454 & 3,-,-\\
f18 &3179.519 (9) & 314.5130 (9) & 0.0738 (80) & & 3,-,-\\
f19 &3874.668 (16) & 258.0866 (11) & 0.0388 (79) & 2.035 & 4,0,0\\
f20 &3876.703 (16) & 257.9511 (11) & 0.0393 (79) & 1.213 & 4,2,-2\\
f21 &3877.916 (16) & 257.8705 (10) & 0.0404 (79) & & 4,2,0\\ \hline
\end{tabular}
\end{table*}

\begin{landscape}
\begin{table}
\label{hzglist}
\caption{Period list for \hzcnc\, $g$ modes.}
\begin{tabular}{|lcccc|cc|cc|cc|cc|cc|cc|} \hline
&   &   &  &  &  C5 & C5&  C16&  C16&  C18&  C18& S44&  S44 & S45 & S45&  S46 & S46 \\
ID & Freq &  $\sigma$Freq  & Period & Mode& $\sigma$F$_{\rm fit}$ & Amp & $\sigma$F$_{\rm fit}$ & Amp & $\sigma$F$_{\rm fit}$ & Amp & $\sigma$F$_{\rm fit}$ & Amp & $\sigma$F$_{\rm fit}$ & Amp & $\sigma$F$_{\rm fit}$ & Amp \\
        & [$\mu$Hz] & [$\mu$Hz] & [s] & $\ell,n$ & [$\mu$Hz] & [ppt] & [$\mu$Hz] &[ppt] & [$\mu$Hz] & [ppt] & [$\mu$Hz] & [ppt] & [$\mu$Hz] & [ppt] & [$\mu$Hz] & [ppt] \\
	fA & 84.496 & 0.062 & 11834.88 & --- & 0.009 & 0.21 & --- &  & 0.01 & 0.22 &--- & --- & --- & --- & --- & --- \\
	fB & 112.137 & 0.006 & 8917.70 & 1,34 & --- & --- & 0.007 & 0.20 & 0.01 & 0.28& --- & --- & --- & --- & --- & --- \\
	fC & 118.899 & 0.010 & 8410.50 & 1,32 & --- & --- & 0.006 & 0.43 & 0.005 & 0.46 &--- & --- & --- & --- & --- & --- \\
	fD & 126.700 & 0.031 & 7892.66 & 1,30 & --- & --- & --- & --- & --- & ---& --- & --- & --- & --- & 0.031 & 1.43 \\
	fE & 131.049 & --- & 7630.73 & 1,29 & 0.008 & 0.23 & --- &  & --- & ---& --- & --- & --- & --- & --- & --- \\
	fF & 137.322 & --- & 7282.15 & --- & --- & --- & 0.033 & 0.05 & --- & --- &--- & --- & --- & --- & --- & --- \\
	fG & 157.816 & 0.017 & 6336.48 & 1,24 & 0.012 & 0.15 & --- & 0.20 & 0.007 & 0.35 &--- & --- & --- & --- & --- & --- \\
	fH & 164.319 & 0.153 & 6085.74 & 1,23 & --- & --- & 0.026 & 0.07 & 0.009 & 0.29 &--- & --- & --- & --- & --- & --- \\
	fI & 171.125 & 0.018 & 5843.70 & 1,22 & --- & --- & 0.004 & 0.48 & 0.013 & 0.20 &--- & --- & --- & --- & --- & --- \\
fJ & 178.565 & 0.016 & 5600.21 & 1,21/2,37 & 0.001 & 2.07 & 0.001 & 2.25 & 0.002 & 1.46 & 0.038 & 1.43 & --- & --- & --- & --- \\
	fK & 187.448 & 0.040 & 5334.81 & 1,20 & --- &  & 0.005 & 0.34 & 0.015 & 0.18 &--- & --- & --- & --- & --- & --- \\
	fL & 196.171 & --- & 5097.59 & 1,19 & --- & 0.27 & --- & --- & --- & --- &--- & --- & --- & --- & --- & --- \\
	fM & 219.311 & 0.065 & 4559.74 & 1,17/2,30 & 0.003 & 0.65 & 0.002 & 0.89 & 0.004 & 0.68& --- & --- & --- & --- & --- & --- \\
	fN & 226.727 & 0.066 & 4410.60 & 2,29 & --- &  & --- & 0.60 & 0.005 & 0.53 &--- & --- & --- & --- & --- & --- \\
	fO & 232.692 & 0.019 & 4297.52 & 1,16 & 0.001 & 1.56 & 0.001 & 1.23 & 0.002 & 1.21 &--- & --- & 0.058 & 0.94 & 0.052 & 0.87 \\
	fP & 234.671 & 0.012 & 4261.28 & 2,28 & 0.001 & 1.39 & 0.002 & 0.86 & 0.005 & 0.58 &--- & --- & 0.068 & 0.08 & --- & --- \\
	fQ & 243.360 & 0.124 & 4109.14 & 2,27 & --- & 3.25 & --- & 2.50 & --- & 1.25& 0.013 & 4.26 & 0.036 & 1.51 & 0.027 & 1.65 \\
	fR & 249.042 & 0.030 & 4015.39 & 1,15 & --- & 1.36 & --- & 1.19 & 0.002 & 1.34 &0.034 & 1.59 & --- & --- & 0.031 & 1.46 \\
	fS & 251.884 & 0.056 & 3970.08 & 2,26 & --- & 0.32 & 0.005 & 0.38 & 0.006 & 0.42 &--- & --- & --- & --- & --- & --- \\
	fT & 254.578 & --- & 3928.07 & --- & --- &  & --- & 0.00 & --- & --- &--- & --- & --- & --- & --- & --- \\
	fU & 254.758 & 0.026 & 3925.30 & --- & 0.003 & 0.56 & 0.003 & 0.59 & --- & 0.50 &--- & --- & --- & --- & --- & --- \\
	fV & 266.225 & 0.041 & 3756.23 & 1,14 & --- & 1.11 & 0.001 & 1.52 & 0.001 & 2.61 &0.045 & 1.32 & 0.036 & 1.50 & 0.031 & 1.44 \\
	fW & 273.170 & 0.064 & 3660.72 & 2,24 & 0.004 & 0.53 & --- & 0.44 & 0.005 & 0.42 &--- & --- & --- & --- & --- & --- \\
	fX & 287.259 & 0.039 & 3481.18 & 1,13 & --- & 0.85 & 0.001 & 2.67 & 0.002 & 1.54 &0.05 & 1.01 & --- & --- & 0.041 & 1.11 \\
	fY & 290.544 & --- & 3441.82 & --- & --- & 0.26 & --- & --- & --- & --- &--- & --- & --- & --- & --- & --- \\
	fZ & 296.253 & --- & 3375.49 & 2,22 & --- & 0.16 & --- & --- & --- & --- &--- & --- & --- & --- & --- & --- \\
	fAA & 311.564 & 0.006 & 3209.62 & 1,12/2,21 & 0.001 & 1.70 & 0.001 & 2.21 & 0.002 & 1.18 0.024 & 2.32 & 0.040 & 1.36 & 0.049 & 0.91 &--\\
	fBB & 337.790 & --- & 2960.42 & --- & 0.013 & 0.13 & --- & --- & --- & --- &--- & --- & --- & --- & --- & --- \\
	fCC & 358.780 & --- & 2787.22 & --- & 0.015 & 0.15 & --- & --- & --- & --- &--- & --- & --- & --- & --- & --- \\
	fDD & 469.422 & --- & 2130.28 & --- & --- & --- & --- & --- & --- & --- &--- & --- & --- & --- & 0.046 & 0.98 \\
	fEE & 490.612 & --- & 2038.27 & --- & --- & --- & --- & --- & --- & --- &--- & --- & --- & --- & 0.052 & 0.87 \\
\end{tabular}
\end{table}
\end{landscape}

\begin{figure*}
\centerline{\psfig{figure=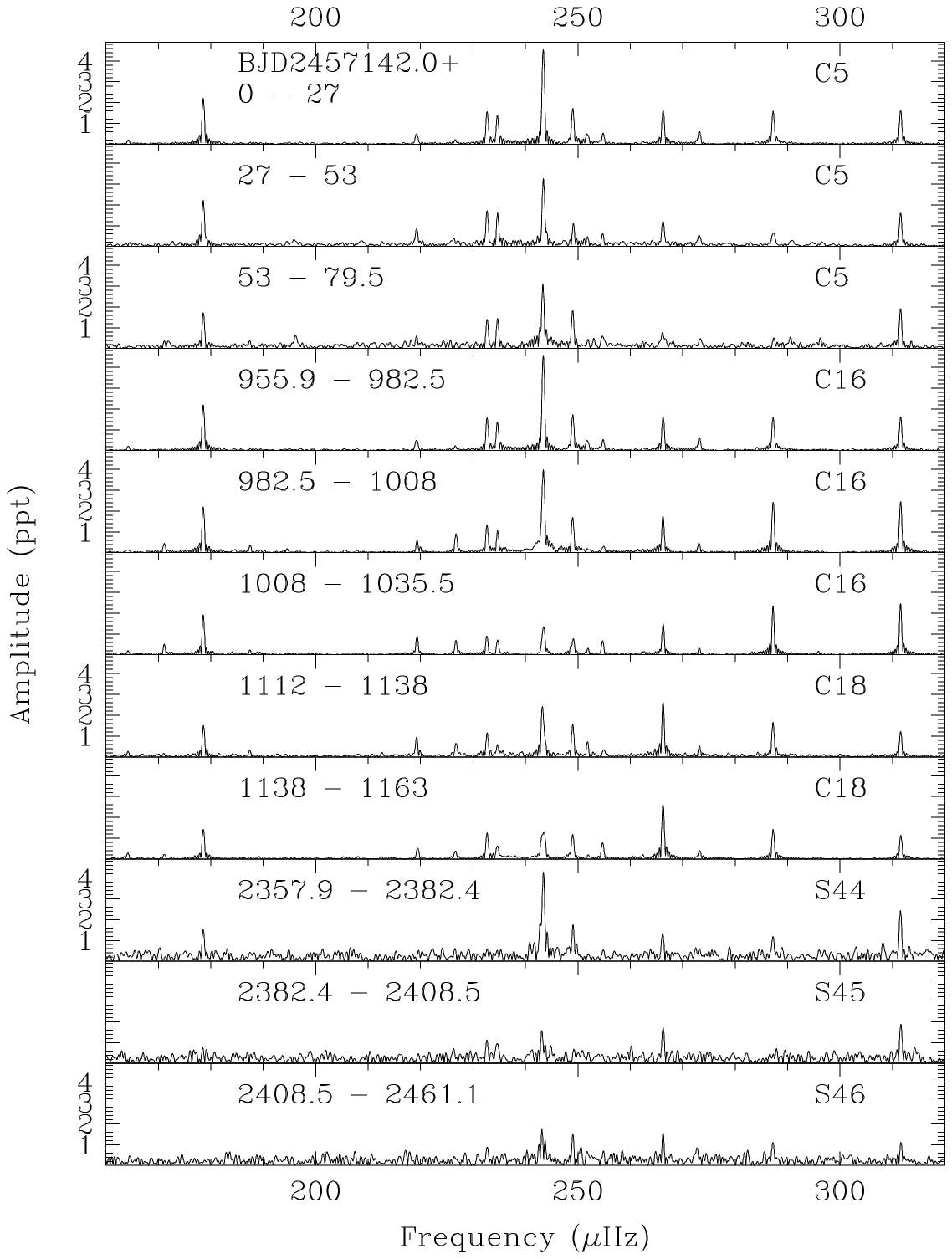,width=6.5in}}
    \caption{FT of \hzcnc\, divided into TESS sector-length pieces. Dates are labelled on the left of
        each panel and K2 Campaigns or TESS sectors labelled on the right.}
    \label{hzFTMO}
\end{figure*}

\begin{figure*}
\centerline{\psfig{figure=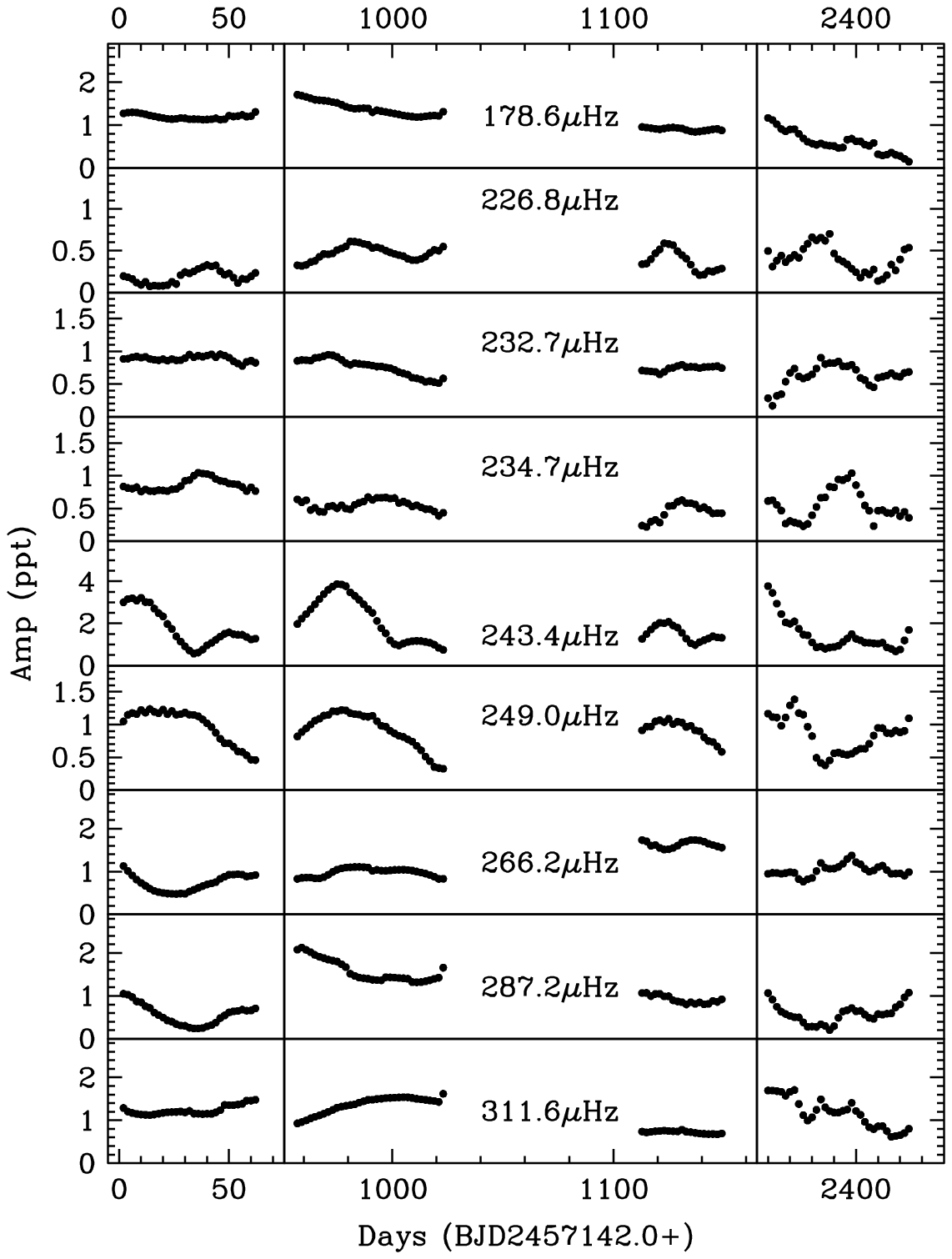,width=\textwidth}}
    \caption{Amplitudes of the nine highest-amplitude pulsations in \hzcnc\, corresponding to the SFT
sampling of Fig.\,\ref{hzSFT}.}
    \label{hzAmp}
\end{figure*}

\begin{table*}
\label{hzplist}
\caption{Period list for \hzcnc\, $p$ modes. Frequencies which were Lorentzian fitted do not have listed frequency
uncertainties.}
\begin{tabular}{|lcccc|cc|cc|cc|c|} \hline
        &   &   &  &   C5 & C5&  C16&  C16&  C18&  C18 & $n$\\
ID  & Freq & $\sigma$Freq & Period & $\sigma$Fit & Amp & $\sigma$Fit & Amp & $\sigma$Fit & Amp &  \\
  & [$\mu$Hz] & [$\mu$Hz] & [s] & [$\mu$Hz] & [ppt] & [$\mu$Hz] & [ppt] & [$\mu$Hz] & [ppt] &  \\
f1 & 1056.904 & 0.028 & 946.160 & --- & --- & 0.014 & 0.094 & 0.016 & 0.109 & $0$  \\
f2 & 1070.246 & --- & 934.365 & --- & --- & 0.029 & 0.046 & --- & --- &   $0$\\
f3 & 1175.660 & 0.057 & 850.586 & --- & --- & --- & 0.057 & --- & 0.036 &   $0$\\
f4 & 1212.170 & --- & 824.967 & --- & --- & --- & --- & 0.035 & 0.049 &  $0$ \\
f5 & 2575.917 & 0.028 & 388.211 & 0.030 & 0.052 & --- & --- & 0.03 & 0.057 &  $4$ \\
f6 & 2585.454 & 0.251 & 386.779 & 0.015 & 0.100 & --- & 0.033 & --- & 0.050 &   $4$\\
f7 & 2586.010 & --- & 386.696 & --- & 0.032 & --- & --- & --- & --- &   $4$\\
f8 & 2629.740 & 0.033 & 380.266 & 0.036 & 0.044 & --- & --- & 0.02 & 0.085 &  $4$ \\
f9 & 2630.800 & --- & 380.113 & --- & 0.048 & --- & --- & --- & --- &   $4$\\
f10 & 3427.457 & 0.181 & 291.762 & 0.036 & 0.043 & 0.027 & 0.049 & --- & --- &  $5$ \\
f11 & 3428.015 & 0.106 & 291.714 & --- & --- & --- & 0.043 & --- & 0.063 &  $5$ \\
f12 & 3438.900 & --- & 290.791 & --- & --- & --- & --- & --- & 0.084 &   $5$\\
f13 & 3441.076 & 0.060 & 290.607 & --- & 0.037 & 0.016 & 0.082 & --- & 0.054 &  $5$ \\
f14 & 3443.523 & 0.087 & 290.400 & --- & 0.040 & 0.021 & 0.063 & --- & 0.050 &   $5$\\
f15 & 3459.820 & --- & 289.032 & --- & --- & --- & --- & --- & 0.043 &   $5$\\
f16 & 3462.250 & --- & 288.830 & --- & --- & --- & --- & --- & 0.038 &   $5$\\
f17 & 3463.429 & 0.013 & 288.731 & --- & 0.045 & --- & --- & 0.046 & 0.037 &  $5$\\
f18 & 8289.450 & --- & 120.635 & --- & --- & --- & 0.038 & --- & --- & 9  \\ \hline
\end{tabular}
\end{table*}

\begin{table*} \label{pgnlist} \caption{Period list for \pgn. }
\begin{tabular}{|lcccc|cc|cc|cc|c|} \hline
 ID & freq  & per  & amp  & Mode & $\Delta$P \\
  & $\mu$Hz & s & ppt  & $\ell,n,m$ &  \% \\  \hline
        f01 & 101.248 (8) & 9876.74 (77) & 0.092 (9) & 1,38 & 12.0 \\
        f02 & 110.271 (6) & 9068.54 (48) & 0.124 (9) & 1,35 & -7.6 \\
        f03 & 132.099 (9) & 7570.08 (49) & 0.085 (9) & 1,29 & -0.2 \\
        f04 & 147.346 (6) & 6786.76 (30) & 0.113 (9) & 1,26,0 & -9.9 \\
        f05 & 148.068 (8) & 6753.65 (34) & 0.097 (9) & 1,26,+1 & \\
        f06 & 180.510 (10) & 5539.87 (2) & 1.086 (10) & 1,21/2,37 & -3.0/0.7 \\
        f07 & 207.892 (10) & 4810.18 (22) & 0.075 (9) & 1,18/2,32 & 8.5/0.9 \\
        f08 & 220.268 (3) & 4539.92 (6) & 0.265 (10) & 1,17,0 & 1.6 \\
        f09 & 220.969 (11) & 4525.53 (22) & 0.075 (10) & 1,17,+1/2,30,- & -4.1/6.0 \\
        f10 & 275.667 (10) & 3627.56 (13) & 0.079 (10) & 2,24 & -9.1 \\
        f11 & 288.383 (9) & 3467.62 (11) & 0.088 (10) & & \\
        f12 & 311.111 (8) & 3214.29 (8) & 0.101 (10) & 2,21 & 7.9 \\
        f13 & 337.973 (4) & 2958.82 (3) & 0.213 (10) & 4,36 & \\
        f14 & 343.380 (6) & 2912.22 (5) & 0.135 (10) & 2,19 & 1.0 \\
        f15 & 360.643 (8) & 2772.82 (6) & 0.098 (10) & 1,10/2,18/3,26 & 2.8/5.5/-7.9 \\
        f16 & 372.363 (2) & 2685.55 (2) & 0.349 (10) & 3,25 & \\
        f17 & 380.627 (2) & 2627.24 (1) & 0.379 (10) & 2,17/4,32 & 5.8/-8.3 \\
        f18 & 410.240 (6) & 2437.60 (4) & 0.133 (10) &   & \\
        f19 & 592.519 (12) & 1687.71 (3) & 0.061 (9) & 4,20,-1 & 16.8 \\
        f20 & 593.795 (15) & 1684.08 (4) & 0.049 (9) & 4,20,0 & 12.2 \\
        f21 & 748.435 (17) & 1336.12 (3 & 0.042 (9) & 3,12,0 & 0.5 \\
        f22 & 749.750 (13) & 1333.78 (2) & 0.055 (9) & 3,12,+1 & -1.8 \\
        f23 & 923.620 (12) & 1082.70 (1) & 0.062 (9) & -,-,-2 & - \\
        f24 & 924.932 (15) & 1081.16 (2) & 0.047 (9) & -,-,-1 &   \\
        f25 & 926.197 (15) & 1079.68 (2) & 0.047 (9) & -,-,0 &   \\
        f26 & 930.075 (13) & 1075.18 (1) & 0.057 (9) & -,-,+3 &   \\
        f27 & 970.734 (9) & 1030.15 (1 & 0.079 (9) & 4,11,-3 & 5.3 \\
        f28 & 972.053 (9) & 1028.75 (1) & 0.076 (9) & 4,11,-2 & 3.6 \\
        f29 & 973.352 (13) & 1027.38 (1) & 0.055 (9) & 4,11,-1 & 1.9 \\
        f30 & 974.616 (8) & 1026.05 (1) & 0.086 (9) & 4,11,0 & 0.2 \\
        f31 & 1014.006 (10) & 986.19 (1) & 0.071 (9) & & \\       \hline
\end{tabular}
\end{table*}

\begin{table*}
\label{pbFlist}
        \caption{Period list for \pb in the same format as Table\,\ref{tabltcncg}.
$^{\dagger}$ These frequencies are closer than the 1.5/T resolution of $0.22\mu$Hz
yet were independently prewhitened. $^a$t01 is only detected in TESS data.}
\begin{tabular}{lcccccc} \hline
ID & Frequency & Period  & amp & ell, n & $\Delta$P/P & $\delta\nu$ \\
   & [$\mu$Hz] & [s]  & [ppt] & & [\%] & [$\mu$Hz] \\ \hline
f01 & 108.297 (17) & 9233.9 (1.4) & 2.84 (0.59) & 1,35 & 4.3 & 0.823\\
f02 & 109.120 (16) & 9164.2 (1.3) & 3.09 (0.59) & - & - & -\\
f03 & 114.822 (16) & 8709.1 (1.2) & 3.08 (0.59) & 1,33 & 0.6 & -\\
f04 & 117.894 (15) & 8482.2 (1.1) & 3.66 (0.66) & 1,32 & 12.5 & 0.521\\
f05 & 118.415 (09) & 8444.8 (0.6) & 6.18 (0.67) & 1,32/2,56 & -2.0/-2.2 & 0.524\\
f06 & 118.939 (13) & 8407.7 (0.9) & 4.23 (0.66) & 1,32 & -16.5 & -\\
f07 & 125.765 (08) & 7951.4 (0.5) & 6.50 (0.67) & 1,30 & 6.4 & 0.393\\
f08 & 126.158 (08) & 7926.6 (0.5) & 6.70 (0.66) & 1,30 & -3.2 & -\\
f09 & 163.572 (12) & 6113.5 (0.5) & 4.42 (0.66) & 1,23 & -7.0 & -\\
f10 & 170.681 (19) & 5858.9 (0.7) & 2.82 (0.66) & 1,22 & -5.9 & -\\
f11 & 186.297 (04) & 5367.8 (0.1) & 12.59 (0.66) & 1,20 & 3.5 & -\\
f12 & 193.470 (16) & 5168.8 (0.4) & 2.65 (0.54) & 2,34 & -4.9 & -\\
f13 & 210.722 (19) & 4745.6 (0.4) & 2.31 (0.54) & 2,31 & 10.5 & -\\
f14$^{\dagger}$ & 217.825 (37) & 4590.8 (0.8) & 4.46 (1.86) & 1,17/2,30 & 1.9/6.5 & 0.128\\
f15$^{\dagger}$ & 217.953 (60) & 4588.2 (1.3) & 3.24 (1.71) & 1,17/2,30 & 0.8/4.7 & 0.193\\
f16$^{\dagger}$ & 218.146 (20) & 4584.1 (0.4) & 3.20 (0.62) & 1,17/2,30 & -0.8/1.9 & -\\
f17 & 229.697 (16) & 4353.6 (0.3) & 2.87 (0.54) & 1,16 & 9.7 & 0.304\\
f18 & 230.001 (16) & 4347.8 (0.3) & 2.90 (0.54) & 1,16 & 7.5 & -\\
f19$^{\dagger}$ & 245.644 (16) & 4070.9 (0.3) & 6.99 (1.15) & 1,15 & 0.0 & 0.128\\
f20$^{\dagger}$ & 245.772 (26) & 4068.8 (0.4) & 4.30 (1.15) & 1,15 & -0.8 & -\\
t01$^a$ & 259.028 (27) & 3860.6 (0.4) & 2.05 (0.46) & - & - & -\\
f21 & 262.426 (09) & 3810.6 (0.1) & 5.71 (0.58) & 1,14 & -1.0 & 0.285\\
f22 & 262.711 (15) & 3806.5 (0.2) & 3.24 (0.58) & 1,14 & -2.6 & -\\
f23 & 281.837 (11) & 3548.1 (0.1) & 4.26 (0.61) & 1,13/2,23 & -2.9/5.4 & 0.298\\
f24 & 282.135 (17) & 3544.4 (0.2) & 3.00 (0.60) & 1,13/2,23 & -4.4/2.9 & -\\
f25$^{\dagger}$ & 308.536 (13) & 3241.1 (0.1) & 4.24 (0.59) & 2,21 & -1.0 & 0.151\\
f26$^{\dagger}$ & 308.687 (14) & 3239.5 (0.1) & 4.01 (0.59) & 2,21 & -2.1 & -\\ \hline
\end{tabular}
\end{table*}

\begin{table}
\caption{\label{specobs}\small
Log of spectroscopic observations. The first column lists the mid-exposure times in UT. Radial velocities are in km/s.}
\centering
\renewcommand{\arraystretch}{1.0}


\begin{tabular}{l@{ ~}c@{ }c@{ }r@{ ~}r@{ }r}

\multicolumn{6}{c}{EPIC 220376019 == PG 0101+039} \\

 DATE-AVG             &t$_{\rm expo}$& BJD --2450000  &   S/N                \\
2016-12-16T20:22:57.0 &  150   & 7739.35136    & 141.6                \\
2017-01-24T20:15:57.4 &  150   & 7778.34265    & 119.9                \\
2017-08-21T04:53:13.8 &  150   & 7986.70753    &  97.8                \\
\\
\\
\multicolumn{6}{c}{EPIC 211433013 == PG 0907+123  == LT Cnc} \\

 DATE-AVG             &  t$_{\rm expo}$ & BJD --2450000  &   S/N &      RV  & $\sigma_{RV}$\\
2018-10-21T05:02:07.3 &  400   & 8412.70798    &  94.9 &    65.9  &   3.6    \\
2018-11-07T05:20:06.4 &  400   & 8429.72214    & 119.6 &   115.5  &   6.4    \\
2019-01-08T02:49:32.6 &  400   & 8491.62272    & 146.7 &   107.2  &   7.0    \\
2019-01-10T01:30:51.0 &  400   & 8493.56817    & 159.1 &     5.9  &   2.1    \\
2019-01-15T01:38:36.3 &  400   & 8498.57376    & 113.8 &    61.2  &   3.2    \\
2019-01-21T00:11:55.6 &  400   & 8504.51377    &  95.8 &    62.4  &   3.9    \\
2019-01-23T01:46:55.6 &  400   & 8506.57979    &  68.3 &     1.3  &   5.9    \\
2019-01-24T01:47:53.6 &  400   & 8507.58049    &  93.1 &    47.1  &   5.2    \\
2019-01-25T04:24:51.7 &  400   & 8508.68952    &  97.2 &   102.1  &   7.3    \\
2019-01-26T01:47:07.3 &  400   & 8509.58000    & 118.8 &   116.8  &   5.1    \\
2019-02-05T00:40:53.7 &  400   & 8519.53411    &  70.0 &    26.6  &  10.9    \\
2019-02-10T22:59:04.8 &  400   & 8525.46339    &  95.6 &    30.5  &   6.2    \\
2019-02-12T00:49:12.9 &  400   & 8526.53986    & 103.3 &    90.8  &   6.6    \\
2019-02-21T00:02:18.1 &  400   & 8535.50712    & 114.8 &    42.9  &   6.7    \\
2019-02-22T21:24:33.0 &  400   & 8537.39753    &  67.8 &     3.9  &   7.7    \\
2019-03-07T01:51:12.6 &  400   & 8549.58224    & 108.2 &     3.0  &   4.6    \\
2019-03-16T00:46:22.8 &  400   & 8558.53673    &  94.0 &   112.9  &   5.8    \\
\\
\\
\multicolumn{6}{c}{EPIC 211765471 == HZ Cnc} \\

  DATE-AVG           &t$_{\rm expo}$ & BJD --2450000 &    S/N &  RV  &   $\sigma_{RV}$   \\
2017-01-17T03:15:02.5 &  400  &  7770.64096  &   108.2  &   47.6 &    4.7    \\
2017-01-18T01:12:36.6 &  400  &  7771.55596  &    83.5  &   51.3 &    5.8    \\
2017-03-14T22:35:28.7 &  400  &  7827.44550  &   102.1  &   50.3 &    5.4    \\
2017-03-22T23:14:28.8 &  400  &  7835.47200  &    64.5  &   52.6 &    7.1    \\
2017-03-25T23:41:19.7 &  400  &  7838.49041  &   109.4  &   27.4 &    5.0    \\
2017-04-03T22:30:05.7 &  400  &  7847.44019  &    84.3  &   12.9 &    5.2    \\
2017-04-20T23:01:19.3 &  400  &  7864.46030  &    92.5  &   33.5 &    4.2    \\
2017-04-21T21:03:27.3 &  400  &  7865.37835  &   107.4  &   37.0 &    3.2    \\
2017-04-22T21:14:37.0 &  400  &  7866.38601  &    89.5  &   29.3 &    5.9    \\
2017-04-23T20:43:36.6 &  400  &  7867.36438  &    83.8  &   16.4 &    5.8    \\
2017-04-30T20:58:03.1 &  400  &  7874.37372  &   105.7  &   17.8 &    8.7    \\
2018-01-10T02:56:19.4 &  400  &  8128.62775  &    88.8  &   22.1 &    5.7    \\
2018-01-20T04:24:05.6 &  400  &  8138.68899  &    94.8  &   59.5 &   11.1    \\
2018-01-27T21:59:23.6 & 259.9 &  8146.42194  &    34.4  &   15.0 &   11.5    \\
2018-01-27T22:37:24.5 & 497.2 &  8146.44834  &    61.7  &   14.6 &    9.5    \\
2018-02-13T00:00:14.0 &  400  &  8162.50574  &    75.5  &   65.0 &   12.2    \\
2018-02-15T00:13:24.0 &  400  &  8164.51484  &   102.1  &   71.5 &    9.0    \\
2018-03-23T23:38:11.6 &  400  &  8201.48843  &    68.5  &   16.5 &    7.4    \\
2018-03-30T01:22:20.1 &  400  &  8207.56026  &    39.2  &    1.9 &   13.7    \\
2018-04-15T21:26:53.4 &  400  &  8224.39524  &    78.2  &   61.6 &   12.9    \\
2018-04-25T23:00:52.4 &  400  &  8234.45954  &    78.6  &    1.2 &   11.2    \\
2018-04-28T20:50:23.3 &  400  &  8237.36864  &    74.7  &   13.2 &    7.9    \\
2018-04-30T20:54:59.0 &  400  &  8239.37163  &    75.8  &   19.1 &    6.5    \\
2018-05-29T21:07:07.4 &  400  &  8268.37731  &    44.5  &   28.6 &    7.2    \\

\end{tabular}
\end{table}

\begin{table}
\caption{
\small
Log of spectroscopic observations.}
\centering
\renewcommand{\arraystretch}{1.0}

\begin{tabular}{l@{ ~}c@{ }c@{ }r@{ ~}r@{ }r}
\multicolumn{6}{c}{EPIC 211437457 == PG0902+124} \\

 DATE-AVG       &      t$_{\rm expo}$ & BJD -2450000   &  S/N  &    RV  &   $\sigma_{RV}$\\
2018-11-07T05:34:43.2 &  600 &     8429.73240  &   118.9  &  -40.9  &   7.2    \\
2019-01-03T07:09:22.5 &  600 &     8486.80297  &   112.5  &  -19.7  &   4.1    \\
2019-01-08T02:35:01.0 &  600 &     8491.61269  &    99.9  &  -45.0  &   7.1    \\
2019-01-10T01:16:40.0 &  600 &     8493.55837  &   104.9  &   17.9  &   5.7    \\
2019-01-15T01:24:40.8 &  600 &     8498.56413  &    76.9  &  -44.7  &   6.5    \\
2019-01-20T23:57:49.1 &  600 &     8504.50400  &    73.9  &   14.0  &   6.3    \\
2019-01-23T01:32:06.0 &  600 &     8506.56952  &    42.6  &  -48.6  &  11.4    \\
2019-01-24T01:32:24.9 &  600 &     8507.56976  &    53.2  &  -17.0  &   4.8    \\
2019-01-25T04:10:36.1 &  600 &     8508.67964  &    74.0  &    0.6  &   5.3    \\
2019-01-26T01:32:59.4 &  600 &     8509.57020  &   115.0  &  -22.6  &   4.9    \\
2019-02-05T00:56:42.5 &  600 &     8519.54509  &    61.6  &   21.2  &   9.0    \\
2019-02-10T22:43:14.7 &  600 &     8525.45238  &    69.4  &   -8.9  &   4.5    \\
2019-02-12T00:34:37.7 &  600 &     8526.52972  &    80.4  &   33.1  &   5.8    \\
2019-02-20T23:47:09.2 &  600 &     8535.49657  &    83.5  &  -25.4  &   5.0    \\
2019-02-22T21:10:28.7 &  600 &     8537.38772  &    67.9  &   23.9  &   7.3    \\
2019-03-07T01:35:05.5 &  600 &     8549.57099  &    90.9  &   19.8  &   6.5    \\
2019-03-15T23:52:02.5 &  600 &     8558.49892  &    82.3  &  -49.2  &   7.8    \\
2019-03-23T22:20:39.6 &  600 &     8566.43493  &   101.5  &    7.2  &   5.1    \\
\\
\\
\multicolumn{6}{c}{EPIC 220188903 == PB 6373} \\

 DATE-AVG       &      t$_{\rm expo}$&  BJD --2450000 &    S/N          \\
2017-01-04T20:01:16.6 &  450   &   7758.33463  &    43.1            \\
2017-08-20T04:38:08.5 &  450   &   7985.69681  &    54.2            \\
2017-08-21T05:21:58.0 &  450   &   7986.72732  &    42.1            \\
2017-12-06T00:37:53.7 &  450   &   8093.52954  &    38.6            \\
2017-12-31T20:09:20.6 &  450   &   8119.34066  &    53.4            \\
2018-01-09T21:41:52.6 &  450   &   8128.40401  &    65.6            \\
\end{tabular}
\end{table}

\end{document}